\documentclass[11pt,a4paper]{article}
\usepackage[margin=2.5cm]{geometry}
\usepackage{amsmath,amssymb,amsthm,bm,graphicx}
\usepackage[T1]{fontenc}\usepackage{lmodern}
\usepackage[colorlinks=true,linkcolor=blue,citecolor=blue,urlcolor=blue]{hyperref}
\usepackage{booktabs,multirow,enumitem,xcolor,microtype,caption,subcaption}
\usepackage{parskip,titlesec,natbib,fancyhdr,tcolorbox,array,tablefootnote}
\tcbuselibrary{skins,breakable}

\newtheorem{definition}{Definition}[section]

\newtheorem{remark}{Remark}[section]

\newcommand{\sigS}{\sigma_{\!\mathrm{s}}}
\newcommand{\beff}{\bm{b}_{\mathrm{eff}}}
\newcommand{\ceff}{\bm{c}_{\mathrm{eff}}}
\newcommand{\Var}{\mathrm{Var}}

\newtcolorbox{algobox}[2][]{colback=gray!4,colframe=black!55,
  fonttitle=\small\bfseries,title=#2,breakable,#1}

\titleformat{\section}{\large\bfseries}{\thesection}{0.7em}{}
\titleformat{\subsection}{\normalsize\bfseries}{\thesubsection}{0.6em}{}
\title{\textbf{Variational Quantum Conditional Boltzmann Machines for Time-Series Forecasting: Architectures, Symmetric Hyperparameter Evaluation, and a Nonlinear Benchmark}}
\author{
  Gerhard Hellstern\\
  \textit{Centre of Finance}\\
  \textit{DHBW Stuttgart}\\
  Stuttgart, Germany\\
  gerhard.hellstern@dhbw-stuttgart.de
  \and
  Danyal Maheshwari\\
  \textit{Centre of Finance}\\
  \textit{DHBW Stuttgart}\\
  Stuttgart, Germany\\
  danyal.maheshwari@dhbw-stuttgart.de
  \and 
  Martin Zaefferer\\
  \textit{Zentrum für Digitale Innovationen}\\
  \textit{DHBW Ravensburg}\\
  Ravensburg, Germany\\
  zaefferer@dhbw-ravensburg.de
  \and
  Martin Braun\\
  \textit{DATEV eG}\\
  Nürnberg, Germany\\
  Martin.Braun@datev.de
  \and
  Tanja Döhler\\
  \textit{DATEV eG}\\
  Nürnberg, Germany\\
  Tanja.Doehler@datev.de
}

\date{July 2026}

\begin{document}
\maketitle\thispagestyle{empty}

\begin{abstract}
In this study, we developed and evaluated four conditional energy-based forecasting architectures a
classical Gaussian--Bernoulli CRBM, a hybrid quantum-classical QCRBM, a full-register
QQRBM, and a lag-feature QFeatureQRBM with complete derivations of their conditional
distributions, Contrastive-Divergence gradients, and hybrid training, bridging the
energy-based formulation and the implementation-level quantum computation.  Unlike
prior comparisons, our evaluation enforces \emph{symmetric hyperparameter
optimisation}: classical and quantum-specific hyperparameters receive an equally
thorough grid search across thirteen structured experiments.  We test on two data
classes, a Gaussian-process dataset (GP) generated with real financial data and 
the input-driven NARMA-10 nonlinear benchmark.

Across both regimes we find \emph{no systematic evidence of a quantum advantage at the available
sample size}: no quantum architecture improves on the best classical baseline.  The
fully quantum QQRBM and QFeatureQRBM are significantly \emph{worse}, whereas the hybrid QCRBM is statistically
indistinguishable from the strongest classical CRBM on both datasets.  A power analysis bounds this
null result: at $n{=}12$ only medium-to-large effects are detectable, so small advantages cannot be excluded.  An
iso-parameter (matched-budget) comparison reaches the same conclusion the classical
CRBM is lowest at three of the four budgets and no CRBM-vs-QCRBM difference is
significant at any budget.

\end{abstract}

\tableofcontents\newpage

\section{Introduction}\label{sec:intro}

Time-series forecasting predicting future observations from a sequence of
past values is a fundamental challenge in machine learning with applications
spanning finance, energy systems, climate modelling, and physiology.
Among probabilistic approaches, \emph{energy-based models} (EBMs) assign to
each configuration a scalar energy, inducing a Boltzmann distribution
$p_\theta(\bm{v}) = e^{-E_\theta(\bm{v})}/Z_\theta$ that can be learned from
data without specifying the normalising constant $Z_\theta$ in closed form.
The Restricted Boltzmann Machine (RBM) \citep{hinton2002cd} is the canonical EBM
with tractable inference: its bipartite structure between visible variables
$\bm{v}$ and binary hidden units $\bm{h}$ yields exact conditional distributions
$p(\bm{h}|\bm{v})$ and $p(\bm{v}|\bm{h})$, enabling efficient Contrastive
Divergence (CD) training \citep{hinton2002cd,hinton2010guide}.
Taylor and Hinton's Conditional RBM (CRBM) \citep{taylor2009fcrbm} extends this
framework to sequential data by conditioning the effective biases on a context
window $\bm{u}$ of past observations, encoding temporal dynamics in
$p(\bm{v}_t|\bm{u}_t)$ and enabling autoregressive multi-step forecasting.

Despite their theoretical appeal, classical CRBMs face an intrinsic representational
ceiling: the hidden--visible interaction is bilinear, so model capacity scales with
the number of hidden units $H$ rather than with higher-order combinations of inputs.
On tasks with strong nonlinear memory, such as the NARMA-10 benchmark
\citep{atiya2000narma}, whose response depends nonlinearly on a 10-step
recurrence, this bilinear bottleneck becomes the binding constraint on
forecasting accuracy, and increasing $H$ yields only incremental improvements
at proportionally growing parameter cost.

Quantum computing offers a principled route to richer feature representations
within the Boltzmann framework. Parameterized quantum circuits (PQCs) can encode
classical inputs into an exponentially large Hilbert space via amplitude encoding,
process them with entangling layers, and project back to real-valued features via
Pauli-$Z$ expectation values.  The resulting feature maps are nonlinear functions
of the input that cannot be expressed by the classical bilinear hidden field at
comparable parameter count \citep{zoufalVQBM2021,cerezo2021vqa}.  This suggests
a natural hybrid design: embed a PQC inside the CRBM to provide input-dependent
nonlinear corrections to the hidden-unit activations, while retaining the tractable
Gaussian visible structure and CD training algorithm. The question is whether this
potential translates into measurable predictive improvement and under which
conditions.

Two open problems have prevented a definitive answer.  The first is
\emph{methodological}: comparative QML studies routinely tune classical model
parameters extensively while fixing quantum-specific parameters such as the circuit
scale $\alpha$ and the quantum learning rate $\eta_Q$ at nominal defaults.
Bowles, Ahmed, and Schuld \citep{bowles2024benchmarking} identify this asymmetric
hyperparameter optimisation as a primary source of misleading conclusions: a quantum
model evaluated at miscalibrated defaults may appear inferior not because of
architectural limitations, but because it is never properly trained.
The second problem is \emph{scientific}: existing hybrid quantum--classical studies
evaluate on a single data regime or omit multiple-comparison correction, making it
impossible to determine whether reported advantages are data-regime-specific or
genuinely general.  On near-linear data, the nonlinear quantum features may add
noise rather than signal; on strongly nonlinear tasks, they may provide exactly the
inductive bias that the classical CRBM lacks.  These two regimes demand separate
controlled evaluation.

This paper addresses both problems simultaneously.  We make four contributions.

\begin{enumerate}[leftmargin=*]
\item \textbf{Comprehensive architecture derivations.}
  We present four conditional energy-based architectures with complete derivations
  of energy functions, CD gradient formulas, pathwise hybrid training procedures,
  and autoregressive forecasting mechanisms: a classical Gaussian--Bernoulli CRBM,
  a hybrid QCRBM (quantum correction to hidden logits), a full-register QQRBM
  (three-register entanglement), and a lag-feature QFeatureQRBM (compact temporal
  feature extractor).  The derivations establish a clear taxonomy of the three
  design choices for where quantum computation enters the Boltzmann structure.

\item \textbf{Fully symmetric hyperparameter optimisation.}
  All twelve hyperparameters across all model families classical ($H$, $k$,
  $\sigma$, $\eta_\mathrm{cl}$, $\lambda_\mathrm{wd}$) and quantum ($\alpha$,
  $\eta_Q$, $k_Q$, $\lambda_a$, $n_\mathrm{layers}$, $L_Q$), are searched by
  validation RMSE with equal budget concerning the hyperparameters.  This is the primary methodological
  contribution and, to our knowledge, the most complete symmetric evaluation of this model family to date.

\item \textbf{Two-regime controlled evaluation.}
  Two independently generated series per regime are evaluated on a 
  Gaussian-process benchmark (GP) and the NARMA-10 nonlinear benchmark, with a pairwise t-test and Holm--Bonferroni correction \citep{holm1979simple} applied to all simultaneous significance tests.  The two-regime design directly tests the hypothesis that quantum advantage is data-dependent.

\item \textbf{Iso-parameter (matched-budget) comparison.}
  Experiment~10 constructs matched model pairs evaluated at four feasible budget
  points $P\in\{84,120,156,192\}$ (QCRBM cannot be matched below $P{=}84$
  because amplitude encoding forces $2^{H}\ge V{+}U$), separating the contribution
  of quantum computation from the confounding effect of parameter count.
\end{enumerate}

The principal finding is that, under symmetric hyperparameter optimisation,
no quantum architecture improves on the best classical baseline on either
data regime. i.e.\ no evidence of a quantum advantage at the available sample
size ($n{=}12$; only medium-to-large effects $d\gtrsim d_\mathrm{min}{\approx}0.89$
are detectable).  On NARMA-10, the strongest classical model is CRBM$(3H^*)$
($\mathrm{RMSE}{=}0.056\pm0.013$); QQRBM ($\mathrm{RMSE}{=}0.119\pm0.028$) and QFeatureQRBM
($\mathrm{RMSE}{=}0.105\pm0.009$) are statistically significantly \emph{worse}
(Holm-adjusted $p_\mathrm{adj}{=}6.3\times10^{-5}$ and $4.9\times10^{-7}$,
$n{=}12$ paired observations), while QCRBM ($\mathrm{RMSE}{=}0.056\pm0.014$) is
statistically indistinguishable from the classical baseline
($p_\mathrm{adj}{=}1.0$).  On GP, QQRBM is significantly \emph{worse} than
CRBM$(H^*)$ ($\mathrm{RMSE}{=}1456\pm803$ vs.\ $\mathrm{RMSE}{=}584\pm511$, $p_\mathrm{adj}{=}0.024$);
QCRBM ($\mathrm{RMSE}{=}607\pm537$) is again indistinguishable from the classical baseline
($p_\mathrm{adj}{=}0.187$).
The iso-parameter matched-budget comparison ($P\in\{84,120,156,192\}$) reinforces
this result: the classical CRBM is lowest at three of the four budgets and no
CRBM-vs-QCRBM difference is statistically significant at any budget, so no
consistent quantum advantage emerges.
Gradient variance analysis reveals that the fitted decay base at circuit depth $L_Q{=}3$ is
$\hat{b}_3{=}0.520\pm0.013$, below the local-cost mitigation threshold
$2^{-1/2}\approx0.707$, indicating earlier-than-predicted barren-plateau onset.

The paper is organised as follows.  Section~\ref{sec:related} reviews related
work and situates this paper within the literature.  Section~\ref{sec:quantum}
provides the quantum computational foundations shared by all three quantum model
families.  Section~\ref{sec:models} derives each architecture in full.
Section~\ref{sec:experiments} presents the thirteen structured experiments.
Section~\ref{sec:discussion} discusses the findings, and
Section~\ref{sec:conclusions} concludes.

\section{Related Work}\label{sec:related}

\subsection{Conditional Restricted Boltzmann Machines for Sequential Data}

The Restricted Boltzmann Machine \citep{hinton2002cd} establishes the bipartite
visible--hidden structure from which all models in this paper derive.
Contrastive Divergence \citep{hinton2002cd} approximates the intractable
log-likelihood gradient by running short Gibbs chains; Hinton's practical guide
\citep{hinton2010guide} documents the implementation choices that stabilise
training for real-valued visible variables, including the $\sigma^{-1}$ and
$\sigma^{-2}$ gradient scaling factors that appear in the Gaussian--Bernoulli
CRBM.  The theoretical foundations of CD convergence are analysed in
\citep{carreira2005contrastive}; Persistent CD \citep{tieleman2008pcd} provides
less-biased negative-phase estimates at the cost of maintaining a persistent
Markov chain.  The representational power of RBMs relative to deeper models is
characterised in \citep{le_roux2008representational}, establishing that finite
RBMs can approximate any discrete distribution.
Xavier uniform initialisation \citep{glorot2010}, used throughout this paper,
addresses activation-variance collapse in sigmoid networks.

Taylor and Hinton \citep{taylor2009fcrbm} introduced the \emph{Conditional RBM}
for motion-style modelling by conditioning visible and hidden biases on a context
vector encoding recent history.  Sutskever et al.\ \citep{sutskever2008rtrbm}
extended this to the Recurrent Temporal RBM (RTRBM), integrating recurrent
hidden dynamics into the autoregressive structure.  This paper takes the
Gaussian--Bernoulli CRBM as the classical baseline and extends it with three
quantum variants, retaining the CD learning algorithm for all classical parameters
so that each quantum model is a strict architectural extension of the CRBM.

\subsection{Quantum Boltzmann Machines}

The idea of replacing the classical Gibbs state by a quantum Gibbs state
$\rho\propto e^{-H/T}$ was formalised by Amin et al.\ \citep{aminQBM2018},
who showed that a transverse-field Ising Hamiltonian
$H=\sum_{ij}J_{ij}\sigma_z^{(i)}\sigma_z^{(j)}+\sum_i\Gamma_i\sigma_x^{(i)}$
induces quantum correlations in the hidden layer.
Computing gradients of this quantum log-likelihood requires evaluating
$\partial_\theta e^{-H/T}$, which expands into an infinite commutator series
intractable on near-term devices without quantum hardware that can directly
sample thermal states.  Benedetti et al.\ \citep{benedetti2017qhm} explored a
semi-classical approximation (the quantum-assisted Helmholtz machine) that
uses quantum hardware for negative-phase sampling while retaining classical
inference; relatedly, Adachi and Henderson \citep{adachi2015application} train
deep networks by sampling a Boltzmann distribution directly on quantum-annealing
hardware.  Kieferov\'a and Wiebe \citep{kieferova2017tomography} analyse the
sample complexity of QBM learning via quantum state tomography; Crawford et al.\
\citep{crawford2018rlqbm} demonstrate QBMs in a reinforcement-learning setting.

Zoufal, Lucchi, and Woerner \citep{zoufalVQBM2021} introduced the
\emph{variational QBM}, replacing exact Gibbs-state preparation by a PQC and
training via the parameter-shift rule \citep{paramshift2019,mitarai2018qnn}.
This variational paradigm is computationally tractable and forms the basis
of the quantum components in the QCRBM, QQRBM, and QFeatureQRBM studied here.

Two structural results bound the scope of what variational QBMs can achieve
relative to their classical counterparts.  Demidik et al.\ \citep{demidik2025sqrbm}
introduced the \emph{semi-quantum RBM} (sqRBM) a model with a Hamiltonian
diagonal in the visible subspace and proved the equivalence
$\mathrm{sqRBM}(m)\equiv\mathrm{RBM}(3m)$: a quantum hidden representation
achieves the same expressive power as three times as many classical hidden units
when using Gibbs states.  Coopmans and Benedetti \citep{coopmans2024sample}
study the sample complexity of fully-visible QBM learning.  These results apply
to Gibbs-state models; the PQC-based models studied here use pure-state variational
features rather than thermal quantum statistics, and the equivalence does not
transfer directly (see Section~\ref{sec:demidik_caveat}).  

\subsection{Trainability of Parameterized Quantum Circuits}

Gradient-based training of PQCs faces a fundamental obstacle: McClean et al.\
\citep{mcclean2018barren} proved that for random-initialised circuits with global
cost functions, gradient variance vanishes exponentially as $O(2^{-n})$,
rendering training infeasible beyond a few dozen qubits.  Cerezo et al.\
\citep{cerezo2021cost} showed that this scaling improves to $O(2^{-n/2})$ when
using \emph{local} cost functions, those depending on single-qubit
observables and providing a quantitative rationale for the Pauli-$Z$ expectation
values used throughout this paper.  Holmes et al.\ \citep{holmes2022connecting}
formalise the connection between circuit expressibility and gradient magnitudes:
circuits that approach approximate $t$-designs are most susceptible to barren
plateaus, linking the expressibility analysis of Sim et al.\
\citep{sim2019expressibility} to the trainability concern.

Abbas et al.\ \citep{abbas2021power} introduce the effective dimension as a
capacity measure for quantum neural networks, providing evidence for potential
advantages over classical models of comparable size.  Cerezo et al.\
\citep{cerezo2021vqa} survey the broader landscape of variational quantum
algorithms and their error-mitigation requirements.  Larocca et al.\
\citep{larocca2024review} provide a comprehensive review of barren-plateau
mitigation strategies, including layerwise pre-training and problem-informed
initialisations.  P\'erez-Salinas et al.\ \citep{perezSalinas2020} propose
data re-uploading as a hardware-compatible encoding alternative that reduces
gate complexity from $O(2^n)$ to $O(n)$.  Experiment~6
(Section~\ref{sec:exp6}) provides an empirical characterisation of the
barren-plateau onset for the specific ansatz and local-observable choice
used in this paper, directly measuring cost-function gradient variance across
$n\in\{2,4,6,8,10\}$ qubits and $L_Q\in\{1,3,5\}$ layers, i.e. depth of the circuit.

\subsection{Benchmarking Quantum Machine Learning}

The reliability of quantum--classical comparisons has come under increasing
scrutiny.  Bowles, Ahmed, and Schuld \citep{bowles2024benchmarking} demonstrate
that classical models frequently outperform quantum classifiers on small structured
tasks, and that entanglement removal does not
consistently degrade performance and suggesting that quantum correlations may not
be the operative mechanism in many published QML results.  A central diagnosis is
\emph{asymmetric hyperparameter optimisation}: quantum models are typically
evaluated at fixed default parameter values while classical models receive extensive
tuning, artificially deflating observed quantum performance.  This paper directly
addresses this critique by applying validation-guided search to all
tunable parameters of all models, including quantum-specific quantities $\alpha$,
$\eta_Q$, $k_Q$, and $n_\mathrm{layers}$.  The use of Holm--Bonferroni correction
\citep{holm1979simple} across all simultaneous significance tests provides
family-wise error rate control absent from most prior QML comparison studies.

\subsection{Quantum Methods for Time-Series Forecasting}

Fujii and Nakajima \citep{fujii2017reservoir} demonstrated that the disordered
dynamics of a quantum system can serve as a fixed nonlinear reservoir for
machine-learning tasks, establishing the NARMA-10 benchmark \citep{atiya2000narma}, a tenth-order nonlinear autoregressive recurrence driven by uniform noise,
evaluated here in its standard input-driven formulation (the models observe the
drive history and predict the output; Section~\ref{sec:data}) as the canonical evaluation task for quantum reservoir computing.  Unlike
reservoir approaches, where the quantum component is fixed and only a linear
readout is trained, all quantum parameters in the models studied here are
actively optimised via gradient methods.  Chittoor et al.\
\citep{chittoor2024qultsf} extend the quantum time-series scope to long-horizon
forecasting through hybrid transformer quantum architectures on real-world
datasets.  The Mackey--Glass chaotic system \citep{mackeyglass1977} provides a
further benchmark with longer memory demands, deferred to future work.

To our knowledge, this paper is the first to evaluate quantum CRBMs on the NARMA-10
benchmark with full hyperparameter symmetry, iso-parameter fairness analysis, and
Holm--Bonferroni-corrected statistical testing across independently generated series.

\section{Quantum Computational Foundations}\label{sec:quantum}

All three quantum model families, QCRBM, QQRBM, and QFeatureQRBM shares
a common set of circuit primitives: amplitude encoding maps classical inputs into
quantum states, \texttt{Variational layer} applies trainable unitary
transformations, and Pauli-$Z$ measurements project back to real-valued features
that enter the Boltzmann model.  Training gradients are computed by the
parameter-shift rule, and the trainability of the resulting circuits is governed
by the barren-plateau bounds reviewed in Section~\ref{sec:related}.
This section collects the precise mathematical definitions of these primitives
as a self-contained reference; Section~\ref{sec:models} then builds each
architecture from these building blocks.

\subsection{Qubits, States, and Single-Qubit Gates}

A single qubit exists in a superposition of computational basis states
$|0\rangle$ and $|1\rangle$:
\begin{equation}
  |\psi\rangle = \alpha|0\rangle + \beta|1\rangle,\quad
  |\alpha|^2 + |\beta|^2 = 1,\quad \alpha,\beta\in\mathbb{C}.
\end{equation}
An $n$-qubit register spans the $2^n$-dimensional Hilbert space
$\mathcal{H}=(\mathbb{C}^2)^{\otimes n}$.  States are manipulated by unitary
operators.  The single-qubit rotations used throughout are
\begin{equation}
  R_z(\phi)=e^{-i\phi\sigma_z/2},\quad
  R_y(\theta)=e^{-i\theta\sigma_y/2},\quad
  R(\phi,\theta,\omega)=R_z(\omega)\,R_y(\theta)\,R_z(\phi),
  \label{eq:rotgates}
\end{equation}
where $\sigma_x,\sigma_y,\sigma_z$ are the Pauli matrices, and $R(\phi,\theta,\omega)$
follows the PennyLane \texttt{Rot} gate convention \citep{pennylane2018} used
throughout the implementations.
Two-qubit entanglement is introduced by the CNOT gate:
$\mathrm{CNOT}|c,t\rangle = |c,\,t\oplus c\rangle$.

\subsection{Amplitude Encoding}\label{sec:amp_encoding}

Classical data $\bm{a}\in\mathbb{R}^{2^n}$ with $\|\bm{a}\|_2=1$ can be encoded
as a pure quantum state via
\begin{equation}
  |\psi\rangle_{\bm{a}}
  = \sum_{k=0}^{2^n-1} a_k\,|k\rangle,
  \label{eq:amplitude_encoding}
\end{equation}
implemented amplitude encoding using the PennyLane framework \citep{plAmplitude}.
Amplitude encoding is \emph{exponentially compact}: $2^n$ real values are stored
in $n$ qubits.  The state preparation requires $O(2^n)$ two-qubit gates via the
M\"{o}tt\"{o}nen decomposition \citep{mottonen2005,plesch2011}; this is exact in
simulation but places the method outside the NISQ-compatible regime for
$n\gtrsim10$.
We stress that this exponentially large state space confers \emph{expressivity},
not a computational speed-up: under the classical \texttt{pennylane statevector}  
simulator used throughout, the encoding and its $O(2^n)$ preparation cost are evaluated
explicitly, so every reference below to an ``exponentially large Hilbert space''
denotes representational capacity rather than any runtime advantage.

Because the encoding requires input length exactly $2^n$, all implementations
pad classical feature vectors to this dimension before doing the encoding.
In the QCRBM, the concatenation $[\bm{v};\bm{u}]$
is padded to $2^H$.  In the QQRBM, visible and context inputs are
prepared on separate amplitude registers of size $2^V$ and $2^U$, respectively.
In the QFeatureQRBM, the lag vector is padded to $2^{n_q}$ with
$n_q=\lceil\log_2 L\rceil$.

\subsection{Parameterized Quantum Circuits and StronglyEntanglingLayers}

A Parameterized Quantum Circuit (PQC) is a unitary $U(\bm\theta)$ with
classically tunable parameters $\bm\theta$. The variational layer from PennyLane framework \citep{plSEL} organizes $L_Q$
successive layers; each layer applies three-parameter single-qubit rotations
$R(\phi,\theta,\omega)$ to every qubit, followed by strided CNOT entanglers
connecting qubits at strides $s=1,2,3,\ldots$.
For $n$ qubits and $L_Q$
layers, the parameter count is $3nL_Q$.  The resulting family of unitaries
is dense in $SU(2^n)$ for large enough $L_Q$, forming an approximate
$t$-design whose expressibility as a function of depth and qubit count is
characterised in \citep{sim2019expressibility}.  All three quantum models use
this template as their variational ansatz.

\subsection{The Parameter-Shift Rule}\label{sec:param_shift}

For quantum gates of the form $G(\theta)=e^{-i\theta P/2}$, where $P$ is
a Hermitian operator with eigenvalues $\{-1,+1\}$, any expectation value
$f(\theta)=\langle\psi|\,G^\dagger(\theta)\,O\,G(\theta)\,|\psi\rangle$
satisfies the exact gradient formula \citep{paramshift2019,mitarai2018qnn}:
\begin{equation}
  \frac{\partial f}{\partial\theta}
  = \frac{f(\theta+\tfrac{\pi}{2}) - f(\theta-\tfrac{\pi}{2})}{2}.
  \label{eq:param_shift}
\end{equation}
This \emph{parameter-shift rule} provides analytic gradients using only
circuit evaluations, enabling gradient-based optimisation of PQCs on hardware
and in simulation alike.  Because it requires two additional circuit evaluations
per parameter, the training cost of a hybrid model scales as $2|\bm\theta|$
circuit evaluations per gradient step.  In the simulation-based implementation,
gradients are computed via backpropagation (for $n_\text{wires}\leq12$) or the adjoint differentiation method
(for $n_\text{wires}>12$), both of which
provide mathematically equivalent exact gradients at lower cost than repeated
parameter-shift circuit evaluations.  Parameter-shift serves as a fallback
and would be the required method on real quantum hardware.

\subsection{Barren Plateau Bounds}\label{sec:barren}

As reviewed in Section~\ref{sec:related}, gradient variance of a random-initialised
PQC with a global cost function vanishes as $O(2^{-n})$ \citep{mcclean2018barren}.
For the \emph{local} cost functions used in this paper, the MSE losses depending on
single-qubit Pauli-$Z$ expectations Cerezo et al.\ \citep{cerezo2021cost}
establish the improved scaling:
\begin{equation}
  \Var\!\left[\frac{\partial\mathcal{L}}{\partial\theta_k}\right]
  = O\!\left(2^{-n/2}\right).
  \label{eq:barren_local}
\end{equation}
This bound applies to each quantum model in this paper and motivates the choice
of local Pauli-$Z$ observables over, for example, global fidelity-based losses.
We verify this scaling empirically in Experiment~6 (Section~\ref{sec:exp6}) by
measuring the cost-function gradient $\partial\mathcal{L}_Q/\partial\theta_0$
directly across $n\in\{2,4,6,8,10\}$ qubits and $L_Q\in\{1,3,5\}$ layers,
and fitting exponential decay models $\hat\sigma^2 = a\cdot b^n$ to establish
the practical trainability boundary.

\section{Model Architectures}\label{sec:models}

\subsection{Notation}
Table~\ref{tab:notation} presents a comprehensive summary of the notation used throughout this paper. The table serves as a convenient reference for the symbols, variables, and operators employed in the proposed methodology and theoretical analysis.
\begin{table}[ht]\centering
\caption{Principal notation.}\label{tab:notation}
\small\begin{tabular}{ll}\toprule
Symbol & Description\\\midrule
$\bm{v}\in\mathbb{R}^V$ & Visible variables; $V=1$ (univariate)\\
$\bm{h}\in\{0,1\}^H$ & Binary hidden units\\
$\bm{u}\in\mathbb{R}^U$ & Context vector, fixed at $U{=}L{=}10$ in all main\\
 & experiments. GP: the series' own lag history;\\
 & NARMA-10: the exogenous drive window $u(t{-}9..t)$\\
 & (input-driven). Exp.~3 varies $L$ (standalone study only).\\
$\bm\ell\in\mathbb{R}^L$ & Lag vector (QFeatureQRBM)\\
$\bm{W},\bm{W}_{cv},\bm{W}_{ch}$ & Weight matrices\\
$\bm{b}\in\mathbb{R}^V,\;\bm{c}\in\mathbb{R}^H$ & Biases\\
$\sigma>0$ & Visible std.\ deviation (scalar, fixed in CRBM/QCRBM/QQRBM;\\
 & vector-valued and learnable as $\sigma_i{=}\mathrm{softplus}(\rho_{v,i}){+}10^{-6}$\\
 & in QFeatureQRBM)\\
$\alpha\in\mathbb{R}$ & Quantum scaling parameter. In QCRBM (scalar) and QQRBM (scalars\\
 & $\alpha_h,\alpha_v$): linear coefficient; the quantum term is $\alpha\cdot\bm{q}$.\\
 & In QFeatureQRBM (vectors $\bm\alpha_h\!\in\!\mathbb{R}^H, \bm\alpha_v\!\in\!\mathbb{R}^V$):\\
 & gate logit; the applied gate is $\sigS(\bm\alpha)$, so the term is $\sigS(\bm\alpha)\cdot\bm{q}$.\\
$\eta_Q$ & Adam learning rate for PQC parameters\\
$\lambda_a$ & Alignment loss weight (QFeatureQRBM)\\
$\bm\theta$ & PQC parameters\\
$\bm{q}(\cdot;\bm\theta)\in[-1,1]^H$ & Pauli-$Z$ expectation values\\
$B,\;k,\;L_Q$ & Batch size; CD steps; PQC layer count\\
$\sigS$ & Logistic sigmoid\\
\bottomrule\end{tabular}\end{table}

\subsection{Classical Gaussian--Bernoulli CRBM}\label{sec:crbm}

\subsubsection{Model Formulation and Conditional Distributions}

The CRBM energy conditioned on context $\bm{u}$ is \citep{taylor2009fcrbm}:
\begin{equation}
E(\bm{v},\bm{h}|\bm{u})
  =\sum_i\frac{(v_i-b_{\mathrm{eff},i})^2}{2\sigma^2}
  -\sum_j c_{\mathrm{eff},j}\,h_j
  -\sum_{ij}\frac{v_i}{\sigma}\,W_{ij}\,h_j,
\label{eq:crbm_energy}
\end{equation}
where the context modulates the effective biases
\begin{equation}
  \beff(\bm{u})=\bm{b}+\bm{u}\bm{W}_{cv},\qquad
  \ceff(\bm{u})=\bm{c}+\bm{u}\bm{W}_{ch}.
  \label{eq:crbm_beff}
\end{equation}
The bipartite structure yields factored conditionals.  Given context $\bm{u}$:
\begin{align}
p(h_j=1|\bm{v},\bm{u})&=\sigS\!\Bigl(\tfrac{1}{\sigma}\textstyle\sum_iv_iW_{ij}+c_{\mathrm{eff},j}\Bigr),
\label{eq:crbm_ph}\\
p(\bm{v}|\bm{h},\bm{u})&=\mathcal{N}\!\bigl(\beff+\sigma\bm{W}^\top\bm{h},\;\sigma^2\mathbf{I}\bigr).
\label{eq:crbm_pv}
\end{align}
The $\sigma^{-1}$ factor in~\eqref{eq:crbm_ph} is critical for correct
gradient scaling in the Gaussian case and directly matches the implementation
of \texttt{CRBM.sample\_hidden} \citep{hinton2010guide}.
All weight matrices are initialised with Xavier uniform \citep{glorot2010}:
\begin{equation}
  a = \sqrt{\tfrac{6}{\mathrm{fan\_in}+\mathrm{fan\_out}}},\qquad
  W_{ij}\sim\mathcal{Uniform}(-a,\,a).
  \label{eq:xavier}
\end{equation}
Biases $\bm{b}$ and $\bm{c}$ are initialised to zero.

\subsubsection{Contrastive Divergence Learning}

\begin{definition}[Contrastive Divergence]
Starting from a data sample $(\bm{v}_0,\bm{u}_0)$, CD-$k$ approximates the
model expectation by running $k$ steps of alternating block Gibbs sampling:
\begin{align*}
\bm{h}_0 &\sim p(\bm{h}|\bm{v}_0,\bm{u}_0),\\
\text{for }t=1,\ldots,k:\quad
\bm{v}_t &\sim p(\bm{v}|\bm{h}_{t-1},\bm{u}_0),\quad
\bm{h}_t \sim p(\bm{h}|\bm{v}_t,\bm{u}_0).
\end{align*}
\end{definition}

The log-likelihood gradients are approximated by substituting positive-phase
expectations at $(\bm{v}_0,\bm{h}_0)$ and negative-phase expectations at
$(\bm{v}_k,\bm{h}_k)$ \citep{hinton2010guide,taylor2009fcrbm}:
\begin{align}
\frac{\partial\log p(\bm{v}_0|\bm{u}_0)}{\partial W_{ij}}
  &= \frac{1}{\sigma}\bigl(\langle v_i h_j\rangle_0 - \langle v_i h_j\rangle_k\bigr),
\label{eq:crbm_dW}\\
\frac{\partial\log p}{\partial b_i}
  &= \frac{1}{\sigma^2}\bigl(v_{0,i} - v_{k,i}\bigr),
\label{eq:crbm_db}\\
\frac{\partial\log p}{\partial c_j}
  &= h_{0,j} - h_{k,j},
\label{eq:crbm_dc}\\
\frac{\partial\log p}{\partial (W_{cv})_{ki}}
  &= \frac{u_{0,k}}{\sigma^2}\bigl(v_{0,i}-v_{k,i}\bigr),
\label{eq:crbm_dWcv}\\
\frac{\partial\log p}{\partial (W_{ch})_{kj}}
  &= u_{0,k}\bigl(h_{0,j}-h_{k,j}\bigr).
\label{eq:crbm_dWch}
\end{align}
In mini-batch form (batch size $B$), the implementation computes:
\begin{align}
\Delta\bm{W}     &\propto \frac{1}{B}\!\left(\frac{\bm{v}_0^\top}{\sigma}\bm{h}_0
  - \frac{\bm{v}_k^\top}{\sigma}\bm{h}_k \right),&
\Delta\bm{b}     &\propto \frac{\bm{v}_0-\bm{v}_k}{B\sigma^2},\notag\\
\Delta\bm{c}     &\propto \frac{\bm{h}_0-\bm{h}_k}{B},&
\Delta\bm{W}_{cv}&\propto \frac{\bm{u}_0^\top(\bm{v}_0-\bm{v}_k)}{B\sigma^2},\notag\\
\Delta\bm{W}_{ch}&\propto \frac{\bm{u}_0^\top(\bm{h}_0-\bm{h}_k)}{B}.
\label{eq:crbm_grads}
\end{align}
The $\sigma^{-1}$ factor in $\Delta\bm{W}$ and the $\sigma^{-2}$ factors in
$\Delta\bm{b}$ and $\Delta\bm{W}_{cv}$ arise from the Gaussian energy
term~\eqref{eq:crbm_energy} and are essential for numerical stability.

\subsubsection{Mean-Field Prediction and Autoregressive Forecasting}

For forecasting, stochastic Gibbs sampling is replaced by a deterministic
\emph{mean-field} approximation.  Starting from the context-derived estimate
$\bm{v}^{(0)}=\beff(\bm{u})$, the iteration
\begin{align}
\bm{h}^{(t)} &\leftarrow \sigS\!\Bigl(\frac{\bm{v}^{(t-1)}}{\sigma}\bm{W}
  + \ceff(\bm{u})\Bigr),\\
\bm{v}^{(t)} &\leftarrow \beff(\bm{u}) + \sigma\,\bm{W}^\top\bm{h}^{(t)}
\end{align}
is repeated for a small number of steps until approximate convergence to a
fixed point $\bm{v}^* \approx \bm{v}^{(t)}$ \citep{hinton2010guide}.
Multi-step autoregressive forecasting updates the context in a sliding-window
fashion after each predicted step:
\begin{equation}
  \bm{u}_{t+1} = \operatorname{concat}\!\bigl(\bm{u}_t[V:],\;\bm{v}_t^*\bigr),
  \label{eq:crbm_rollout}
\end{equation}
dropping the oldest $V$ values and appending the new prediction $\bm{v}_t^*$
\citep{taylor2009fcrbm,sutskever2008rtrbm}.

\begin{algobox}{Algorithm 1 --- CRBM Training and Autoregressive Prediction}
\textbf{Input:} training set $\{(\bm{v}_0^{(n)},\bm{u}_0^{(n)})\}_{n=1}^N$ and
hyperparameters $\sigma,\eta,k$.

\textbf{Training (per epoch, per mini-batch $(\bm{V}_0,\bm{U}_0)$):}
\begin{enumerate}[leftmargin=*,label=\textbf{\arabic*.}]
\item Initialise $\bm{W},\bm{W}_{cv},\bm{W}_{ch}$ Xavier-uniform; set $\bm{b}=\bm{c}=\bm{0}$.
\item Sample $\bm{H}_0\sim p(\bm{H}|\bm{V}_0,\bm{U}_0)$ via~\eqref{eq:crbm_ph} (positive phase).
\item Run $k$ Gibbs steps to obtain $(\bm{V}_k,\bm{H}_k)$ via~\eqref{eq:crbm_ph}--\eqref{eq:crbm_pv}.
\item Compute gradients~\eqref{eq:crbm_dW}--\eqref{eq:crbm_dWch} and update parameters.
\end{enumerate}

\textbf{Prediction (horizon $T$, context $\bm{u}_0$):}
\begin{enumerate}[leftmargin=*,label=\textbf{\arabic*.}]
\item Set $\bm{v}^{(0)}\leftarrow\beff(\bm{u}_0)$ and $\bm{u}\leftarrow\bm{u}_0$.
\item For $t=1,\ldots,T$: compute $\bm{v}^*\leftarrow\text{MeanField}(\bm{u},\bm{v}^{(t-1)})$.
\item Record $\bm{v}^*$; update $\bm{u}\leftarrow\operatorname{concat}(\bm{u}[V:],\bm{v}^*)$.
\end{enumerate}
\end{algobox}

\subsection{Hybrid Quantum-Classical QCRBM}\label{sec:qcrbm}

\subsubsection{Motivation and Architecture}

The QCRBM extends the classical CRBM by inserting a variational quantum circuit (VQC)
whose output modulates the logits of the hidden units.  The motivation is
twofold.  First, a PQC can express correlations among the hidden units that the
classical bilinear interaction $(\bm{v}/\sigma)\bm{W}$ cannot capture without
enlarging $H$.  Second, amplitude encoding maps the concatenated input
$\bm{x}=[\bm{v};\bm{u}]$ into the exponentially large Hilbert space, potentially
providing nonlinear basis functions beyond a classical shallow readout.  The
architecture augments the hidden conditional with a quantum term while keeping
the visible conditional and the context mechanism fully classical, so the QCRBM
is best understood as \emph{a classical CRBM with a learned, input-dependent
nonlinear correction to the hidden field}.

\subsubsection{Amplitude Encoding of the Input}

For each training sample $(\bm{v},\bm{u})$, both vectors are concatenated or
padded to length $2^H$, and $\ell_2$-normalised:
\begin{equation}
  \bm{x} = \operatorname{pad}\!\bigl([\bm{v};\bm{u}],\;2^H\bigr)\in\mathbb{R}^{2^H},
  \quad\tilde{\bm{x}} = \bm{x}/\|\bm{x}\|_2,
  \label{eq:qcrbm_encoding}
\end{equation}
$|\psi\rangle = \sum_{k=0}^{2^H-1}\tilde{x}_k|k\rangle$ from Eq. \ref{eq:amplitude_encoding} is prepared, the input must be encoded without discarding information, the implementation requires $2^{H}\ge V{+}U$; with $V{=}1$ and the fixed context
$U{=}10$ this forces $H\ge H_Q{=}\lceil\log_2(V{+}U)\rceil{=}4$.  Consequently the
QCRBM always carries at least as many parameters as the classical CRBM at the same
hidden width, so the parameter-matched comparison of Experiment~10
(Section~\ref{sec:exp10}) is the apples-to-apples test rather than the equal-$H$
comparison.
A \texttt{variational layer} PQC of depth $L_Q$ acts on $|\psi\rangle$,
returning Pauli-$Z$ quantum features:
\begin{equation}
  \bm{q}(\bm{x};\bm\theta)
  = \bigl(\langle Z_1\rangle,\ldots,\langle Z_H\rangle\bigr)
  \in[-1,1]^H.
  \label{eq:qcrbm_qout}
\end{equation}
This construction realises a quantum feature map: the amplitude-encoded state is
lifted into an exponentially large Hilbert space and read out through measured
observables, in the spirit of quantum-enhanced feature spaces
\citep{havlicek2019supervised,schuld2021machine}.
The circuit is wrapped as a PennyLane QNode and evaluated on a classical simulator, with batched evaluation minimising
device transfers.

\subsubsection{Augmented Hidden Conditional}

The quantum output $\bm{q}$ shifts the hidden-unit logits by an additive term
scaled by a learnable coefficient $\alpha\in\mathbb{R}$.  The modified hidden
conditional is
\begin{equation}
p(h_j=1|\bm{v},\bm{u})
  =\sigS\!\bigl(\tfrac{1}{\sigma}\textstyle\sum_iv_iW_{ij}+c_{\mathrm{eff},j}
  +\alpha\,q_j(\bm{x};\bm\theta)\bigr).
\label{eq:qcrbm_ph}
\end{equation}
By mapping the input into Hilbert space and projecting back to $[-1,1]^H$ via
Pauli-$Z$ expectations, the linear shift $\alpha\,\bm{q}$ introduces a family of
nonlinear basis functions that cannot be expressed by the original affine hidden
field alone.  The visible conditional remains Gaussian, identical
to~\eqref{eq:crbm_pv}.

\begin{remark}[Conservative extension]\label{rem:cons}
At $\alpha=0$ the quantum term $\alpha\,\bm{q}$ vanishes and~\eqref{eq:qcrbm_ph} recovers
the classical CRBM exactly~\eqref{eq:crbm_ph}.  In the implementation, the classical parameter
updates $\{W,b,c,W_{cv},W_{ch}\}$ are applied by a momentum-free SGD
optimiser that receives the
negated CD gradients.  The update is
numerically identical to the CD rule and does not pass through the PQC
computation graph; thus the CD trajectory for classical parameters is
unaffected by the quantum training step even at finite $\alpha$.
\end{remark}

\subsubsection{Hybrid Training via a Positive-Phase Soft Surrogate Loss}

Training uses two simultaneous but independent optimisation paths.

\paragraph{Classical path.}
The parameters $\{W,b,c,W_{cv},W_{ch}\}$ are updated with the same CD-$k$
formulas~\eqref{eq:crbm_dW}--\eqref{eq:crbm_dWch} as the pure CRBM, computed
in a block.  This requires only stochastic binary samples
$\bm{h}_0,\bm{h}_k$ and provides no gradients with respect to the circuit.

\paragraph{Quantum path.}
Propagating gradients into $\bm\theta$ and $\alpha$ encounters a fundamental
obstacle: the Bernoulli sampling $\bm{h}_t\sim p(\bm{h}|\bm{v}_t,\bm{u}_0)$
is non-differentiable, breaking standard backpropagation.  The implementation
sidesteps this with a \emph{positive-phase soft surrogate loss}: rather than
differentiating through the discrete Gibbs samples, it computes a soft
reconstruction loss through the \emph{positive-phase}
probabilistic hidden activations $\bm{p}_{h,0}\in[0,1]^H$.  (Principled
continuous relaxations of discrete sampling, such as the Concrete /
Gumbel-softmax estimator \citep{maddison2017concrete}, offer an alternative route
to differentiable discrete variables; here the positive-phase probabilities
suffice because the sampled (negative-phase) term is handled by the classical
CD gradients.)  Using the
positive-phase probabilities (rather than the end-of-chain $\bm{p}_{h,k}$)
means the circuit is trained to reconstruct data from its data-driven hidden
representation, consistent with the QFeatureQRBM convention:
\begin{align}
\tilde{\bm{v}}(\bm{p}_{h,0},\bm{u})
  &= \beff(\bm{u}) + \sigma\,\bm{W}^\top\bm{p}_{h,0},
  \label{eq:qcrbm_vtilde}\\
\mathcal{L}_Q(\bm\theta,\alpha)
  &= \frac{1}{B}\sum_{n=1}^B\|\bm{v}_0^{(n)}-\tilde{\bm{v}}^{(n)}\|_2^2.
  \label{eq:qcrbm_qloss}
\end{align}
Because $\bm{p}_{h,0}$ is a continuous sigmoid output derived from the
quantum-augmented logits~\eqref{eq:qcrbm_ph}, the computational graph remains
intact.  Backpropagation flows
$\mathcal{L}_Q\to\bm{p}_{h,0}\to\bm{q}(\bm{x};\bm\theta)\to\bm\theta$
using the parameter-shift rule~\eqref{eq:param_shift}.  The quantum parameters
$\bm\theta$ and $\alpha$ are updated jointly \citep{kingma2015adam}
with rate $\eta_Q$.

\begin{algobox}{Algorithm 2 --- QCRBM Training Step (one mini-batch)}
\textbf{Input:} batch $(\bm{V}_0,\bm{U}_0)$ and parameters
$\bm{W},\bm{b},\bm{c},\bm{W}_{cv},\bm{W}_{ch},\bm\theta,\alpha$.
\begin{enumerate}[leftmargin=*,label=\textbf{\arabic*.}]
\item Compute positive phase: $\bm{P}_{h,0},\bm{H}_0\leftarrow p(\bm{H}|\bm{V}_0,\bm{U}_0)$
  via~\eqref{eq:qcrbm_ph}.
\item Run $k$ Gibbs steps:
\begin{enumerate}[leftmargin=1.5em,label=\textbf{\alph*.}]
  \item Sample $\bm{V}_t\sim p(\bm{V}|\bm{H}_{t-1},\bm{U}_0)$ via~\eqref{eq:crbm_pv}.
  \item Sample $\bm{P}_{h,t},\bm{H}_t\leftarrow p(\bm{H}|\bm{V}_t,\bm{U}_0)$
    via~\eqref{eq:qcrbm_ph}.
\end{enumerate}
\item Compute classical CD gradients~\eqref{eq:crbm_dW}--\eqref{eq:crbm_dWch}.
\item Form soft reconstruction loss $\mathcal{L}_Q$~\eqref{eq:qcrbm_qloss}
  through $\bm{P}_{h,0}$ (positive-phase probabilities).
\item Quantum update: \textbf{opt\_q}.zero\_grad()~$\to$~backward($\mathcal{L}_Q$)~$\to$
  Adam step for $(\bm\theta,\alpha)$ via parameter-shift.
\item Classical update: \textbf{opt\_cl}.zero\_grad()~$\to$~assign negated CD
  gradients~$\to$~SGD step for $\{W,b,c,W_{cv},W_{ch}\}$ (see Remark~\ref{rem:cons}).
\end{enumerate}
\end{algobox}

\paragraph{Relation to sqRBM and Demidik et al.}\label{sec:demidik_caveat}
The sqRBM of Demidik et al.\ \citep{demidik2025sqrbm} uses quantum Gibbs
states $\rho\propto e^{-H/T}$ as hidden distributions.  The QCRBM, by
contrast, uses PQC expectation values $\langle Z_j\rangle$ on pure states.
These are fundamentally different quantum objects: the sqRBM exploits thermal
quantum statistics, whereas the QCRBM uses variational pure-state features.
The equivalence $\mathrm{sqRBM}(m)\equiv\mathrm{RBM}(3m)$ is therefore not
directly applicable to the QCRBM.  We include $\mathrm{CRBM}(H=12)$ as a
\emph{heuristic upper bound} on the classical capacity that might correspond
to the quantum hidden representation, motivated by Demidik et al.

\subsection{Quantum--Quantum QQRBM}\label{sec:qqrbm}

\subsubsection{Architecture and Three-Register Design}

The QQRBM takes the quantum approach further by assigning separate amplitude
registers to the visible data $\bm{v}$ and the context $\bm{u}$, and entangling
\emph{all three} registers visible, hidden, and context through a shared PQC.
The design aims to route the context exclusively through the quantum circuit, creating a fully quantum-mediated context dependency, and to
introduce separate learnable projection heads that map raw $Z$-expectations to the
correct visible and hidden dimensions.  The crucial architectural shift relative to
QCRBM is that quantum computation no longer acts as a single additive correction
to the hidden field alone; instead, the circuit becomes the primary path through
which context influences both hidden and visible updates.

The quantum device has $n_\text{wires}=V+H+U$ qubits, partitioned into three
registers: $\mathcal{V}=\{0,\ldots,V-1\}$, $\mathcal{H}=\{V,\ldots,V+H-1\}$,
$\mathcal{U}=\{V+H,\ldots,V+H+U-1\}$.  For each sample $(\bm{v},\bm{u})$, two
independent amplitude embeddings are prepared:
\begin{align}
|\psi_v\rangle &= \text{AmplitudeEmbedding}(\tilde{\bm{v}}_{\mathrm{amp}},\;\mathcal{V}),
  \quad \tilde{\bm{v}}_{\mathrm{amp}}\in\mathbb{R}^{2^V},\\
|\psi_u\rangle &= \text{AmplitudeEmbedding}(\tilde{\bm{u}}_{\mathrm{amp}},\;\mathcal{U}),
  \quad \tilde{\bm{u}}_{\mathrm{amp}}\in\mathbb{R}^{2^U}.
  \label{eq:qqrbm_encoding}
\end{align}
The hidden register $\mathcal{H}$ is initialised to $|0\rangle^{\otimes H}$.
A single \texttt{Variational Layers} circuit acts on all $V+H+U$ qubits,
entangling the three registers.  The circuit then measures $Z$-expectations:
\begin{align}
\bm{q}_v^{\mathrm{raw}} &= \bigl(\langle Z_w\rangle\bigr)_{w\in\mathcal{V}}\in[-1,1]^V,\\
\bm{q}_h^{\mathrm{raw}} &= \bigl(\langle Z_w\rangle\bigr)_{w\in\mathcal{H}}\in[-1,1]^H.
\label{eq:qqrbm_rawout}
\end{align}
This three-register view is essential: the visible and context inputs are encoded
independently and mixed only by the shared entangling dynamics; the hidden wires
start uninformative and acquire structure through joint unitary evolution.
Any dependence of the hidden activation on context in quantum mode is therefore
generated by the entangled circuit, not by an explicit classical affine
context term.  The implementation requires $V\ge 2$, as a single visible qubit
leaves the amplitude register degenerate at $|0\rangle$ and the visible PQC then
carries no information.  At $(V,H,U)=(2,2,3)$, which we adopt as the reference
configuration, the circuit has $n_q=V+H+U=7$ qubits.

\subsubsection{Projection Heads and Conditional Distributions}

Raw $Z$-expectations are projected through learnable linear maps:
\begin{equation}
  \bm{q}_v = \bm{q}_v^{\mathrm{raw}}\,\bm{Q}_v\in\mathbb{R}^V,\qquad
  \bm{q}_h = \bm{q}_h^{\mathrm{raw}}\,\bm{Q}_h\in\mathbb{R}^H,
  \label{eq:qqrbm_heads}
\end{equation}
where $\bm{Q}_h\in\mathbb{R}^{H\times H}$ and $\bm{Q}_v\in\mathbb{R}^{V\times V}$
are optimised jointly.  The
projection heads perform a learned linear re-basing from raw quantum observables
to model-relevant correction vectors, making the QQRBM more expressive than
component-wise scaling, at the cost of additional classical post-processing.

In the quantum mode, classical context terms
$\bm{u}\bm{W}_{ch}$ and $\bm{u}\bm{W}_{cv}$ are disabled; the context
enters only through the circuit:
\begin{align}
p(h_j=1|\bm{v},\bm{u}) &= \sigS\!\bigl(
  \tfrac{1}{\sigma}\bm{v}\bm{W}_{vh} + \bm{c} + \alpha_h\,(\bm{q}_h)_j
\bigr),
\label{eq:qqrbm_phid}\\
\mathbb{E}[v_i|\bm{h},\bm{u}] &=
  b_i + \sigma\,(\bm{h}\bm{W}_{vh}^\top)_i + \alpha_v\,q_{v,i}(\bm{v},\bm{u};\bm\theta),
\label{eq:qqrbm_pvis}
\end{align}
where the model uses a Gaussian visible layer, consistent with CRBM and QCRBM.  The circuit thus influences \emph{both} upward
and downward Gibbs passes the defining difference from QCRBM, where the
visible conditional remains fully classical.

\begin{remark}[Single interaction matrix]
Only a single weight matrix $\bm{W}_{vh}\in\mathbb{R}^{V\times H}$ is stored.
The visible conditional uses its transpose, $\bm{W}_{vh}^\top$, directly, so
the symmetry of the standard RBM energy is preserved by construction without
maintaining a separate $\bm{W}_{hv}$ tensor.  This matches the standard
symmetric RBM energy $E(\bm{v},\bm{h})=-\bm{v}^\top\bm{W}_{vh}\bm{h}-\bm{b}^\top\bm{v}-\bm{c}^\top\bm{h}$.
\end{remark}

\begin{remark}[Scalar quantum scaling]
The scalars $\alpha_h,\alpha_v\in\mathbb{R}$ in QQRBM are linear coefficients
that directly scale the projected quantum features $\alpha_h\,\bm{q}_h$ and $\alpha_v\,\bm{q}_v$.
At $\alpha_h=\alpha_v=0$ the quantum terms vanish and the classical conditionals are recovered.
Per-unit adaptation is delegated to the projection heads $\bm{Q}_h,\bm{Q}_v$.
\end{remark}

\subsubsection{Training}

Classical parameters $\{W_{vh},b,c\}$ are updated via
CD-$k$ with Gaussian sigma-scaling (identical to CRBM); $W_{cv}$ and $W_{ch}$ 
remain frozen in the default quantum mode:
\begin{equation}
  \Delta\bm{W}_{vh} \propto \tfrac{1}{B}
  \bigl(\tfrac{\bm{v}_0^\top}{\sigma}\bm{h}_0 - \tfrac{\bm{v}_k^\top}{\sigma}\bm{h}_k\bigr),\quad
  \Delta\bm{b} \propto \tfrac{\bm{v}_0-\bm{v}_k}{B\sigma^2},\quad
  \Delta\bm{c} \propto \tfrac{\bm{h}_0-\bm{h}_k}{B}.
\end{equation}
Quantum parameters $\{\bm\theta,\bm{Q}_h,\bm{Q}_v,\alpha_h,\alpha_v\}$ are
updated via a pathwise soft reconstruction loss through the deterministic
positive-phase visible conditional expectation $\hat{\bm{v}}_0$, which depends
on the data-driven hidden probabilities $\bm{p}_{h,0}$ and on the visible
quantum features evaluated at the data $\bm{v}_0$:
\begin{equation}
  \mathcal{L}_Q = \frac{1}{B}\sum_n\|\bm{v}_0^{(n)}-\hat{\bm{v}}_0^{(n)}\|_2^2,
  \quad \hat{\bm{v}}_0=\bm{b}+\sigma\,\bm{p}_{h,0}\bm{W}_{vh}^\top+\alpha_v\,\bm{q}_v(\bm{v}_0,\bm{u};\bm\theta).
  \label{eq:qqrbm_qloss}
\end{equation}
This positive-phase formulation matches the QCRBM convention~\eqref{eq:qcrbm_qloss}:
the circuit is trained to reconstruct each data point from its own data-driven
hidden representation, rather than from end-of-chain Gibbs samples.  The loss
gradient flows continuously from $\hat{\bm{v}}_0$ through $\bm{q}_v$ and $\bm{q}_h$
into the QNode, seamlessly invoking the parameter-shift rule.

\begin{algobox}{Algorithm 3 --- QQRBM Training Step (one mini-batch)}
\textbf{Input:} batch $(\bm{V}_0,\bm{U}_0)$ and full QQRBM parameter set.
\begin{enumerate}[leftmargin=*,label=\textbf{\arabic*.}]
\item Compute $\bm{P}_{h,0}\leftarrow p(\bm{H}|\bm{V}_0,\bm{U}_0)$
  via~\eqref{eq:qqrbm_phid}; sample $\bm{H}_0$.
\item For $t=1,\ldots,k$:
\begin{enumerate}[leftmargin=1.5em,label=\textbf{\alph*.}]
  \item Evaluate $\bm{P}_{v,t}\leftarrow\mathbb{E}[\bm{V}|\bm{H}_{t-1},\bm{U}_0]$
    via~\eqref{eq:qqrbm_pvis}, using $\bm{V}_{t-1}$ as input to the quantum circuit for $\bm{q}_v$; sample $\bm{V}_t$.
  \item Update $\bm{P}_{h,t},\bm{H}_t\leftarrow p(\bm{H}|\bm{V}_t,\bm{U}_0)$.
\end{enumerate}
\item Compute classical CD gradient, e.g.\
  $\Delta\bm{W}_{vh}=(\bm{V}_0^\top\bm{H}_0 - \bm{V}_k^\top\bm{H}_k)/(B\sigma)$
  (analogous formulas for $\bm{b},\bm{c}$; $\bm{W}_{cv},\bm{W}_{ch}$ not updated in default quantum mode).
\item Form soft loss $\mathcal{L}_Q\leftarrow\|\bm{V}_0-\hat{\bm{V}}_0\|^2/B$
  through the positive-phase reconstruction $\hat{\bm{V}}_0$~\eqref{eq:qqrbm_qloss}.
\item Backpropagate $\mathcal{L}_Q$; Adam update for
  $\{\bm\theta,\bm{Q}_h,\bm{Q}_v,\alpha_h,\alpha_v\}$.
\item Apply manual SGD updates to $\{\bm{W}_{vh},\bm{b},\bm{c}\}$
  in a \texttt{no\_grad} block ($\bm{W}_{cv},\bm{W}_{ch}$ remain frozen in the default quantum mode);
  clear any spurious autograd gradients accumulated
  on the classical parameters by the quantum backward pass.
\end{enumerate}
\end{algobox}

\paragraph{Heuristic nature of the quantum-feedback Gibbs chain.}
Because the visible and hidden conditionals are augmented by quantum features
that are trained through the positive-phase surrogate of
Eq.~\eqref{eq:qqrbm_qloss} rather than derived from a single symmetric energy,
the block-Gibbs sweep in Algorithm~3 does not satisfy detailed balance and is not
guaranteed to relax to a well-defined stationary distribution.  We therefore
treat the $k$-step quantum-feedback chain as a \emph{heuristic}
Contrastive-Divergence sampler for parameter estimation and one-step inference,
consistent with the long-standing practice of using CD as a biased-but-effective
gradient estimator in the (hybrid) RBM literature
\citep{hinton2002cd,carreira2005contrastive}.  We make no claim that the
quantum-augmented chain converges to a true model distribution; the same caveat
applies to the QCRBM Gibbs updates.
More precisely, the CD-$k$ update is a \emph{biased} estimator of the
log-likelihood gradient: the bias shrinks as $k$ grows but does not vanish at the
finite $k$ used here \citep{carreira2005contrastive}.  In the hybrid models this
classical CD bias compounds with the additional stochasticity of the
parameter-shift / soft-surrogate gradients of the quantum path
\citep{arrasmith2022equivalence}, so all (hybrid) quantum-RBM training in this work
is best read as biased-but-effective heuristic optimisation rather than exact
maximum-likelihood learning.

\subsection{Lag-Feature QFeatureQRBM}\label{sec:qfeatqrbm}

\subsubsection{Motivation and Lag-Space Design}

The QFeatureQRBM takes a different approach to integrating quantum features:
instead of encoding the current observation $\bm{v}$, it encodes a
\emph{temporal lag vector} $\bm\ell\in\mathbb{R}^L$ containing the $L$ most
recent values of the time series.  This is motivated by the observation that
temporal dependencies are best captured in a dedicated lag representation that
can be efficiently encoded into a small quantum state, leaving the current
observation $\bm{v}$ to be processed by the classical parameters.

This choice changes the role of the quantum circuit substantially.  In QCRBM
the circuit refines the hidden field of the current state; in QQRBM it mediates
between present visible and contextual information; in QFeatureQRBM it acts as
a compact temporal feature extractor whose input is a lag summary prepared
\emph{outside} the current Gibbs state.  The model is therefore the most
forecasting-oriented of the three quantum variants: quantum resources are used
primarily to summarise recent history rather than to redefine the instantaneous
visible--hidden interaction.

\subsubsection{Lag Encoding and Quantum Features}

For $L$ lag values, the number of qubits is $n_q=\lceil\log_2 L\rceil$,
so the quantum state dimension is $d_q=2^{n_q}\geq L$.  The lag vector
is padded and normalised:
\begin{equation}
  \tilde{\bm\ell}
  = \operatorname{pad}\!\bigl(\bm\ell/\|\bm\ell\|_2,\;d_q\bigr)\in\mathbb{R}^{d_q}.
\end{equation}
A PQC with $n_q$ qubits and $n_\mathrm{layers}$ \texttt{Variational Layers}
produces quantum features
\begin{equation}
  \bm{z}(\bm\ell;\bm\theta)
  = \bigl(\langle Z_1\rangle,\ldots,\langle Z_{n_q}\rangle\bigr)\in[-1,1]^{n_q}.
  \label{eq:qfeat_zout}
\end{equation}
These are projected into the hidden and visible dimensions via learnable heads
$\bm{Q}_h\in\mathbb{R}^{n_q\times H}$ and $\bm{Q}_v\in\mathbb{R}^{n_q\times V}$:
\begin{equation}
  \bm{q}_h = \bm{z}\,\bm{Q}_h\in\mathbb{R}^H,\qquad
  \bm{q}_v = \bm{z}\,\bm{Q}_v\in\mathbb{R}^V.
  \label{eq:qfeat_proj}
\end{equation}
For $L\leq8$ we have $n_q\leq3$, giving a circuit footprint of at most
$9n_\mathrm{layers}$ parameters the most compact quantum variant.

\subsubsection{Conditional Distributions}

In quantum mode, the per-unit gate logits $\bm\alpha_h\in\mathbb{R}^H$ and
$\bm\alpha_v\in\mathbb{R}^V$ (learnable vectors, unlike the scalar $\alpha$ in QCRBM)
scale the quantum contributions with bounded gains $\sigS(\alpha_{h,j}),\sigS(\alpha_{v,i})\in(0,1)$.
The visible standard deviation is learnable and parameterised as
$\sigma_i=\mathrm{softplus}(\rho_{v,i})+10^{-6}$, with $\bm\rho_v\in\mathbb{R}^V$
initialised to $-1$.  In the Gaussian visible mode:
\begin{align}
p(h_j=1|\bm{v},\bm{u},\bm\ell) &= \sigS\!\bigl(
  \bm{c} + \tilde{\bm{v}}\,\bm{W}_{vh}
  + [\bm{u}\bm{W}_{ch}]_{\mathrm{opt}}
  + \sigS(\alpha_{h,j})\,q_{h,j}
\bigr),\label{eq:qfeat_ph}\\
\mathbb{E}[v_i|\bm{h},\bm{u},\bm\ell] &=
  b_i + \sigma_i\,(\bm{h}\,\bm{W}_{vh}^\top)_i
  + [\bm{u}\bm{W}_{cv}]_{i,\mathrm{opt}}
  + \sigS(\alpha_{v,i})\,q_{v,i},\label{eq:qfeat_pv}
\end{align}
where $\tilde{v}_i=v_i/\sigma_i$ and $[\,]_{\mathrm{opt}}$ denotes
optional context terms when. The $\sigma_i$
factor in~\eqref{eq:qfeat_pv} is the conditional mean implied by the
standard Gaussian-Bernoulli energy $\tfrac{(v_i-b_i)^2}{2\sigma_i^2}-\sum_j\tfrac{v_i}{\sigma_i}W_{ij}h_j$
\citep{hinton2010guide}; it generalises the fixed scalar $\sigma$ of the
CRBM (Eq.~\eqref{eq:crbm_ph}) to a learnable per-visible scale $\sigma_i$ and is
mandatory for consistency with the gradient rules of the classical CD update.
In the Bernoulli visible mode, the same expressions
hold with $\tilde{\bm v}$ replaced by $\bm v$, $\sigma_i$ replaced by $1$, and
the visible conditional wrapped in a sigmoid.
During training a small heuristic decay
$\bm\rho_v\leftarrow\bm\rho_v(1-\eta\cdot10^{-3})$ gently relaxes the visible scale
towards $\sigma_i\to\mathrm{softplus}(0)+10^{-6}\approx0.693$, stabilising it early
in optimisation; it is a regulariser, not a Contrastive-Divergence gradient.

\subsubsection{Training and Knowledge Distillation}

Training combines a CD objective for the classical parameters with a joint
quantum objective.  The quantum loss combines reconstruction, knowledge-distillation
alignment, and an $L_2$ penalty on the quantum projection heads
\citep{hinton2015distilling}:
\begin{align}
\mathcal{L}_{\mathrm{recon}} &= \mathrm{MSE}(\hat{\bm{v}}_Q,\bm{v}_0), \\
\mathcal{L}_{\mathrm{align}} &= \mathrm{MSE}\!\bigl(\hat{\bm{v}}_Q,\;\mathrm{stopgrad}(\hat{\bm{v}}_{NQ})\bigr),\\
\mathcal{L}_Q &= \mathcal{L}_{\mathrm{recon}}
  + \lambda_a\,\mathcal{L}_{\mathrm{align}}
  + 10^{-4}\bigl(\overline{\bm{Q}_h^{\odot 2}}+\overline{\bm{Q}_v^{\odot 2}}\bigr),
\label{eq:qfeat_loss}
\end{align}
where $\hat{\bm{v}}_Q$ and $\hat{\bm{v}}_{NQ}$ are reconstructions with and
without the quantum circuit, respectively (both passed through the same
$p$-hidden/visible-mean chain that defines~\eqref{eq:qfeat_ph}--\eqref{eq:qfeat_pv}),
and $\overline{\bm{Q}^{\odot 2}}$ denotes the elementwise-square mean of the
projection head $\bm{Q}\in\{\bm{Q}_h,\bm{Q}_v\}$.  The regulariser is applied
to the unbounded projection heads rather than to the gate logits $\bm\alpha_h,\bm\alpha_v$:
the gates are already bounded in $(0,1)$ by the sigmoid, so shrinking
$\bm\alpha$ toward zero would merely push every gate toward $0.5$ without
suppressing the quantum branch's contribution.  It is ensured 
that the gradient of the alignment term flows
on within each step.  The alignment term acts as \emph{knowledge
distillation} \citep{hinton2015distilling}: the faster, more stable classical
path serves as a teacher and the quantum path as the student.  By regularising
the variational circuit toward the classical solution, the alignment loss bounds
parameter-space exploration and mitigates the barren-plateau risk often
encountered in over-parameterised PQCs \citep{mcclean2018barren}, while
simultaneously encouraging the quantum path to search for additive predictive
value rather than drifting to a qualitatively different solution.  This makes
QFeatureQRBM the only architecture in the paper that is quantum-regularised
by design, rather than merely quantum-enhanced.  In practice, the alignment
term is disabled for the first $n_{\mathrm{warmup}}$ epochs and the gate logits $\bm\alpha_h,\bm\alpha_v$
are assigned a higher learning rate than the remaining quantum parameters
(default $10\times$), preventing the optimiser from freezing the gates at
their initial values before the projection heads have had a chance to learn.

Multi-step forecasting uses a lag-update function that shifts the lag vector and
appends each new prediction:
$\bm\ell_{t+1}=\text{lag\_update\_fn}(\bm\ell_t,\bm{v}_t^*)$.
QFeatureQRBM is therefore the only model in which quantum features are
\emph{recursively fed by the model's own predictions} during multi-step rollout.

\begin{algobox}{Algorithm 4 --- QFeatureQRBM Training Step}
\textbf{Input:} batch $(\bm{V}_0,\bm\Lambda_0)$ and optionally $\bm{U}_0$.
\begin{enumerate}[leftmargin=*,label=\textbf{\arabic*.}]
\item Compute $\bm{H}_0^{\mathrm{pos}}\leftarrow\mathrm{sample\_hidden}(\bm{V}_0;\,
  \text{quantum=True},\bm\Lambda_0)$.
\item For $t=1,\ldots,k$:
\begin{enumerate}[leftmargin=1.5em,label=\textbf{\alph*.}]
  \item Update $\bm{V}_t\leftarrow\mathrm{sample\_visible}(\bm{H}_{t-1};\,
    \text{quantum=True},\bm\Lambda_0)$.
  \item Update $\bm{H}_t\leftarrow\mathrm{sample\_hidden}(\bm{V}_{t};\,
    \text{quantum=True},\bm\Lambda_0)$.
\end{enumerate}
\item Apply classical CD update to $\{W_{vh},b,c\}$ (and $\{W_{cv},W_{ch}\}$ when context
  is active); apply $\sigma$-scale heuristic decay $\bm\rho_v\leftarrow\bm\rho_v(1-\eta\cdot0.001)$
  to $\bm\rho_v$ (not a CD gradient).
\item Call \textbf{opt\_q}.zero\_grad().
\item Evaluate $\mathcal{L}_Q\leftarrow\operatorname{QuantumLoss}(\bm{V}_0,\bm\Lambda_0)$
  via~\eqref{eq:qfeat_loss}.
\item Backpropagate $\mathcal{L}_Q$; apply quantum-parameter update for
  $\{\bm\theta,\bm{Q}_h,\bm{Q}_v,\bm\alpha_h,\bm\alpha_v\}$.
\end{enumerate}
\end{algobox}

\subsection{Comparative Architecture Analysis and Transition to Experiments}
\label{sec:arch_comparison}

The three quantum variants represent three distinct answers to the design
question: \emph{where should a parameterised quantum circuit enter a conditional
Boltzmann model, and through what pathway?}

In the QCRBM, quantum computation acts as a targeted enrichment of the hidden
field only.  The classical Gaussian visible conditional, the autoregressive
context mechanism, and the CD update formulas are fully preserved; the circuit
contributes a nonlinear logit correction $\alpha\,\bm{q}(\bm{x};\bm\theta)$ that
vanishes at $\alpha=0$.  This conservative placement maximises interpretability and
enables a formal conservative-extension property, but restricts quantum resources
to the hidden inference step and excludes them from visible generation and context
routing.

In the QQRBM, all three sub-registers, visible, hidden, and context are jointly
entangled in a single circuit, making quantum computation the primary mediator
between context and latent state in quantum mode.  The price of this richer
coupling is a larger qubit count ($n_q=V+H+U$), the introduction of classical projection heads on the quantum output, and
greater susceptibility to barren plateaus at scale.

In the QFeatureQRBM, quantum resources are used neither to enrich the hidden
field nor to entangle all sub-registers simultaneously; instead, the circuit
processes a dedicated lag vector, returning a compact temporal feature that
enters both hidden and visible conditionals as an additive correction.  The
circuit is small ($n_q\leq3$ for $L\leq8$), making this the most practically
tractable variant, but its contribution depends critically on the adequacy of the
lag representation.  The knowledge-distillation alignment loss further
distinguishes this model by making classical agreement an explicit training
objective.

\begin{table}[ht]\centering
\caption{Comparative summary of the four Boltzmann-model families. Note, in all models the hidden 
variables are binary.}\label{tab:comparison}
\small
\begin{tabular}{lllllll}\toprule
\textbf{Model} & \textbf{Visible} & \textbf{Context} & \textbf{Quantum input} & \textbf{Training}\\\midrule
CRBM          & Gaussian &  Classical & None                & CD-$k$\\
QCRBM         & Gaussian &  Classical & $\bm{v};\bm{u}$  & CD-$k$ + Adam\\
QQRBM         & Gaussian &  Quantum   & $\bm{v}$,$\bm{u}$ & CD-$k$ + Adam\\
QFeatureQRBM  & Gaussian/Binary      &  Optional  & lag $\bm\ell$       & CD-$k$ + Adam\\
\bottomrule
\end{tabular}\end{table}

These architectural distinctions (summarised in Table~\ref{tab:comparison}) determine which CD formulas apply, where
differentiable quantum training must be introduced, and how susceptible each model
is to barren plateaus.  The following experiments probe these differences
empirically.  Two key questions drive the design: (i) does quantum augmentation
provide measurable forecasting gains, and if so, on which data regimes? and (ii)
are observed gains attributable to the quantum computation or to the additional
parameters it introduces?  We address question~(i) directly in Experiments~2
and~7, and question~(ii) in the iso-parameter comparison of Experiment~10.

\section{Experiments}\label{sec:experiments}

\subsection{Data}\label{sec:data}

For the experiments, we use two data types: (i) synthetic continuations 
of real-world financial time series (GP data), and (ii) the NARMA-10 benchmark task.

\paragraph{GP data.}  
The original raw data stems from a real-world financial data set, where revenue-related measures 
are recorded for a larger set of products. For the present analysis, we retain 20 sufficiently complete 
and non-zero product series. For each retained product, this leaves us with a univariate monthly time series 
covering roughly eight years (96 observations). 
As this is a comparatively short history for studying long-range forecasting behavior, we extend 
it artificially by generating synthetic continuations that preserve important characteristics of the raw data. 
Gaussian Processes (GP) \cite{RasmussenWilliams2006} are usually well suited for this purpose, 
because they are able to represent fast changes, slower trends and even periodic behavior. 
We used GPyTorch \cite{GardnerEtAl2018} to implement the GP model. 
However, one important characteristic of the raw data is that many 
products appear to occasionally alternate between a lower and a higher level, seemingly at random. 
A plain GP predictor represents such behavior poorly, largely ignoring this noise component when 
fitted correctly (or worse, overfitting the noise). We therefore augment the GP with a 
two-state hidden Markov model (HMM) trained by Baum-Welch estimation~\cite{Rabiner1989,BaumEtAl1970} 
on the residual process, so that synthetic samples combine smooth GP dynamics with 
state-dependent local level shifts.




Synthetic data are obtained by evaluating the trained GP on an extended query grid, drawing samples from the GP's prior. At the same time, HMM state trajectories are sampled and added to the GP samples. Gaussian observation noise is also added. 
This procedure yields synthetic series that preserve a smooth trend and seasonal structure as well as 
random level shifts.
In the following we use this data to benchmark the model variants discussed in this paper.

While more data can easily be generated, most of our experiments use $T=280$ time steps 
of just three series (more series / time steps are used in a select few experiments, as necessary). 
Each series is chronologically split 70\%/15\%/15\%  
into train/validation/test subsets, and a
standardization scaler is fit on the training prefix only and applied to all three splits, so that no
validation- or test-set statistics leak into training. Errors are reported on the original, non-standardized data scale.

\paragraph{NARMA-10 data.} The NARMA-10 task \citep{atiya2000narma} is defined
by
\begin{equation}
y(t)=0.3\,y(t{-}1)+0.05\,y(t{-}1)\!\sum_{i=1}^{10}y(t{-}i)
  +1.5\,u(t{-}9)\,u(t)+0.1,
\label{eq:narma}
\end{equation}
where $u(t)$ is drawn from a uniform distribution in $[0,0.5]$.  
Equation~\eqref{eq:narma} is the
forward-shifted (contemporaneous-$u$) form of the Atiya--Parlos recurrence
\citep{atiya2000narma}---equivalent up to an index shift to the
$u(t{-}10)u(t{-}1)$ convention.
We generate $N_\mathrm{NR}=3$ series of
$280$ steps each with distinct random seeds and discard the first $50$ transient
steps, leaving $T{=}230$ usable steps per series; as for GP, a
standardization scaler is fit on the training prefix only.  (We discard $50$
transient steps, fewer than the $100$ used by, e.g., Fujii \& Nakajima
\citep{fujii2017reservoir}; since all models see identical series this does not
affect the internal comparison, although absolute RMSE values are not directly
comparable to that protocol.)  The NARMA-10 system has a rich
10th-order memory and strong nonlinearity, providing a task class where models
with nonlinear feature extraction are expected to have an advantage.

\emph{Input-driven evaluation.}  We evaluate NARMA-10 in its standard
input-driven formulation: each model receives the
\emph{drive} window $(u(t{-}9),\dots,u(t))$ as context $\bm{u}$ and predicts the
output $y(t)$; it does \emph{not} observe past values of $y$.  This is the
formulation used in the reservoir-computing literature
(e.g.\ \citep{atiya2000narma,fujii2017reservoir,jaeger2004harnessing}) and is the
protocol that produced every NARMA-10 number in this paper. 
The $L{=}10$ window therefore spans both
factors of the dominant bilinear term $1.5\,u(t{-}9)\,u(t)$.
All NARMA-10 RMSE values are reported in raw (unstandardised) NARMA units.

\textit{Note on sample size and statistical power for the experiments.} 
With $N=3$ series and 4 seeds for retraining each series,
each dataset yields $n{=}12$ paired observations per comparison.  A post-hoc power
analysis \citep{cohen1988statistical} shows that at $n{=}12$, $\alpha{=}0.05$, and
power $0.8$ the minimum detectable paired effect size (two-sided) is
$d_\mathrm{min}{\approx}0.89$: only medium-to-large effects can be reliably
detected, and effects with $|d|{<}0.89$ may go undetected.  This bears directly on
the QCRBM ''tie'' at the \emph{observed} effect sizes the achieved power is
only $0.42$ on GP and $0.06$ on NARMA-10, so the absence of a significant
QCRBM difference reflects limited power for small effects rather than
demonstrated equivalence; small positive (or negative) effects with
$|d|{<}d_\mathrm{min}$ cannot be excluded.  A complementary 19-series paired design
(Experiment~13, Section~\ref{sec:exp13}) probes effects well below this
$d_\mathrm{min}$ floor.  The sample also exceeds the $n\geq5$
minimum of the Wilcoxon signed-rank test; we report the paired $t$-test throughout
and corroborate every conclusion with Shapiro--Wilk normality and Wilcoxon
robustness checks (Section~\ref{sec:stats}).

\paragraph{Context/lag window.}  Throughout Experiments~1--2 and~4--13 the
context/lag window is fixed at $L=U=10$ for \emph{all} models (CRBM, QCRBM,
QFeatureQRBM) to match NARMA-10's 10-step recurrence order and to give every
model an identical input context and amplitude-encoding budget; on NARMA-10 this
window is the input-driven drive history $u(t{-}9..t)$, on GP the series' own
lag history.  Experiment~3
sweeps $L\in\{2,4,6\}$ as a standalone
sensitivity analysis: the resulting $L^*=6$ is reported but \emph{not} propagated
to any other experiment.  The only deliberate exception is QQRBM, which, as a
structural reference at its natural register size ingests only the last few drives
values ($U_{qq}{=}3$ context steps plus the contemporaneous $u(t)$ in its visible
register, i.e.\ $u(t{-}3..t)$) and therefore never observes $u(t{-}9)$, the first
factor of the dominant bilinear term (Section~\ref{sec:qqrbm}; the classical
control of Section~\ref{sec:exp2} quantifies the resulting cost).

\subsection{Hyperparameter Search Protocol}\label{sec:hpsearch}

A key design principle of this paper is symmetric hyperparameter optimisation.
Table~\ref{tab:hpgrid} documents every searched hyperparameter for every model,
with all selection decisions made on validation RMSE.  Classical and quantum
models receive proportionally equivalent search budgets.

\begin{table}[ht]\centering
\caption{Complete hyperparameter search grids.  All selections by validation RMSE.
  Exp.\ column indicates when the parameter was introduced.
  Bold = selected value (GP / NARMA-10 where they differ).}\label{tab:hpgrid}
\small\setlength{\tabcolsep}{4pt}
\begin{tabular}{lll p{4.1cm} p{3.2cm}}\toprule
Model & Exp. & Parameter & Grid & Selected\\\midrule
\multirow{5}{*}{CRBM}
 & 1 & $H$ (hidden units) & $\{1,2,4\}$ & $\mathbf{1}$ (GP) / $\mathbf{2}$ (NARMA-10)\\
 & 1 & $k$ (CD steps) & $\{1,3\}$ & $\mathbf{3}$ (both)\\
 & 1 & $\sigma$ (visible std) & $\{0.1,0.5,1.0\}$ & $\mathbf{0.1}$ (both)\\
 & 11 & $\eta_\mathrm{cl}$ (learning rate) & $\{10^{-4},10^{-3},10^{-2}\}$ & $\mathbf{10^{-3}}$ (both)\\
 & 11 & $\lambda_\mathrm{wd}$ (weight decay) & $\{0,10^{-3},10^{-2},10^{-1}\}$ & see Exp.~11\\\midrule
\multirow{4}{*}{QCRBM}
 & 9 & $L_Q$ (PQC layers) & $\{1,2,3\}$ & $\mathbf{2}$\\
 & 2 & $\alpha$ (quantum scale) & $\{0.01,0.03,0.1,0.3,1.0,3.0\}$ & $\mathbf{3.0}$ (both)\\
 & 11b & $\eta_Q$ (quantum lr) & $\{10^{-4},10^{-3},10^{-2}\}$ & $\mathbf{10^{-3}}$ (both)\\
 & 11b & $\lambda_Q$ (quantum wd) & $\{0,10^{-3},10^{-2}\}$ & $\mathbf{10^{-3}}$ (GP) / $\mathbf{10^{-2}}$ (NARMA-10)\\
 & 12 & $k_Q$ (quantum CD steps) & $\{1,2,3\}$ & $\mathbf{3}$ (GP) / $\mathbf{1}$ (NARMA-10)\\\midrule
\multirow{4}{*}{QFeatureQRBM}
 & 3 & $L$ (lag window) & $\{2,4,6\}$ & see Exp.~3\\
 & 2 & $\lambda_a$ (align weight) & $\{0.0,0.1,0.5,1.0\}$ & $\mathbf{1.0}$ (both)\\
 & 12 & $n_\mathrm{layers}$ (PQC depth) & $\{1,2,3\}$ & $\mathbf{2}$ (GP) / $\mathbf{1}$ (NARMA-10)\\\midrule
\multirow{2}{*}{QQRBM}
 & 2 & $H$ (hidden qubits) & fixed $V{=}2,H{=}2$, $n_q{=}7$ & $\mathbf{2}$ (natural size)\\
 & 2 & $L_Q$ (PQC layers) & fixed $L_Q{=}1$ & $\mathbf{1}$\\\bottomrule
\end{tabular}\end{table}

\subsection{Statistical Protocol}\label{sec:stats}

All models are trained with 4 random seeds on 3 independently generated
series per dataset, yielding $n{=}12$ paired observations per comparison.
Pairwise comparisons use the two-sided \emph{paired $t$-test} with
Holm--Bonferroni correction \citep{holm1979simple}.  The paired $t$-test is
reported throughout; at $n{=}12$ the sample exceeds the $n\geq5$ minimum of
the Wilcoxon signed-rank test.  As a robustness check we additionally compute,
for every model-versus-reference comparison, the Shapiro--Wilk test on the paired
differences and the Wilcoxon signed-rank test.  Shapiro--Wilk flags non-normal
paired differences ($p{<}0.05$) only for QFeatureQRBM on GP; here, 
the Wilcoxon result is the primary inference and it
agrees with the $t$-test.  The two tests agree on every significant call: on
GP, QQRBM is significant (Wilcoxon $p{=}0.002$) while QCRBM is not
($p{=}0.15$); on NARMA-10, QFeatureQRBM, and QQRBM are significant
($p{\leq}0.0005$) while QCRBM is not ($p{=}0.68$).  Holm--Bonferroni is
uniformly more powerful than Bonferroni while controlling the family-wise error
rate.  All tables report mean $\pm$ SD across seeds and
series.  Significance levels: ${}^*\!p<0.05$, ${}^{**}\!p<0.01$,
${}^{***}\!p<0.001$ (all Holm-corrected).

\begin{table}[ht]\centering
\caption{Model variants.  The final column gives \emph{illustrative}
  total-parameter counts at a common reference point $(V{=}1,H{=}4,U{=}4)$, for
  cross-model scale only; they do not correspond to any single run configuration.
  Operative runs use $U{=}10$, and the deployed iso-parameter counts (e.g.\
  QCRBM ${=}84$ at $U{=}10$) are given in Table~\ref{tab:iso_pairs} and
  Section~\ref{sec:exp10}.  $H^*{=}1$ (GP) / $H^*{=}2$ (NARMA-10) (see
  Experiment~1).}\label{tab:models}
\small\begin{tabular}{lllc}\toprule
Label & Architecture & PQC params & Total params (illustr., $H{=}4$)\\\midrule
CNet$(H^*)$ & 2-layer MLP, hid$=H^*$\tablefootnote{The hidden width of
  \textsc{CNet} is set to $\lfloor 3n_bL_Qn_q/(n_q{+}1)\rceil$ (rounded),
  matching the PQC parameter count divided by $(n_q{+}1)$.  This formula is
  empirical: it gives a hidden width that grows roughly linearly with circuit
  depth while remaining smaller than the full PQC parameter count, avoiding
  over-parameterisation of the classical surrogate.  It does not affect any
  published QBM result; \textsc{CNet} is a structural reference only.} & --- & 37\\
CRBM$(H^*)$ & Gaussian--Bernoulli CRBM, dataset-dep.\ $H^*$ & --- & 30\\
CRBM$(3H^*)$ & Gaussian--Bernoulli CRBM, $H{=}3H^*$ & --- & 118\\
QCRBM & Hybrid CRBM, $L_Q{=}2$ & 24 & 55\\
QFeatureQRBM & Lag-feature QRBM, $n_q{\leq}3$ & 9 & 39\\
QQRBM & Full-register, $n_q{=}7$ ($V{+}H{+}U$, $V{=}2$) & 21 & 51\\
\bottomrule\end{tabular}\end{table}

\subsection{Experiment 1:  Hyperparameter Tuning of the CRBM Baseline}\label{sec:exp1}
For the parameters $H, k, \sigma$ of the classical CRBM baseline, we perform a grid search over the 
following values: $H\in\{1,2,4\}$, $k\in\{1,3\}$, $\sigma\in\{0.1,0.5,1.0\}$ (18 configs $\times$
4 seeds $\times$ 20 epochs on each dataset). The parameters learning rate and weight decay are 
fixed at $\eta=10^{-3}$ and $\lambda=0$ for this experiment but will be addressed in Experiment 11. 
The best hyperparameter configuration is selected on validation RMSE:

\textbf{GP:} $H^*{=}1$, $k^*{=}3$, $\sigma^*{=}0.1$.\\
\textbf{NARMA-10:} $H^*{=}2$, $k^*{=}3$, $\sigma^*{=}0.1$.

\begin{figure}[ht]\centering
\includegraphics[width=0.9\linewidth]{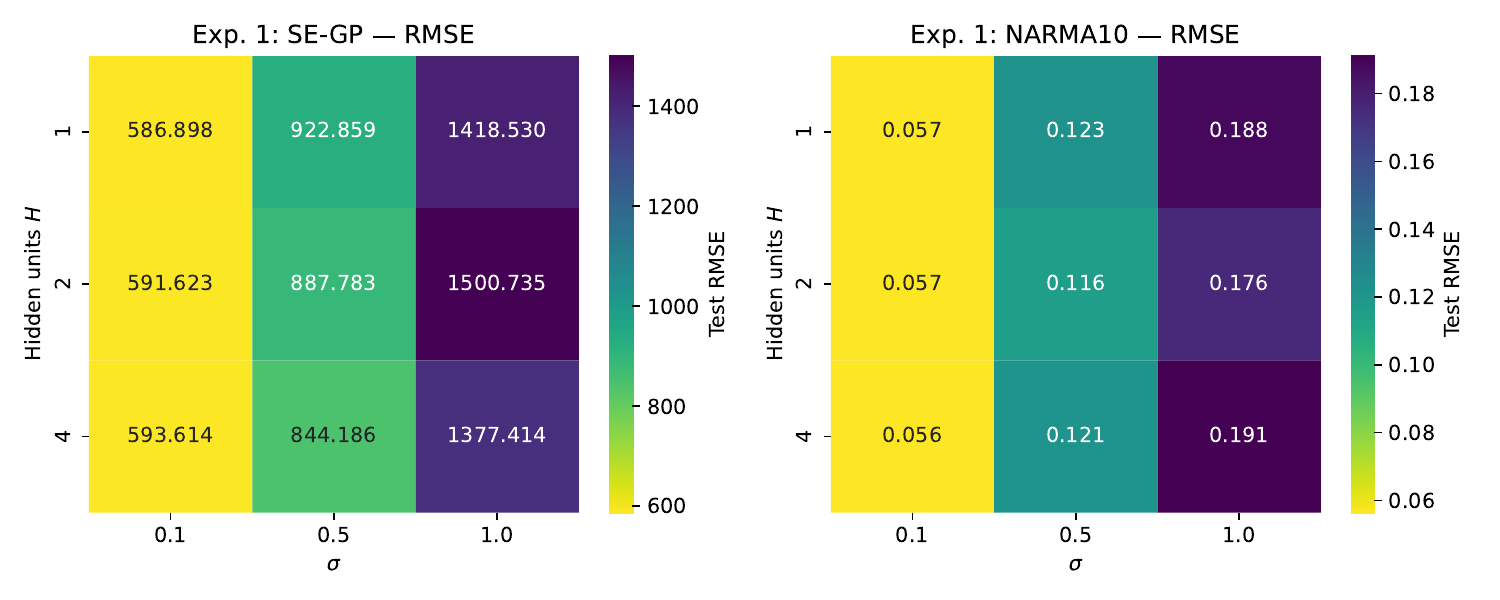}
\caption{Exp.~1: RMSE heatmap $(H,\sigma)$ for GP (left) and NARMA-10 (right) and mean values of $k$.}
\label{fig:exp1}\end{figure}

Both datasets select the smallest visible noise $\sigma^*{=}0.1$ and three CD
steps, but differ in hidden capacity (Fig.~\ref{fig:exp1}): GP prefers the most parsimonious model
($H^*{=}1$), while NARMA-10 prefers $H^*{=}2$.  The dataset-dependence of $H^*$
motivates separate tuning per task.
With limited observations split 70/15/15\% and high output variance,
validation RMSE favours parsimonious models within the tested grid
$H\in\{1,2,4\}$; larger sample sizes or a wider grid would clarify this.

\subsection{Experiment 2:  Quantum Value-Add with Symmetric Search}\label{sec:exp2}

All models are evaluated at their grid-selected hyperparameters (Table~\ref{tab:hpgrid}),
20 epochs, 4 seeds, 3 series per dataset.  The per-dataset reference baseline is the
best-performing classical CRBM: CRBM$(H^*)$ on GP and CRBM$(3H^*)$ on NARMA-10.
\textbf{All three quantum architectures} are compared: QCRBM, QFeatureQRBM,
and QQRBM (full-register, $V{=}2, H{=}2, n_q{=}7$, 51 total params).

\textit{QQRBM parameter note.}  QQRBM registers $V{+}H{+}U = 7$ qubits for
$V{=}2, H{=}2, U{=}3$ (the implementation requires
$V\ge 2$, see Section~\ref{sec:qqrbm}).  It deliberately ingests only the last
$U_{qq}{=}3$ context values rather than the full
10-step window used by the other models, which keeps its register at
$n_{\text{wires}}{=}7$: QQRBM is retained as a structural reference at its natural
quantum-register size, not as a memory-matched competitor, so it is intentionally
not given the full context.  Its 51 total parameters exceed
CRBM$(H^*)$ at the illustrative $H{=}4$ reference (${\approx}30$ params,
Table~\ref{tab:models}), but the full-register quantum encoding
is its defining structural feature, not a tunable knob.  Parameter-matched
comparison is addressed in Experiment~10.

\begin{table}[ht]\centering
\caption{Exp.~2. Mean RMSE~$\pm$~SD. Paired $t$-test $p$-values Holm-corrected
  for 7 comparisons against the per-dataset classical reference (CRBM$(H^*)$ on
  GP, CRBM$(3H^*)$ on NARMA-10; $n{=}12$).
  ${}^*\!p{<}0.05$, ${}^{***}\!p{<}0.001$ (Holm-corrected).}\label{tab:exp2}
\small
\begin{tabular}{lcccc}\toprule
 & \multicolumn{2}{c}{\textbf{GP}} & \multicolumn{2}{c}{\textbf{NARMA-10}}\\\cmidrule(lr){2-3}\cmidrule(lr){4-5}
Model & RMSE & $p_\mathrm{adj}$ & RMSE & $p_\mathrm{adj}$\\\midrule
CRBM$(H^*)$     & $584\pm511$      & ref                  & $0.057\pm0.015$ & $1.000$\\
CRBM$(3H^*)$    & $605\pm521$      & $0.187$              & $0.056\pm0.013$ & ref\\
QCRBM        & $607\pm537$      & $0.187$              & $0.056\pm0.014$ & $1.000$\\
QFeatureQRBM    & $1036\pm419$     & $0.060$              & $0.105\pm0.009$ & $<0.001^{***}$\\
QQRBM ($H{=}2$, $n_q{=}7$) & $1456\pm803$ & $0.024^{*}$ & $0.119\pm0.028$ & $<0.001^{***}$\\
\bottomrule
\multicolumn{5}{l}{\footnotesize QQRBM: 51 params ($V{=}2,H{=}2,n_q{=}7,L_Q{=}1$);
  structural reference, not iso-parameter match.}\\
\multicolumn{5}{l}{\footnotesize GP RMSE in raw units; NARMA-10 RMSE in raw (unstandardised) units.}\\
\multicolumn{5}{l}{\footnotesize NARMA-10 window references (raw units): mean-predictor $\sigma(y){=}0.109$;}\\
\multicolumn{5}{l}{\footnotesize linear-window asymptote $0.064$; nonlinear-window floor $0.056$ (Section~\ref{sec:exp2}).}\\
\end{tabular}\end{table}

\begin{figure}[ht]\centering
\includegraphics[width=\linewidth]{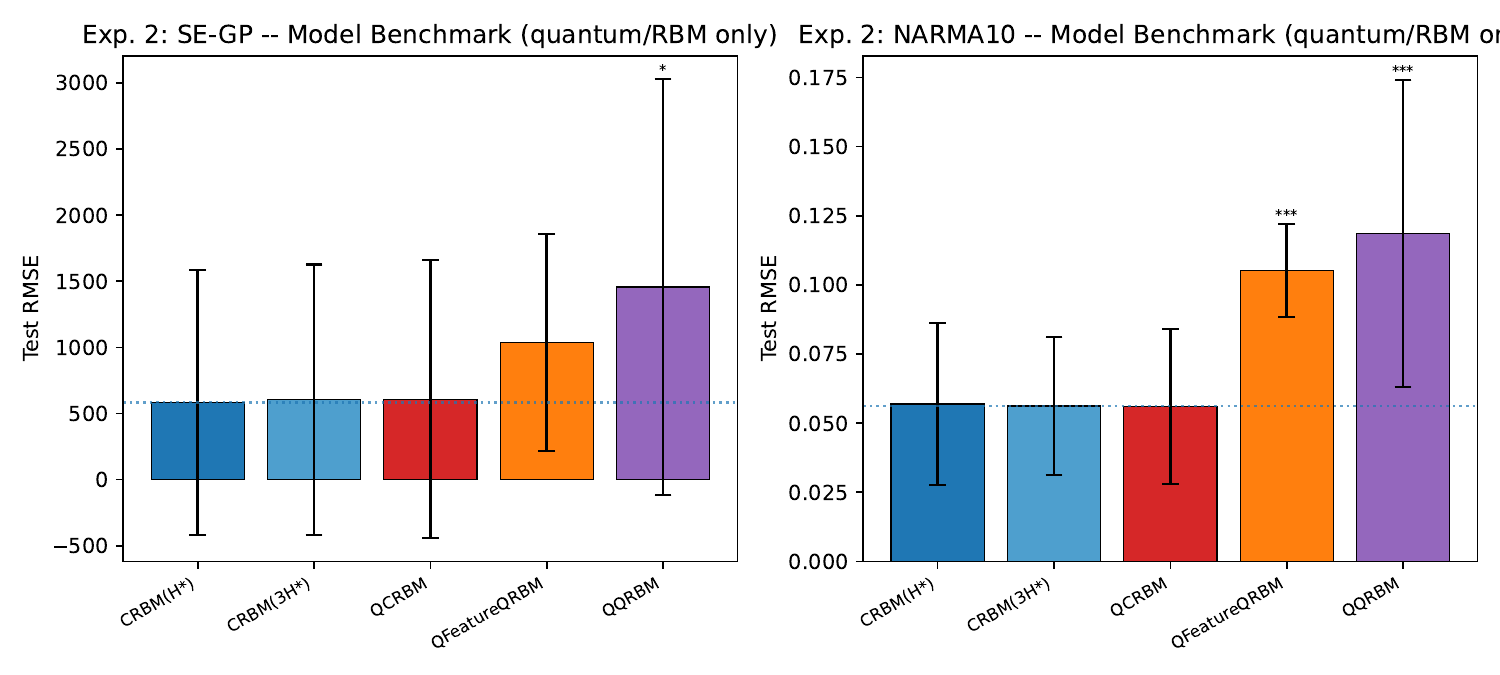}
\caption{Exp.~2: Mean RMSE by model for GP (left) and NARMA-10 (right).
  Error bars: $\pm$1~SD across $n{=}12$ paired observations.
  Holm-corrected significance vs.\ CRBM$(H^*)$: ${}^*\!p{<}0.05$,
  ${}^{**}\!p{<}0.01$, ${}^{***}\!p{<}0.001$ (e.g.\ QQRBM on GP,
  $p_\mathrm{adj}{=}0.024$, is marked~${}^*$). GP RMSE in raw units; NARMA-10 in
  raw (unstandardised) units.  On NARMA-10 the horizontal references mark the
  mean-predictor $\sigma(y){=}0.109$, the linear-window asymptote $0.064$, and the
  empirical nonlinear-window floor $0.056$: the top models sit at the nonlinear
  floor, while QQRBM reaches mean-predictor level.}
\label{fig:exp2}\end{figure}

\textbf{GP results.}  On smooth near-linear GP data (Fig.~\ref{fig:exp2}), CRBM$(H^*)$ ($\mathrm{RMSE}{=}584\pm511$)
achieves the lowest mean RMSE, closely followed by
CRBM$(3H^*)$ ($\mathrm{RMSE}{=}605\pm521$).  All quantum variants produce higher mean RMSE than the
classical baselines.  QQRBM ($\mathrm{RMSE}{=}1456\pm803$) is the only model reaching Holm-corrected
significance, being statistically significantly \emph{worse} than CRBM$(H^*)$
($p_\mathrm{adj}{=}0.024$).  QCRBM($\mathrm{RMSE}{=}607\pm537$) is statistically
indistinguishable from the classical reference ($p_\mathrm{adj}{=}0.187$);
QFeatureQRBM ($\mathrm{RMSE}{=}1036\pm419$) trends worse but does not survive correction
($p_\mathrm{adj}{=}0.060$).  No quantum model improves on the classical baseline.

\textbf{NARMA-10 results.}  On NARMA-10, the classical models cluster tightly at the
top: CRBM$(3H^*)$ ($\mathrm{RMSE}{=}0.056\pm0.013$), QCRBM ($\mathrm{RMSE}{=}0.056\pm0.014$),
and CRBM$(H^*)$ ($\mathrm{RMSE}{=}0.057\pm0.015$) are all statistically indistinguishable
($p_\mathrm{adj}{=}1.0$).  Notably, QCRBM matches the best classical model exactly
to three decimals, with no detectable advantage or disadvantage.  In contrast, both
fully quantum architectures are significantly \emph{worse} than the classical
reference: QFeatureQRBM ($\mathrm{RMSE}{=}0.105\pm0.009$, $p_\mathrm{adj}{<}0.001$) and QQRBM
($\mathrm{RMSE}{=}0.119\pm0.028$, $p_\mathrm{adj}{<}0.001$).

\textbf{Key finding:}  On neither data regime does any quantum architecture improve
on the best classical baseline.  QCRBM, the hybrid model that recovers the
classical CRBM at $\alpha=0$ (the quantum term enters additively as $\alpha\,q_\mathrm{out}$
and vanishes there), is statistically indistinguishable
from the classical reference on both datasets ($p_\mathrm{adj}{=}0.187$ on GP,
$p_\mathrm{adj}{=}1.0$ on NARMA-10).  The fully quantum QQRBM is significantly worse
on both ($p_\mathrm{adj}{=}0.024$ on GP, $p_\mathrm{adj}{<}0.001$ on NARMA-10,
$n{=}12$ paired observations).

\paragraph{Connection of the $h{=}1$ NARMA-10 tie to a window-information ceiling}
The top models cluster at $\mathrm{RMSE}{\approx}0.056$, which raises a fair
question: how close is this to the lowest error achievable from the 10-step drive
window alone?  We answer it for the actual input-driven task with a large
auxiliary simulation (pooling ${\sim}2{\times}10^5$ input--output pairs from
independent NARMA-10 series),
reporting three references in raw units: a mean-predictor reference
$\sigma(y){=}0.109$ (which reproduces the formulation-independent
$\sigma(y){\approx}0.109$); a large-sample \emph{linear}-window asymptote of
$0.064$; and an empirical \emph{nonlinear}-window floor of $0.056$, a
gradient-boosting upper bound on
$\mathbb{E}[\mathrm{Var}(y\mid\text{window})]^{1/2}$.  The study's best
models, CRBM$(3H^*)$, and QCRBM. All cluster at this ${\approx}0.056$
ceiling, comfortably inside the $\pm0.014$ seed spread.  The gap between the
large-sample linear ($0.064$) and nonlinear ($0.056$) asymptotes is small
(${\approx}0.008$), and at the experiment's finite sample size even a linear model
reaches ${\approx}0.056$ (small per-series test windows admit modest small-sample
optimism).  The $h{=}1$ NARMA-10 tie is therefore a
\emph{window-information ceiling}: given only the 10-step drive window,
${\approx}0.056$ is essentially the best attainable, so no model class, classical
or quantum, can separate there.  At the available sample size, linear,
classical-nonlinear, and hybrid models are all indistinguishable at this ceiling;
the ${\approx}0.008$ nonlinear headroom that the large-sample diagnostic reveals is
below what $n{=}12$ can resolve.  It is therefore the quantum-vs-classical contrast
that the ceiling renders uninformative at $h{=}1$.
Discriminative signal instead comes from the models that fail to reach the ceiling
(QQRBM at $0.119{\approx}\sigma(y)$, i.e.\ mean-predictor level; QFeatureQRBM at
$0.105$) and from the GP and data-efficiency evidence.

\paragraph{Connection of QQRBM's NARMA-10 deficit and information starvation}
QQRBM ingests only the last $U_{qq}{=}3$ drive values as context (plus the
contemporaneous $u(t)$ in its visible register, i.e.\ $u(t{-}3..t)$), against the
full $U{=}10$ window of the other models (Section~\ref{sec:qqrbm}); crucially it
never observes $u(t{-}9)$, the first factor of the dominant bilinear term
$1.5\,u(t{-}9)\,u(t)$, so part of its NARMA-10 deficit could be reduced-context
starvation rather than the quantum architecture.  We test this with a classical
control: a CRBM$(H^*)$ restricted to the same last-3 context (\emph{CRBM-U3}),
identical in every other respect ($n{=}12$ paired seed\,$\times$\,series).  On
NARMA-10, CRBM-U3 ($\mathrm{RMSE}{=}0.076\pm0.005$) is significantly worse than the full-context
CRBM$(H^*)$ ($\mathrm{RMSE}{=}0.057\pm0.014$; paired $t$-test $p{=}1.4\times10^{-5}$, Holm-adjusted
within the control's two-comparison family to $2.8\times10^{-5}$; Wilcoxon
$p{=}4.9\times10^{-4}$), so the 3-step window does cost accuracy, yet QQRBM
($\mathrm{RMSE}{=}0.119\pm0.028$) remains significantly worse than CRBM-U3
($p{=}2.3\times10^{-4}$, Holm $2.3\times10^{-4}$; Wilcoxon $4.9\times10^{-4}$).
Decomposing the QQRBM-vs-CRBM$(H^*)$ gap ($0.062$) into additive mean
contributions an approximation, as it assumes the two effects add the reduced
context accounts for approximately $32\%$ ($0.020$) and the quantum architecture
for the remaining ${\sim}68\%$ ($0.042$).  The NARMA-10 $p_\mathrm{adj}{<}0.001$ of
Table~\ref{tab:exp2} therefore reflects, in part, the deliberately reduced context
window and not solely the quantum architecture.  On GP the same control shows
\emph{no} context effect (CRBM-U3 $582$ vs CRBM$(H^*)$ $589$; paired $t$-test
$p{=}0.59$, Wilcoxon $p{=}0.52$), so there the deficit is entirely
architectural.\footnote{The control trains its own CRBM$(H^*)$ reference 
rather than reusing the Experiment~2 records, so its
GP CRBM$(H^*)$ mean ($\mathrm{RMSE}{=}589$) differs from Table~\ref{tab:exp2}'s ($\mathrm{RMSE}{=}584$) by
${\approx}1\%$ well within the seed spread and the two are not expected to be
identical.} This refines without overturning QQRBM's role as a fixed-size
structural reference: its full quantum register comes at the cost of a smaller
usable context, and both contribute to its weaker NARMA-10 forecasts.

\paragraph{Scale-free GP metrics.}
Because GP RMSE is reported in raw units (${\sim}10^4$),
Table~\ref{tab:segp_scalefree} additionally reports a scale-free normalised RMSE
($\mathrm{nRMSE}{=}\mathrm{RMSE}/\sigma(y_\mathrm{test}^\mathrm{raw})$) and
$R^2{=}1{-}\mathrm{nRMSE}^2$ for external comparability; the raw-unit RMSE of
Table~\ref{tab:exp2} remains the primary metric.  LR, CRBM$(H^*)$, and QCRBM
explain ${\sim}26$--$28\%$ of the test variance (nRMSE${\approx}0.78$--$0.81$),
whereas the fully-quantum QFeatureQRBM and QQRBM are worse than the mean predictor
(nRMSE${>}1$, $R^2{<}0$); the ranking matches the raw-RMSE comparison.

\begin{table}[ht]\centering
\caption{Exp.~2 (GP) scale-free metrics for external comparability: normalised
  RMSE (nRMSE${=}$RMSE$/\sigma(y_\mathrm{test}^\mathrm{raw})$, the standard deviation
  of the raw test target) and $R^2{=}1{-}\mathrm{nRMSE}^2$, mean over the $n{=}12$
  runs.  Because $R^2$ is computed per run and then averaged, applying
  $1{-}\mathrm{nRMSE}^2$ to the tabulated \emph{mean} nRMSE does not reproduce the
  tabulated mean $R^2$ (Jensen's inequality).
  Raw-unit RMSE (Table~\ref{tab:exp2}) remains the primary metric.}\label{tab:segp_scalefree}
\small\begin{tabular}{lcc}\toprule
Model & nRMSE & $R^2$\\\midrule
CRBM$(H^*)$   & $0.80$ & $0.26$\\
CRBM$(3H^*)$  & $0.83$ & $0.22$\\
QCRBM      & $0.81$ & $0.26$\\
QFeatureQRBM  & $1.72$ & $-2.17$\\
QQRBM         & $2.30$ & $-5.35$\\
\bottomrule\end{tabular}\end{table}

\subsection{Experiment 3: Lag-Window Sensitivity}\label{sec:exp3}

Lag depth $L\in\{2,4,6\}$ is swept for QFeatureQRBM and CRBM on both data
types ($L{=}8$ dropped as the optimum plateaus by $L{=}6$).
The quantum pathway norm $\|\bm\alpha_h\|_2$ is recorded.

\begin{figure}[ht]\centering
\includegraphics[width=\linewidth]{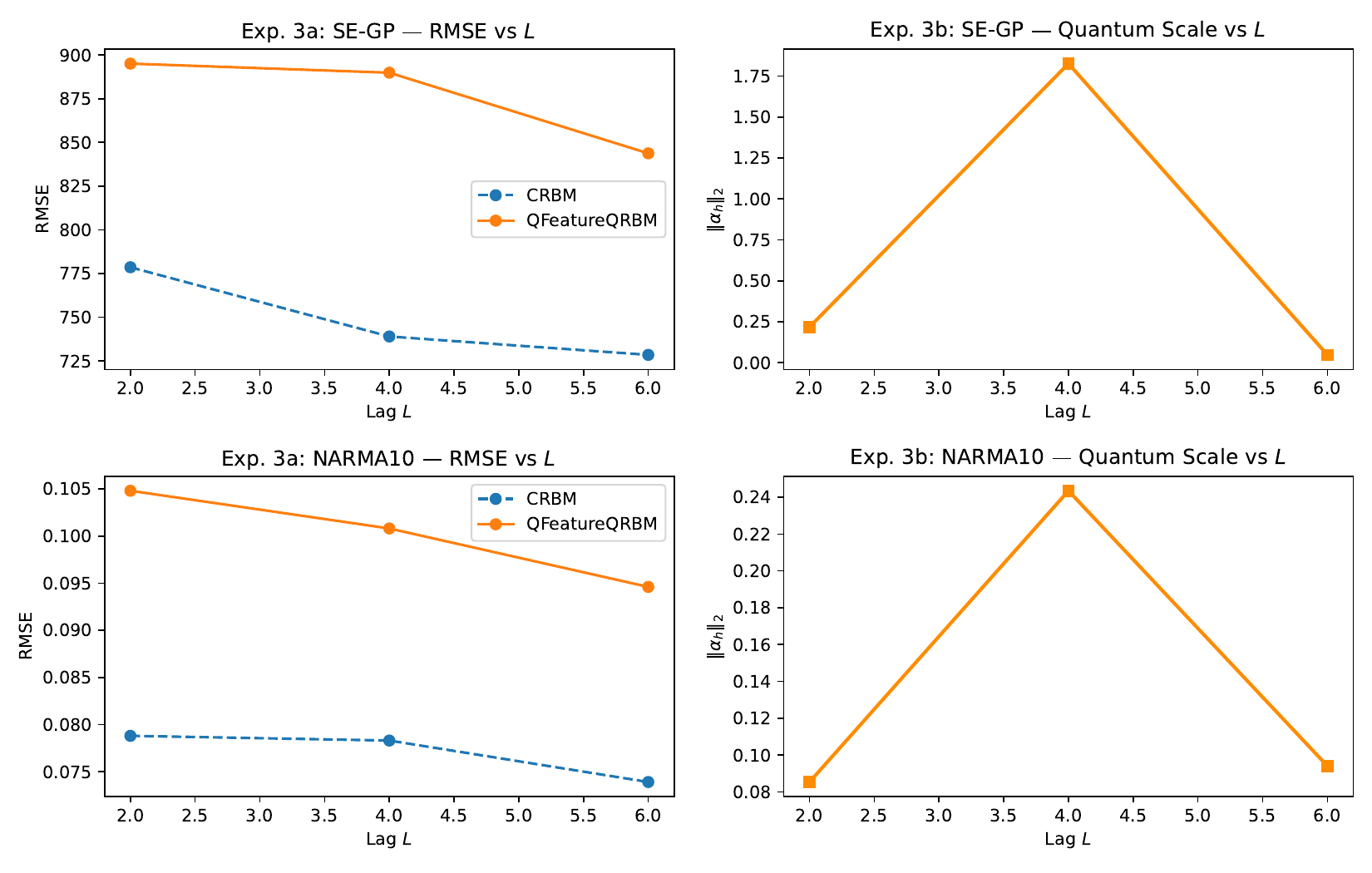}
\caption{Exp.~3: RMSE vs.\ $L$ and $\|\bm\alpha_h\|_2$ vs.\ $L$ for GP
  (top) and NARMA-10 (bottom).}
\label{fig:exp3}\end{figure}

For GP: optimal $L^*=6$ for both models (Fig.~\ref{fig:exp3}); QFeatureQRBM underperforms CRBM
at all $L$ ($p{=}0.0046$ at $L{=}6$, paired $t$-test, raw).  For NARMA-10: optimal
$L^*=6$ for both QFeatureQRBM (covering a substantial portion of NARMA-10's 10-step
nonlinear memory) and CRBM.  The QFeatureQRBM significantly underperforms CRBM at
$L=6$ ($p{=}0.0008$) on NARMA-10.
These $L^*$ values characterise sensitivity only; the main model comparison
(Experiments~1--2, 4--12) fixes $L=10$ for all models and does \emph{not}
propagate $L^*=6$ (see Section~\ref{sec:data}).

\subsection{Experiment 4: Hidden-Unit Scaling}\label{sec:exp4}

\begin{figure}[ht]\centering
\includegraphics[width=\linewidth]{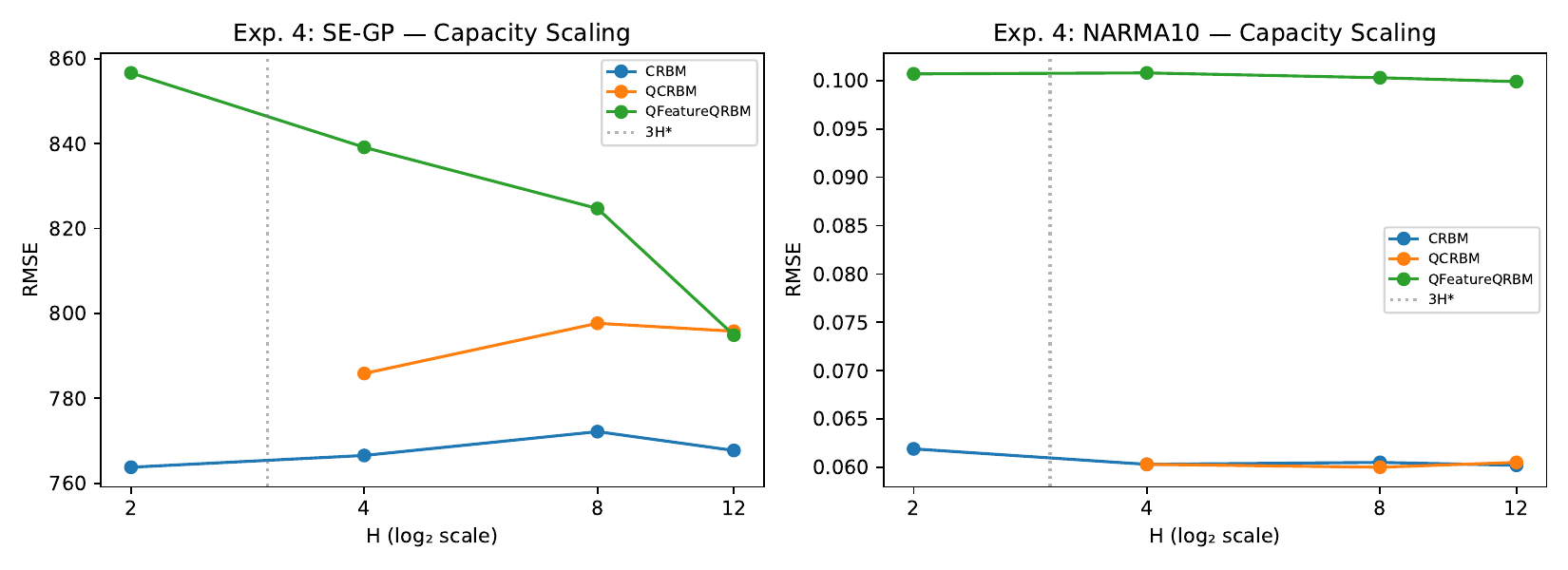}
\caption{Exp.~4: RMSE vs.\ $H$ for GP (left) and NARMA-10 (right).}
\label{fig:exp4}\end{figure}

RMSE as function of $H$ in the regime  $H\in\{2,4,8,12\}$ is recoded for CRBM, QCRBM, QFeatureQRBM
(intermediate $H$ values dropped since the trend is well-captured by
the remaining four points).
The slopes of RMSE vs.\ $H$ are very small (Fig.~\ref{fig:exp4}). On GP the classical slope is
$2.23$ (95\% CI $[-15.4,19.8]$) and the QCRBM slope is $6.87$ (95\% CI
$[-38.6,52.4]$); both intervals span zero, so no capacity-scaling trend can
be established.  On NARMA-10 the slope of CRBM is $-0.0006$ (95\% CI
$[-0.0006,-0.0005]$) and the QCRBM slope is $0.0001$ (95\% CI $[-0.0006,0.0008]$),
i.e.\ both pathways are essentially flat in $H$.  Across all $H$, QCRBM tracks
the classical CRBM almost exactly (e.g.\ NARMA-10: $0.060$ vs.\ $0.060$ at $H{=}8$),
while QFeatureQRBM sits markedly higher ($\approx 0.10$) at every capacity.  There
is no evidence that quantum augmentation alters the (essentially absent) capacity
scaling.

\subsection{Experiment 5: Ablation Study}\label{sec:exp5}

In the following, four QFeatureQRBM variants are evaluated at the best hyperparameters
(Table~\ref{tab:hpgrid}) for the ablation study.  
Variant~A is the full model with trained $\alpha$.
We choose variants B and C to isolate the quantum contribution under matched gating.
In B, the PQC remains trainable but the gate is fixed at $\alpha=2$,
so any improvement must come from a learned quantum feature map rather than from suppressing the quantum branch.
In C, the same fixed gate is kept but the PQC weights are frozen, which provides a 
direct control for whether trained quantum features outperform random quantum features 
at identical gate strength. Variant~D is a classical fallback with $\alpha{\equiv}{-}\infty$.

\begin{table}[ht]\centering
\caption{Exp.~5. QFeatureQRBM ablation on GP and NARMA-10.
  $\Delta$ = RMSE relative to variant~A (full trained model).}\label{tab:exp5}
\small
\begin{tabular}{lcclcc}\toprule
 & \multicolumn{2}{c}{GP} && \multicolumn{2}{c}{NARMA-10}\\\cmidrule(lr){2-3}\cmidrule(lr){5-6}
Variant & RMSE & $\Delta$ && RMSE & $\Delta$\\\midrule
A: Full (trained $\alpha$)          & $852\pm425$ & ---       && $0.102\pm0.009$ & ---\\
B: Forced $\alpha{=}2$ (q+PQC)     & $904\pm496$ & $+52$     && $0.107\pm0.010$ & $+0.005$\\
C: Forced $\alpha{=}2$ (rnd PQC)   & $908\pm501$ & $+56$     && $0.107\pm0.011$ & $+0.006$\\
D: Classical ($\alpha{\equiv}{-}\infty$) & $830\pm390$ & $-23$ && $0.099\pm0.007$ & $-0.003$\\
\bottomrule
\end{tabular}\end{table}

\begin{figure}[ht]\centering
\includegraphics[width=\linewidth]{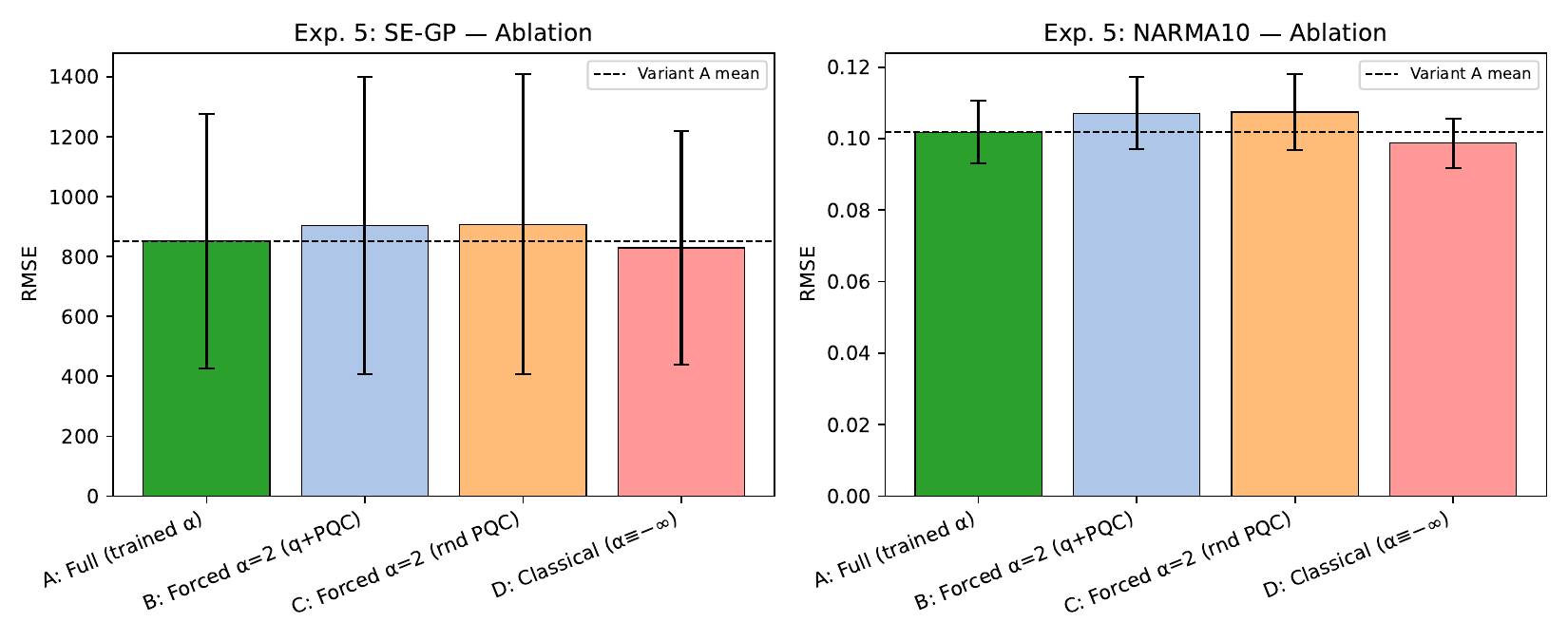}
\caption{Exp.~5: Ablation RMSE for GP (left) and NARMA-10 (right).}
\label{fig:exp5}\end{figure}

On GP (Table~\ref{tab:exp5}, Fig.~\ref{fig:exp5}), variant~D (classical fallback, $\alpha{\equiv}{-}\infty$) achieves lower RMSE
than the full model ($\Delta{=}{-}23$), confirming that quantum features are not helpful
on this data.  Forcing $\alpha{=}2$ (variants B and C) degrades performance
($\Delta{\approx}+52$ to $+56$), so the trained quantum scale does not coincide with this
fixed value.
On NARMA-10, variant~D (classical fallback) also outperforms the full model
($\Delta{=}{-}0.003$), while forcing $\alpha{=}2$ with a trained PQC (B) or random
PQC (C) both degrade RMSE by $+0.005$ to $+0.006$.  The near-identical results of B and C
indicate that the PQC weights contribute little once $\alpha$ is fixed away from its
trained value; on both datasets the classical fallback is the best of the four variants,
reinforcing that the quantum pathway adds no measurable predictive value.

\subsection{Experiment 6: Gradient Variance and Barren Plateau Analysis}
\label{sec:exp6}

To characterise the typical trainability landscape for this kind of model, we analyse the circuit for the QCRBM model with 
$n\in\{2,4,6,8,10\}$ qubits and $L_Q\in\{1,3,5\}$, sampling 50 random
parameter initialisations per $(n, L_Q)$ point and measure $\partial\mathcal{L}_Q/\partial\theta_0$ (the \emph{cost-function}
gradient, consistent with the theoretical bounds of Section~\ref{sec:barren})
via backpropagation.  We fit separate exponential models $\hat\sigma^2=a_\ell\cdot b_\ell^n$
for each depth $\ell=L_Q$.
We emphasise that $\partial\mathcal{L}_Q/\partial\theta_0$ is a single-component
proxy: the rigorous barren-plateau diagnostic is the variance of the full
gradient norm $\|\nabla_{\bm\theta}\mathcal{L}_Q\|$ over all circuit parameters
\citep{mcclean2018barren,sweke2020stochastic}.  The decay rates reported here
should therefore be read as a representative-component estimate of the trainability
trend rather than a full-gradient measurement.

\begin{table}[ht]\centering
\caption{Exp.~6. Empirical gradient variance $\hat\sigma^2$ ($\times10^{-2}$)
  and fitted decay bases $\hat{b}_\ell$ for model $\hat\sigma^2=a_\ell\cdot\hat{b}_\ell^n$.
  $N=50$ samples per $(n,L_Q)$ point.  Final column gives the local-cost reference
  $2^{-n/2}$ ($\times10^{-2}$).}
\label{tab:exp6}
\small
\begin{tabular}{ccccc}\toprule
$n$ & $L_Q=1$ & $L_Q=3$ & $L_Q=5$ & $2^{-n/2}$\\\midrule
2  & $5.68$  & $8.37$  & $9.21$  & 50.0 \\
4  & $2.44$  & $2.37$  & $2.18$  & 25.0 \\
6  & $0.68$  & $0.42$  & $0.65$  & 12.5 \\
8  & $0.18$  & $0.15$  & $0.14$  & 6.25 \\
10 & $0.046$ & $0.036$ & $0.030$ & 3.125 \\
\midrule
Fit: $\hat{b}_\ell$ & $0.622\pm0.022$ & $0.520\pm0.013$ & $0.492\pm0.007$ & \\
\bottomrule
\end{tabular}\end{table}

\begin{figure}[ht]\centering
\includegraphics[width=\linewidth]{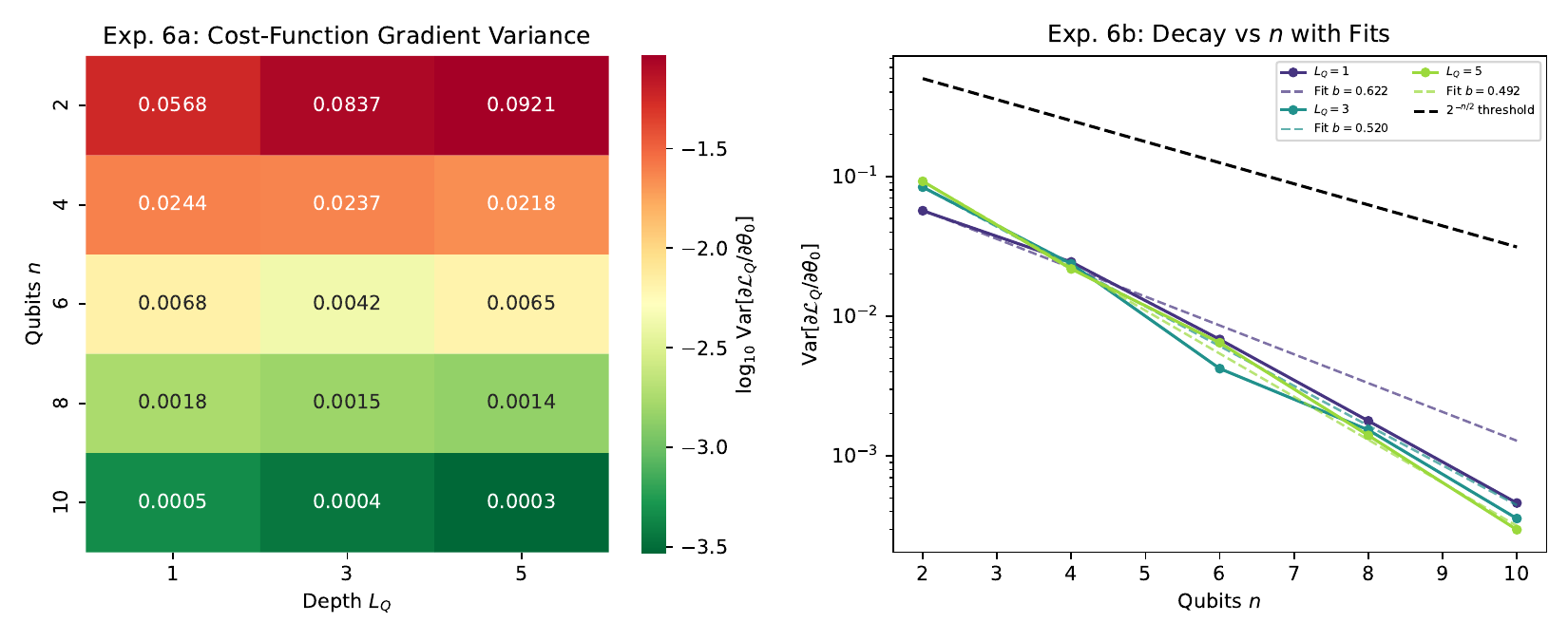}
\caption{Exp.~6: Gradient variance heatmap (left) and decay curves (right)
  with $2^{-n/2}$ reference (dashed) and exponential fits (solid).}
\label{fig:exp6}\end{figure}

Key findings (Table~\ref{tab:exp6}, Fig.~\ref{fig:exp6}):
(i) At every tested depth the gradient variance decays clearly with $n$;
the fitted base $\hat{b}_1=0.622\pm0.022$ at $L_Q=1$ already falls \emph{below}
the local-cost threshold $2^{-1/2}\approx0.707$, so even the shallowest circuit
exhibits faster-than-bound decay at these sizes.
(ii) For $L_Q=3$ the decay deepens: the fitted base
$\hat{b}_3=0.520\pm0.013$ falls well below $2^{-1/2}$, indicating that gradient
variance decays \emph{faster} than the Cerezo et al.\ \citep{cerezo2021cost}
local-cost bound predicts.  The variational layers ansatz approaches a
near-design more rapidly than a generic shallow circuit, so the $O(2^{-n/2})$
bound is not tight for this circuit family.
(iii) At $L_Q=5$, $\hat{b}_5=0.492\pm0.007$, the steepest decay observed,
confirming that the faster-than-predicted decay sharpens monotonically with depth.

\subsection{Experiment 7: Multi-Horizon Forecasting}\label{sec:exp7}

Models are trained at their respective optimal hyperparameters and evaluated
autoregressively at $h\in\{1,5,10\}$ steps (the $h{=}20$ horizon dropped as
deep extrapolation lies outside the test set at the series lengths used here).
All four model types are compared: CRBM, QCRBM, QFeatureQRBM, and QQRBM.

\begin{table}[ht]\centering
\caption{Exp.~7. Test RMSE (mean~$\pm$~SD across the $n{=}12$ seed\,$\times$\,series
  runs) at forecast horizons $h\in\{1,5,10\}$. GP RMSE in raw units; NARMA-10
  in raw (unstandardised) units; no per-horizon significance
  test is conducted.}\label{tab:exp7}
\scriptsize
\setlength{\tabcolsep}{3.5pt}
\begin{tabular}{lccc|ccc}\toprule
 & \multicolumn{3}{c|}{\textbf{GP}} & \multicolumn{3}{c}{\textbf{NARMA-10}}\\\cmidrule(lr){2-4}\cmidrule(lr){5-7}
Model & $h{=}1$ & $h{=}5$ & $h{=}10$ & $h{=}1$ & $h{=}5$ & $h{=}10$\\\midrule
CRBM$(H^*)$  & $609{\pm}442$ & $785{\pm}539$ & $809{\pm}463$ & $0.050{\pm}0.011$ & $0.055{\pm}0.012$ & $0.055{\pm}0.012$\\
CRBM$(3H^*)$ & $594{\pm}447$ & $774{\pm}560$ & $771{\pm}504$ & $0.050{\pm}0.010$ & $0.054{\pm}0.011$ & $0.054{\pm}0.011$\\
QCRBM     & $622{\pm}455$ & $803{\pm}564$ & $812{\pm}480$ & $0.050{\pm}0.011$ & $0.055{\pm}0.012$ & $0.056{\pm}0.012$\\
QFeatureQRBM & $706{\pm}319$ & $857{\pm}404$ & $861{\pm}374$ & $0.080{\pm}0.009$ & $0.087{\pm}0.008$ & $0.087{\pm}0.005$\\
QQRBM        & $763{\pm}488$ & $893{\pm}568$ & $892{\pm}558$ & $0.114{\pm}0.045$ & $0.126{\pm}0.045$ & $0.128{\pm}0.046$\\
\bottomrule
\end{tabular}\end{table}

\begin{figure}[ht]\centering
\includegraphics[width=\linewidth]{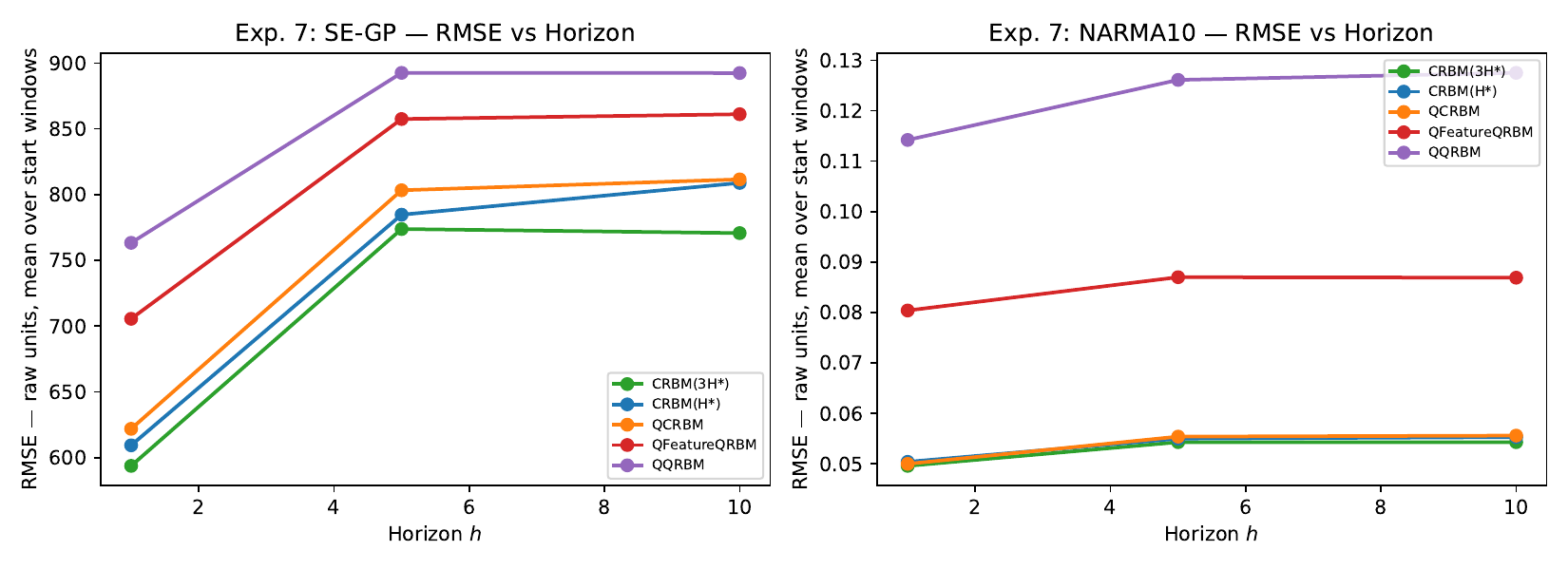}
\caption{Exp.~7: RMSE vs.\ horizon $h$ for GP (left) and NARMA-10 (right).}
\label{fig:exp7}\end{figure}

\textbf{GP:} The RMSE increases with horizon for all models (Table~\ref{tab:exp7}, Fig.~\ref{fig:exp7}), reflecting error accumulation
in autoregressive evaluation.  CRBM$(3H^*)$ (form $594$ to $774$ and $771$) achieves the lowest
multi-step RMSE overall, with CRBM$(H^*)$ (from $609$ to $785$ and $809$) close behind.  QCRBM
(from $622$ to $803$ and $812$) tracks the classical CRBMs closely across all horizons.  QQRBM
(from $763$ to $893$ and $892$) shows the highest RMSE at every horizon, consistent with its
significantly worse single-step RMSE in Exp.~2.

\textbf{NARMA-10:} The classical models and QCRBM cluster tightly at all horizons
(single-step $\approx0.050$, ten-step $\approx0.055$), with CRBM$(3H^*)$ marginally
best.  QCRBM (from $0.050$ to $0.055$ and $0.056$) tracks CRBM$(H^*)$ (from $0.050$ to $0.055$ and $0.055$)
almost exactly.  The fully quantum architectures are clearly worse at every horizon:
QFeatureQRBM (from $0.080$ to $0.087$ and $0.087$) and especially QQRBM (from $0.114$ to $0.126$ and $0.128$),
whose error is more than double the classical baseline.
Statistical significance is assessed via the primary Exp.~2 paired-$t$ test;
no per-horizon significance test is conducted here.

\subsection{Experiment 8: Training Convergence Analysis}\label{sec:exp8}

\begin{figure}[ht]\centering
\includegraphics[width=\linewidth]{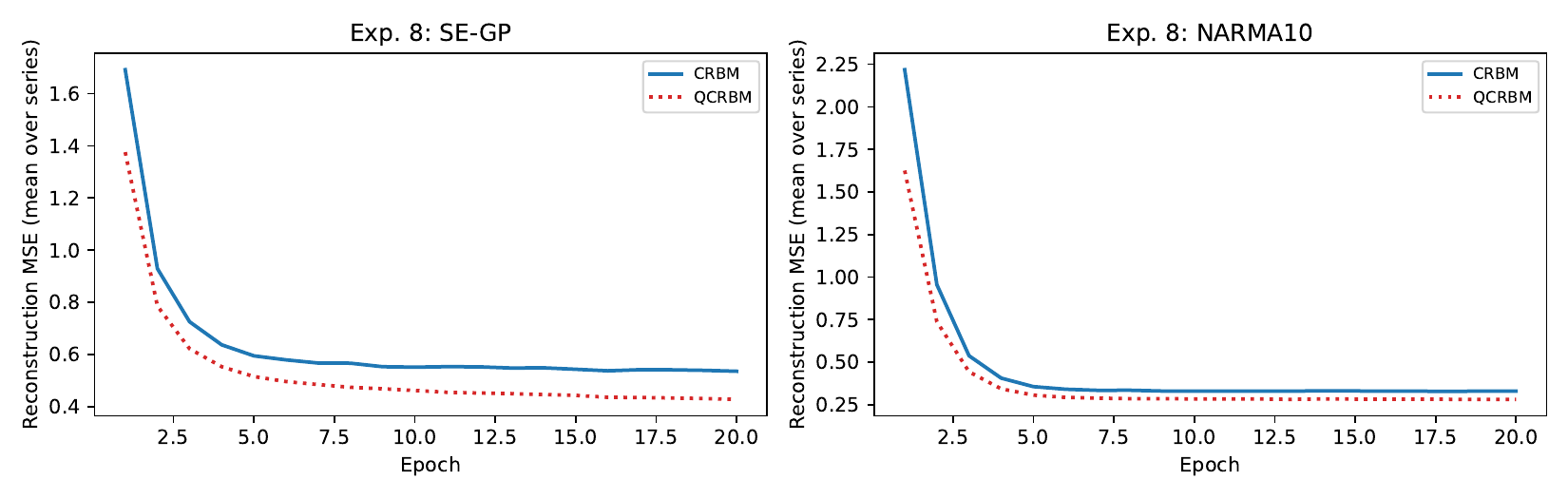}
\caption{Exp.~8: Reconstruction MSE vs.\ epoch for GP (left) and NARMA-10
  (right).  CRBM (solid), QCRBM (dotted).}
\label{fig:exp8}\end{figure}

\paragraph{Reconstruction MSE.}  Reconstruction (training-objective) MSE is
tracked per epoch for 20 epochs (CRBM, QCRBM; Fig.~\ref{fig:exp8}).  Aggregated over the
$n{=}12$ seed\,$\times$\,series runs, the final reconstruction MSE is, on GP,
CRBM $0.536\pm0.369$, QCRBM $0.428\pm0.266$; and on
NARMA-10, CRBM $0.330\pm0.029$, QCRBM $0.281\pm0.030$.
On both datasets the quantum circuits attain \emph{significantly lower}
reconstruction MSE than the classical CRBM (paired $t$-test, Holm-corrected across
the two QCRBM variants: $p_\mathrm{adj}{=}0.013$ on GP and
$p_\mathrm{adj}{<}0.001$ on NARMA-10; the Wilcoxon signed-rank test agrees,
$p{=}0.002$ and $p{<}0.001$ respectively).  Crucially, the better reconstruction MSE of the quantum models
does \emph{not} translate into lower forecast RMSE (Exp.~2), underscoring that a
better fit of the visible--hidden joint does not imply better predictive accuracy.

\begin{figure}[ht]\centering
\includegraphics[width=\linewidth]{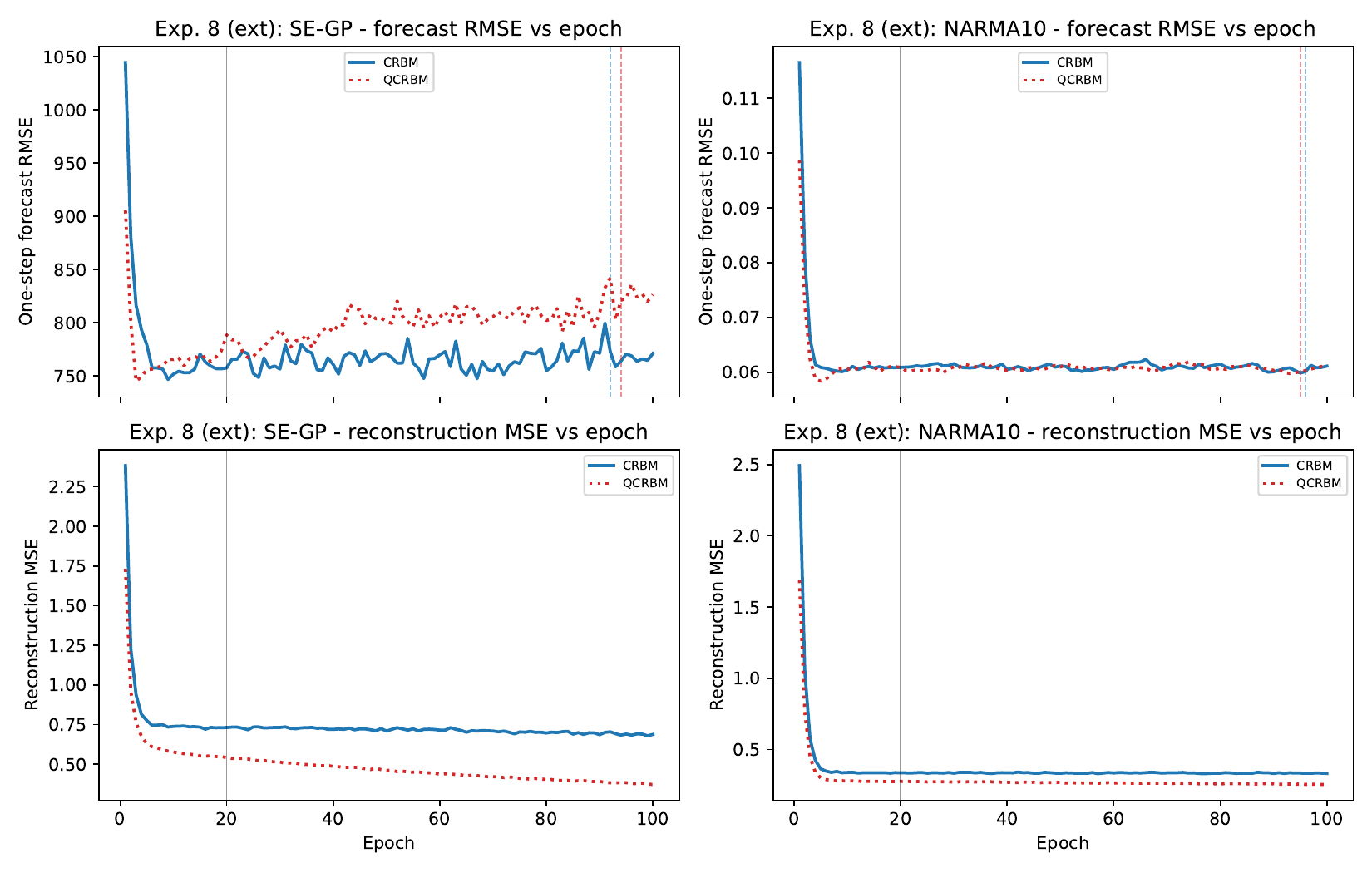}
\caption{Exp.~8: One-step \emph{forecast} RMSE (top) and
  reconstruction MSE (bottom) over 100 training epochs for CRBM (solid) and
  QCRBM (dotted), GP (left) and NARMA-10 (right).  The black vertical line
  marks the 20-epoch budget used throughout the paper; coloured dashed lines mark
  the per-model forecast-RMSE plateau.  Forecast RMSE has plateaued (or, for
  QCRBM on GP, begun to rise) by epoch~20 even where reconstruction MSE
  keeps decreasing, confirming that the 20-epoch budget does not undertrain CRBM
  and QCRBM on the reported metric (the extended analysis covers these two
  models only).}
\label{fig:exp8c}\end{figure}

\paragraph{Justification of the 20-epoch budget for training the CRBM and QCRBM models}
To rule out that the quantum models are merely undertrained at the 20-epoch budget
used throughout, we extend training to 100 epochs for CRBM and QCRBM and track
\emph{both} reconstruction MSE and one-step \emph{forecast} RMSE the reported
metric per epoch (Fig.~\ref{fig:exp8c}).  Forecast RMSE does not improve beyond
epoch~20.  On NARMA-10 it is essentially flat for both models (within $0.4\%$ of
its 100-epoch value already at epoch~20; plateau reached at epoch $\approx95$).
On GP it is at or below its long-run level by epoch~20: for QCRBM it in fact
\emph{rises} by about $5\%$ from epoch~20 ($\approx789$) to epoch~100
($\approx826$), i.e.\ further training overfits rather than helps, even as the
reconstruction MSE keeps falling ($0.54\to0.37$).  The 20-epoch budget therefore
does not disadvantage QCRBM on the reported forecast metric; if anything,
additional epochs would slightly worsen its forecast accuracy.  (This extended
analysis covers CRBM and QCRBM; QQRBM and QFeatureQRBM were not run to 100
epochs, so the claim is restricted to QCRBM.)  The
same divergence reconstruction MSE still decreasing while forecast RMSE has
plateaued or turned upward explains why the lower reconstruction MSE of the
quantum models (above) does not yield a forecasting advantage.  Mechanistically,
this is a bias/variance signature: the additional quantum parameters continue to
fit training-set structure (lower reconstruction MSE) that does not generalise to
held-out forecasts, and the comparatively light weight decay on the quantum
parameters (validation-selected $\lambda_Q{=}10^{-3}$ on GP) does not fully
arrest this, leaving QCRBM more prone to over-fitting than the smaller
classical CRBM.

\subsection{Experiment 9: PQC Depth Sensitivity}\label{sec:exp9}

\begin{table}[ht]\centering
\caption{Exp.~9. For QCRBM RMSE~$\pm$~SD and training time vs.\ $L_Q$ on GP
  and NARMA-10 are shown. Both datasets use $H{=}4$, the QCRBM floor
  $H_Q{=}\lceil\log_2(1{+}U)\rceil{=}4$ enforced because the model requires
  $2^H\ge V{+}U$ (here $V{=}1,U{=}10$, the fixed context window used for all
  models; see Section~\ref{sec:data}); total-parameter counts as reported by
  the run. }\label{tab:exp9}
\small\begin{tabular}{ccccc}\toprule
$L_Q$ & RMSE (GP) & RMSE (NARMA) & Time\,(s) & Total params\\\midrule
1 & $783\pm556$ & $0.061\pm0.015$ & 8.8 & $72$\\
2 & $778\pm550$ & $0.061\pm0.016$ & 12.5 & $84$\\
3 & $782\pm553$ & $0.061\pm0.015$ & 16.4 & $96$\\\midrule
\multicolumn{5}{l}{\footnotesize Training time scales linearly with $L_Q$; RMSE differences are within 1~SD across all depths.}\\
\bottomrule\end{tabular}\end{table}

\begin{figure}[ht]\centering
\includegraphics[width=\linewidth]{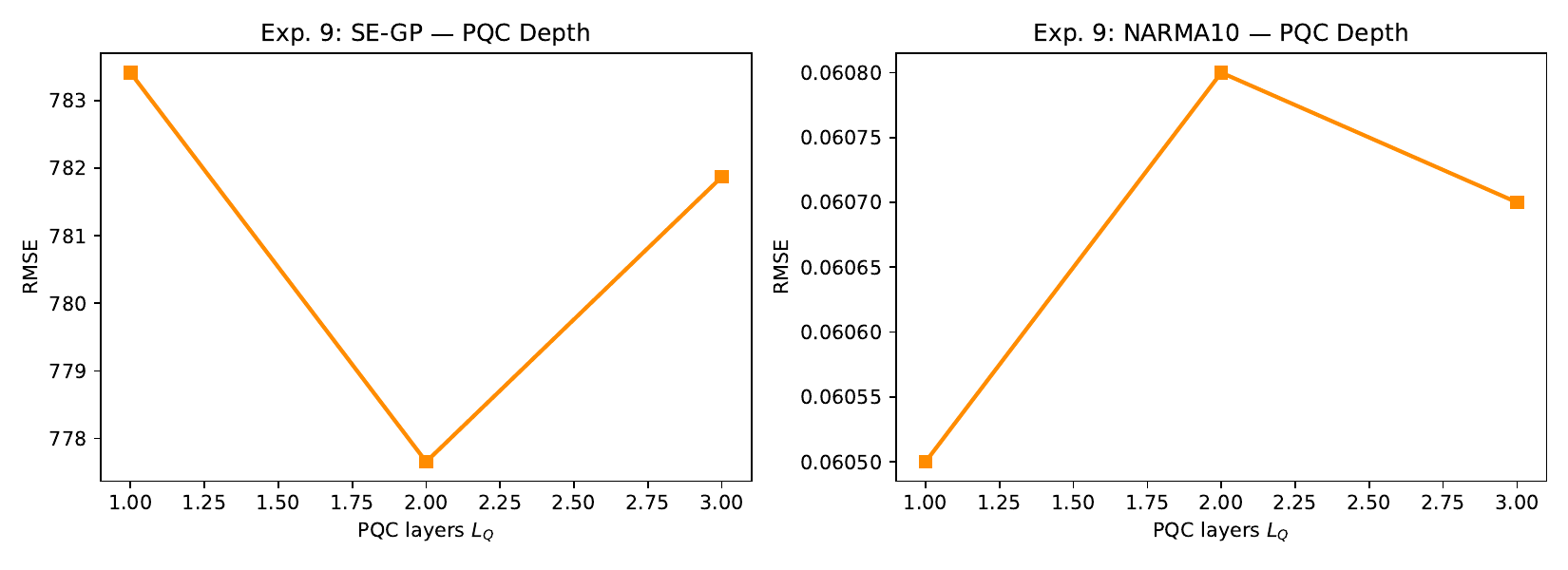}
\caption{Exp.~9: RMSE with 95\% CIs (left axis) and training time bars
  (right axis) vs.\ $L_Q$. Left: GP. Right: NARMA-10.}
\label{fig:exp9}\end{figure}

On GP (Table~\ref{tab:exp9}, Fig.~\ref{fig:exp9}), the depth differences are marginal and well within one standard deviation
at each step ($L_Q{=}1$: $783\pm556$; $L_Q{=}2$: $778\pm550$;
$L_Q{=}3$: $782\pm553$), with $L_Q{=}2$ nominally lowest.  On NARMA-10, all three
depths produce essentially identical RMSE ($\approx0.061$ throughout).  Training time
for NARMA-10 scales linearly with $L_Q$ ($8.8$, $12.5$, $16.4$\,s for
$L_Q{=}1,2,3$ respectively).  As no depth offers a statistically meaningful
improvement, $L_Q{=}2$ is retained as the default.

\subsection{Experiment 10: Iso-Parameter (Matched-Budget) Comparison}\label{sec:exp10}

Experiments~1--9 compare QCRBM at $H^*, L_Q{=}2$ against CRBM at $H^*$,
a potential parameter-count asymmetry. The other hyperparameters are fixed at their per-dataset optima 
(Table~\ref{tab:hpgrid}).
Experiment~10 addresses this by constructing iso-parameter pairs; results are
reported for four feasible budget points $P\in\{84,120,156,192\}$
(Table~\ref{tab:iso_pairs}):
\begin{equation}
  \mathrm{CRBM}(H_\mathrm{cl}) \;\approx\; \mathrm{QCRBM}(H_\mathrm{qc}, L_Q) \;\approx\;
  \mathrm{QFeatureQRBM}(H_\mathrm{qf}, L)
  \label{eq:iso_budget}
\end{equation}
with $|\Delta P|\leq4$.  We emphasise that QCRBM \emph{cannot} be
parameter-matched below $P{=}84$: amplitude encoding forces $2^{H}\ge V{+}U{=}11$,
i.e.\ $H\ge4$, so $P_\mathrm{qc}{=}12H{+}3HL_Q{+}12\ge84$ at $L_Q{=}2$.  No budget
points below $84$ are therefore reported, a continuous downward Pareto sweep
(e.g.\ $P\in\{30,50\}$) is infeasible for this architecture rather than an
omission.  Hidden-unit counts are chosen per architecture so that
total trainable parameters match each target budget (Table~\ref{tab:iso_pairs}).
QQRBM is included as a structural reference at its natural size
($V{=}2, H{=}2, n_q{=}7$, $P_\mathrm{qq}{\approx}51$ params).
Exact iso-parameter matching is not applicable to QQRBM because its qubit
register $n_\mathrm{wires}=V+H+U$ is architecture-fixed and the model requires
$V\ge 2$.

\begin{table}[ht]\centering
\caption{Iso-parameter pairs design for Experiment~10 ($|\Delta P|\leq4$).
  QQRBM included at natural architecture size as structural reference.
  Achieved counts (CRBM/QCRBM/QFeat): $83/84/83$, $119/120/120$, $155/156/155$,
  $191/192/190$.}\label{tab:iso_pairs}
\small\begin{tabular}{rccccc}\toprule
Budget & CRBM & \multicolumn{2}{c}{QCRBM} & \multicolumn{2}{c}{QFeatureQRBM}\\\cmidrule(lr){3-4}\cmidrule(lr){5-6}
$P\approx$ & $H$ & $H$ & $L_Q$ & $H$ & $L$ \\\midrule
 84 &  6 &  4 & 2 &  4 & 4\\
120 &  9 &  6 & 2 & 11 & 3\\
156 & 12 &  8 & 2 & 16 & 3\\
192 & 15 & 10 & 2 & 21 & 3\\\midrule
\multicolumn{6}{l}{\footnotesize QQRBM reference: $V{=}2, H{=}2, n_q{=}7, L_Q{=}1$ $\approx 51$ params}\\
\bottomrule\end{tabular}\end{table}

Both classical and quantum models use their per-dataset optimal learning rates
from Experiment~11.  For each budget point, the test RMSE averaged across seeds and series
($n{=}8$; 4 seeds $\times$ 2 series, the third series excluded for runtime) is reported.  The central question is whether the quantum models achieve
lower RMSE than CRBM at matched parameter budgets.  QQRBM appears as a single
reference point at $P{\approx}51$ (architecture-fixed, not budget-matched).

\begin{figure}[ht]\centering
\includegraphics[width=\linewidth]{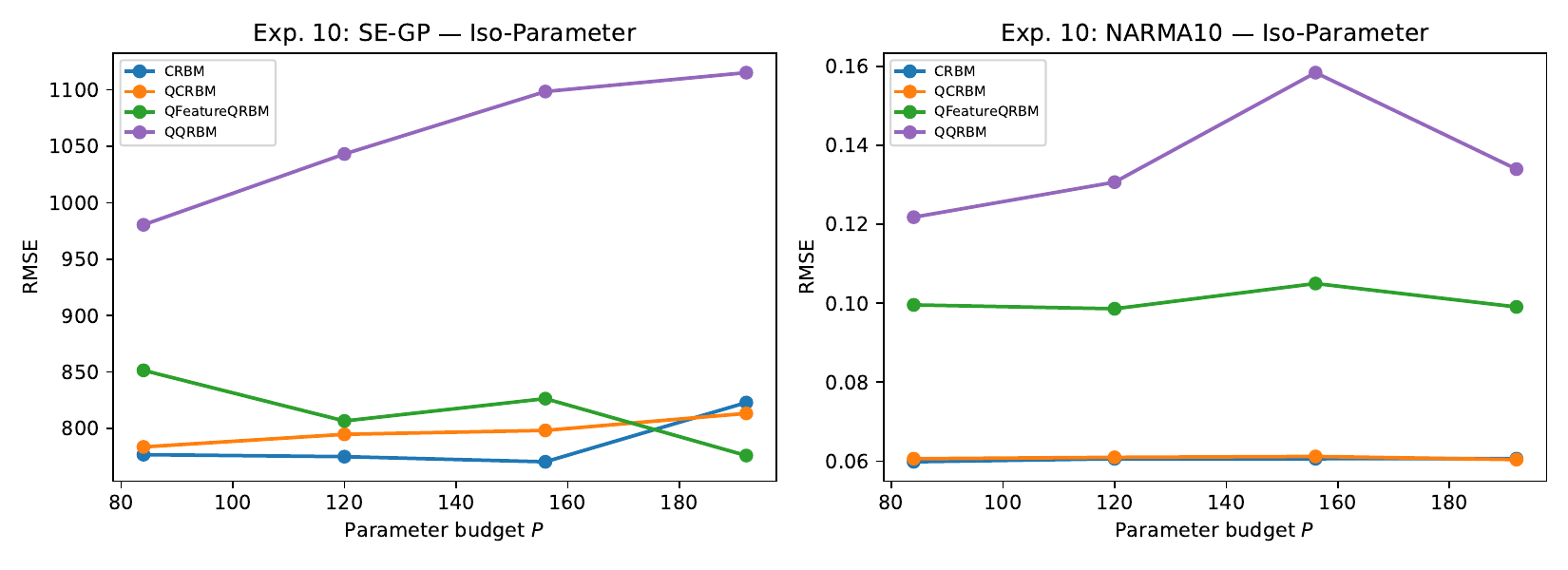}
\caption{Exp.~10: Matched-budget RMSE vs.\ parameter budget
  $P\in\{84,120,156,192\}$ for GP (left) and NARMA-10 (right).  Each point is a
  model trained at the corresponding budget ($n{=}8$ seeds\,$\times$\,series).
  Green shading marks budgets where QCRBM lies below CRBM and red shading the
  reverse; green appears only at the largest budget ($P{=}192$), where the
  difference is not statistically significant and coincides with the classical
  CRBM over-fitting at $H_\mathrm{cl}{=}15$.  Error bars: 95\,\% CI across seeds.}
\label{fig:exp10}\end{figure}

At the three smaller budgets the classical CRBM holds the lowest mean RMSE on both
datasets (Fig.~\ref{fig:exp10}).  On GP: $P{=}84$ CRBM $RMSE=777$ vs.\ QCRBM $RMSE=784$; $P{=}120$ CRBM $RMSE=775$
vs.\ QCRBM $RMSE=795$; $P{=}156$ CRBM $RMSE=771$ vs.\ QCRBM $RMSE=798$.  On NARMA-10 the
CRBM--QCRBM gap is within rounding at these budgets ($0.060$ vs.\ $0.061$).
At the largest budget $P{=}192$ the ordering reverses: on GP the CRBM degrades
to $RMSE=823$ (consistent with over-capacity on the short series at $H_\mathrm{cl}{=}15$)
and both QCRBM ($RMSE=813$) and QFeatureQRBM ($RMSE=776$) fall below it, while on NARMA-10
QCRBM ($0.0604$) marginally edges CRBM ($0.0606$).  \emph{None} of these
differences is statistically significant: the CRBM-vs-QCRBM Wilcoxon $p$ is
$\geq0.15$ at every budget on both datasets ($n{=}8$ per budget; not significant
after Holm correction).  QFeatureQRBM and QQRBM remain well above the classical
CRBM at the three smaller budgets (QQRBM on GP with $RMSE=980/1043/1098$; on NARMA-10
$RMSE=0.122/0.131/0.158$).  We therefore find \emph{no consistent and no statistically
significant quantum advantage at any matched budget}: the classical CRBM is lowest
at three of the four budgets, the apparent reversals at $P{=}192$ are not
significant and coincide with the classical model over-fitting at large width, and
the two fully quantum architectures are uniformly worse.  Whether the
large-budget reversal reflects a genuine inductive-bias advantage of the quantum
models at high capacity or merely classical over-fitting cannot be resolved at
this sample size ($n{=}8$ per budget) and is left to future work.

\begin{figure}[ht]\centering
\includegraphics[width=\linewidth]{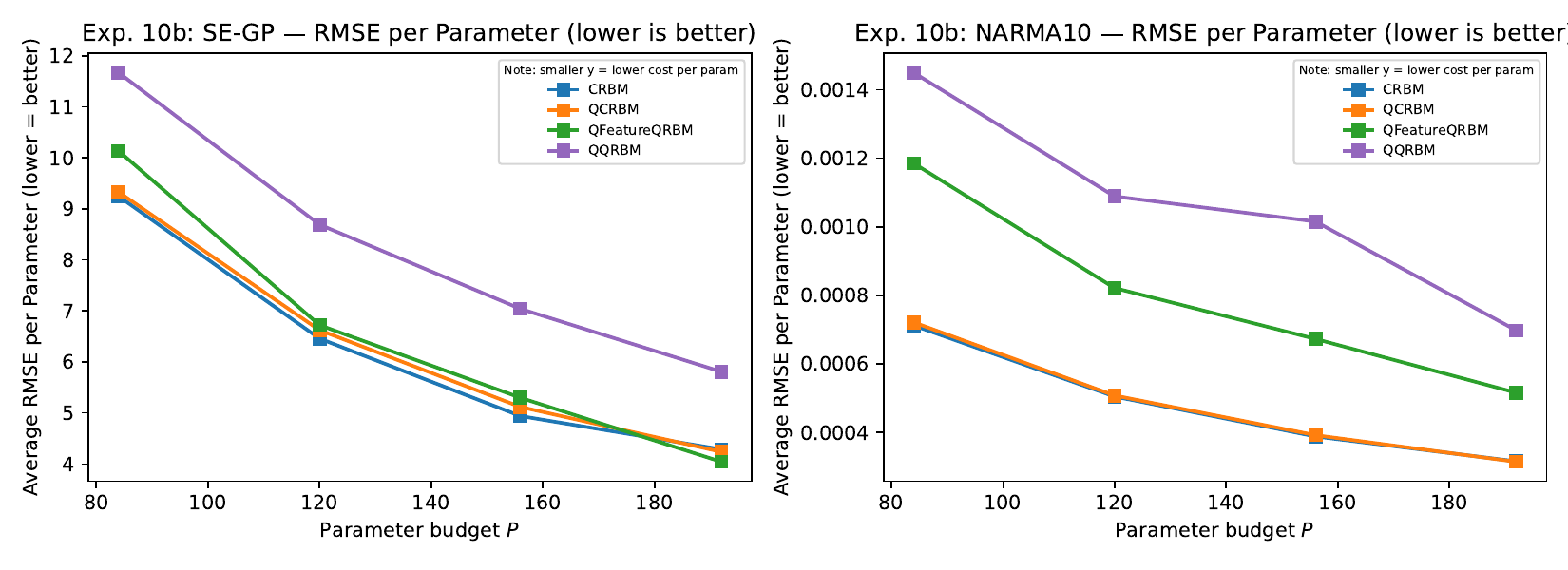}
\caption{Exp.~10b: RMSE per parameter (efficiency) for GP (left) and NARMA-10 (right)
  at each iso-parameter budget point.  Lower is better.
  CRBM (blue) and QCRBM (orange) efficiency curves; QQRBM reference point ($P{\approx}51$)
  shown as a cross.}
\label{fig:exp10b}\end{figure}

\subsection{Experiment 11: Classical Learning Rate Sensitivity}\label{sec:exp11}

In all prior experiments the classical learning rate $\eta_\mathrm{cl}$ is fixed
at $10^{-3}$.  This is the most significant missing parameter in the original
hyperparameter search, since $\eta_\mathrm{cl}$ governs the speed of CD-$k$
weight updates and interacts with the visible standard deviation~$\sigma$ and
the gradient scaling factors $\sigma^{-1}$ and $\sigma^{-2}$
in~\eqref{eq:crbm_ph}.
Experiment~11 sweeps $\eta_\mathrm{cl}\in\{10^{-4},10^{-3},10^{-2}\}$ at the
Exp.~1 optimum $(H^*,k^*,\sigma^*)$ on both data types, together with a weight-decay
sweep $\lambda_\mathrm{wd}\in\{0,10^{-3},10^{-2},0.1\}$ applied as an $L_2$ penalty on
$\bm{W},\bm{W}_{cv},\bm{W}_{ch}$.  The grid (3 learning rates $\times$ 4 weight decay values = 12 configs per dataset)
captures the essential optimum while keeping runtime tractable.

\begin{figure}[ht]\centering
\includegraphics[width=\linewidth]{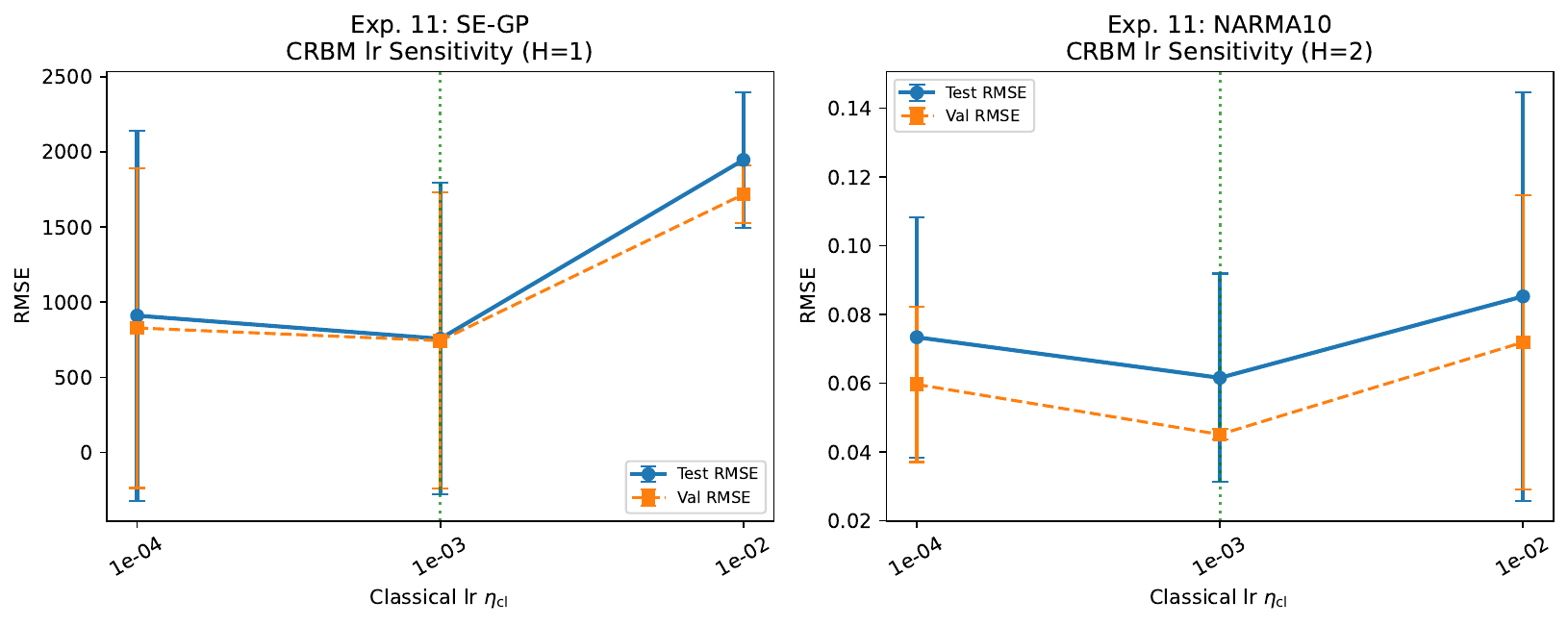}
\caption{Exp.~11: CRBM test and validation RMSE vs.\ $\eta_\mathrm{cl}$ (at
  $\lambda_\mathrm{wd}=0$) on GP (left) and NARMA-10 (right).
  The green vertical line marks $\eta_\mathrm{cl}^*$.}
\label{fig:exp11}\end{figure}

The selection criterion is validation RMSE (Fig.~\ref{fig:exp11}).  The optimal $\eta_\mathrm{cl}^*{=}10^{-3}$
on both GP ($RMSE\approx744$) and NARMA-10 ($RMSE\approx0.045$).
The smaller rate $\eta_\mathrm{cl}{=}10^{-4}$ under-fits (GP $RMSE\approx829$;
NARMA-10 $RMSE\approx0.060$), while the larger rate $\eta_\mathrm{cl}{=}10^{-2}$
degrades sharply on both datasets (GP val $RMSE\approx1717$;
NARMA-10 val $RMSE\approx0.072$), indicating numerical instability at high learning
rates, though without outright divergence.
Weight decay $\lambda_\mathrm{wd}\in\{0,10^{-3},10^{-2},0.1\}$ produces negligible
differences ($|\Delta\mathrm{RMSE}|$ in the fourth decimal or smaller) in all
configurations.  The optimal $\eta_\mathrm{cl}^*$ feeds into the iso-parameter
comparison of Experiment~10.

Now, the same calculation is done for the quantum pathway, sweeping
$\eta_Q\in\{10^{-4},10^{-3},10^{-2}\}$ and quantum weight decay
$\lambda_Q\in\{0,10^{-3},10^{-2}\}$ for the QCRBM PQC parameters~$\bm\theta$,
holding all other hyperparameters at their Exp.~1/11 optima.
The grid (3 learning rates $\times$ 3 weight decay values = 9 configs per dataset) is evaluated at
$(H^*,k_Q^*,L_Q{=}2,\alpha^*)$ on both data types.

\begin{figure}[ht]\centering
\includegraphics[width=\linewidth]{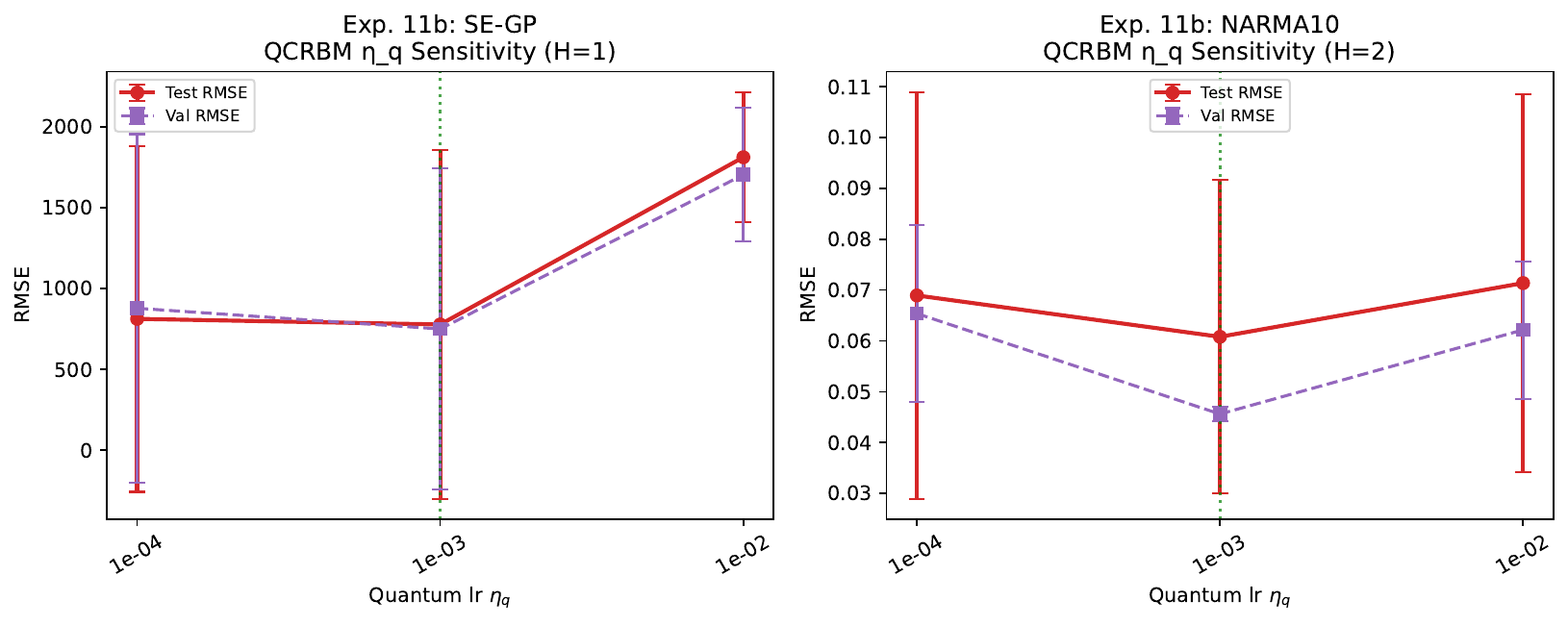}
\caption{Exp.~11b: QCRBM test and validation RMSE vs.\ $\eta_Q$ (at $\lambda_Q{=}0$)
  on GP (left) and NARMA-10 (right).
  The green vertical line marks $\eta_Q^*{=}10^{-3}$.}
\label{fig:exp11b}\end{figure}

The optimal $\eta_Q^*{=}10^{-3}$ on both GP (val RMSE$\approx750$)
and NARMA-10 (val RMSE$\approx0.046$), matching the classical optimum
from Experiment~11.  $\eta_Q{=}10^{-2}$ degrades performance on both datasets
(GP val RMSE$\approx1703$, test$\approx1812$; NARMA-10 val RMSE$\approx0.062$),
indicating instability in the PQC gradient updates at high learning rates
(Fig.~\ref{fig:exp11b}).
Quantum weight decay $\lambda_Q$ produces negligible differences
($|\Delta\mathrm{RMSE}|$ in the fourth decimal on NARMA-10; sub-unit on GP)
across all configurations, consistent with the classical weight-decay finding in
Experiment~11.  By validation RMSE the optimum is $\lambda_Q^*{=}10^{-3}$ (GP)
and $\lambda_Q^*{=}10^{-2}$ (NARMA-10), though the effect is negligible in both cases.

\subsection{Experiment 12: \texorpdfstring{$k_Q$}{k\_Q} and QFeatureQRBM Depth Search}\label{sec:exp12}

Two remaining asymmetries are addressed.  First, the QCRBM uses $k_Q=1$ CD steps;
the classical CRBM uses grid-selected $k^*\in\{1,3\}$.  We sweep
$k_Q\in\{1,2,3\}$ for QCRBM at the Exp.~1 optimum.
Second, the QFeatureQRBM PQC depth has been fixed at $n_\mathrm{layers}=1$; we
sweep $n_\mathrm{layers}\in\{1,2,3\}$.  Both sweeps use validation RMSE for
selection and feed into the final recommended configuration.

\begin{figure}[ht]\centering
\includegraphics[width=\linewidth]{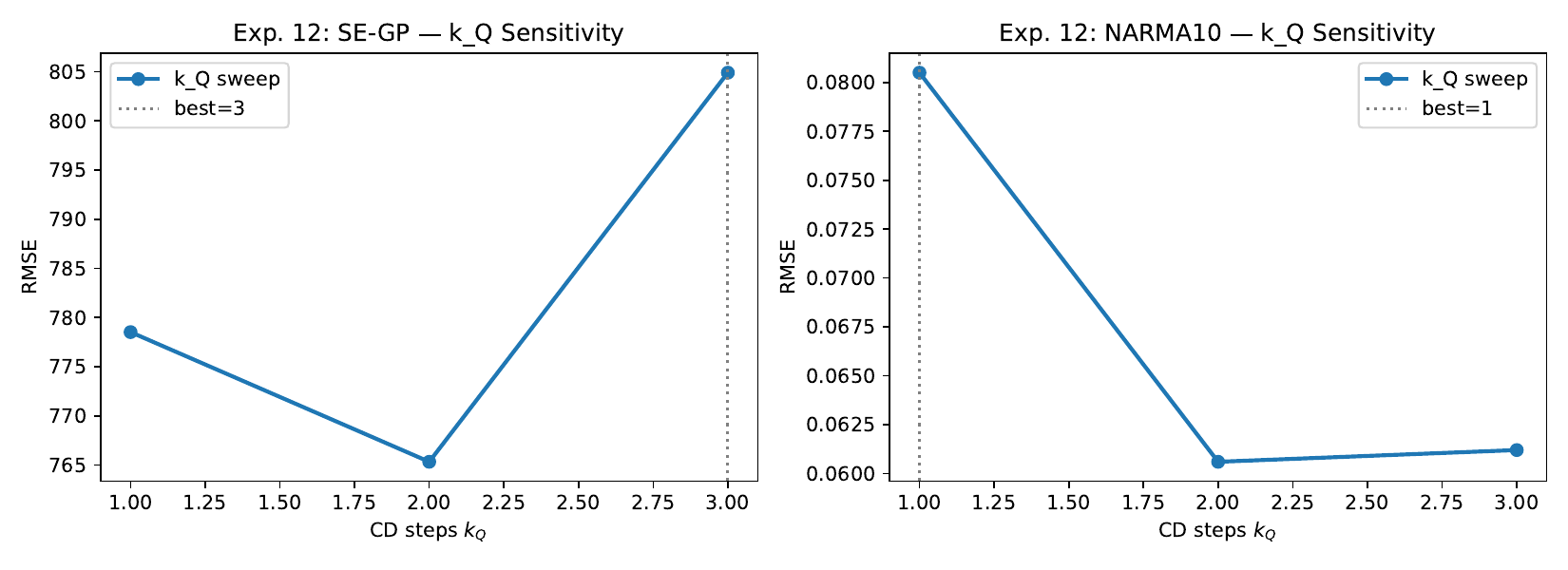}
\caption{Exp.~12: $k_Q$ sensitivity for QCRBM (left column) and $n_\mathrm{layers}$
  sensitivity for QFeatureQRBM (right column) on GP (top) and NARMA-10 (bottom).
  Green dotted line: selected optimum.}
\label{fig:exp12}\end{figure}

Both sweeps use validation RMSE for selection (Fig.~\ref{fig:exp12}).  For QCRBM $k_Q$ search:
GP selects $k_Q^*{=}3$ (mean test RMSE: $k_Q{=}1$: $779$, $k_Q{=}2$: $765$,
$k_Q{=}3$: $805$; note selection is on validation RMSE, which need not coincide
with the test minimum).  NARMA-10 selects $k_Q^*{=}1$ (mean test RMSE: $k_Q{=}1$:
$0.081$, $k_Q{=}2$: $0.061$, $k_Q{=}3$: $0.061$; again selected by validation RMSE).
The two datasets thus select different $k_Q$, and in both cases the validation
optimum does not match the lowest mean test RMSE, underscoring the noise in this
small-sample regime.
For QFeatureQRBM $n_\mathrm{layers}$ search:
GP selects $n_\mathrm{layers}^*{=}2$ (mean test RMSE: $n{=}1$: $856$,
$n{=}2$: $805$, $n{=}3$: $894$); NARMA-10 selects $n_\mathrm{layers}^*{=}1$
(mean test RMSE: $n{=}1$: $0.100$, $n{=}2$: $0.081$, $n{=}3$: $0.104$).
Across both sweeps the QFeatureQRBM remains well above the classical and QCRBM
baselines at every setting, so its depth selection does not close the gap.

\subsection{Experiment 13: Data-Efficiency Learning Curve}\label{sec:exp13}

A natural challenge to the no-advantage conclusion is that a quantum model's
expressivity might help specifically in the \emph{small-data} regime, which the
main comparison (fixed training length, $n{=}12$) does not isolate.%

Experiment~13 tests this directly.  We use the \emph{full} 19-series GP library
(rather than the three series used elsewhere), sweep the training-set size
$N_\mathrm{train}\in\{32,64,128,256,512\}$, and evaluate one-step forecasts on a
fixed held-out window common to all sizes.  Because the 19 series span roughly
four orders of magnitude in amplitude, we report a \emph{scale-free} metric the
RMSE in each series' own train-prefix-standardised units and pair
comparisons by series.  CRBM$(H^*)$ and QCRBM use their paper-selected
hyperparameters and the same $L{=}10$ autoregressive lag window as the main GP
experiments at every size (no per-size retuning, since the hypothesis concerns a
fixed architecture's data efficiency); three seeds are averaged per point. QFeatureQRBM is omitted
because its per-row circuit inference makes a sweep of this size infeasible, and it
was already markedly worse in Experiments~2,~7, and~10.

\begin{table}[ht]\centering
\caption{Exp.~13: median scaled-space RMSE over the 19 GP series at each
  training size, with the median per-series relative difference
  $(\mathrm{QCRBM}{-}\mathrm{CRBM})/\mathrm{CRBM}$ (positive $=$ quantum worse),
  the number of series on which QCRBM is lower, and the paired Wilcoxon $p$.}\label{tab:exp13}
\small\begin{tabular}{rccccc}\toprule
$N_\mathrm{train}$ & CRBM$(H^*)$ & QCRBM & rel.\ diff. & QCRBM lower & $p_\mathrm{Wilcoxon}$\\\midrule
 32 & $1.325$ & $1.313$ & $+0.4\%$ & $9/19$ & $0.74$\\
 64 & $1.049$ & $1.075$ & $+1.8\%$ & $4/19$ & $0.08$\\
128 & $0.959$ & $0.966$ & $+1.6\%$ & $5/19$ & $0.10$\\
256 & $0.927$ & $0.929$ & $+1.5\%$ & $5/19$ & $0.016$\\
512 & $0.885$ & $0.901$ & $+2.8\%$ & $6/19$ & $0.020$\\
\bottomrule\end{tabular}\end{table}

\begin{figure}[ht]\centering
\includegraphics[width=\linewidth]{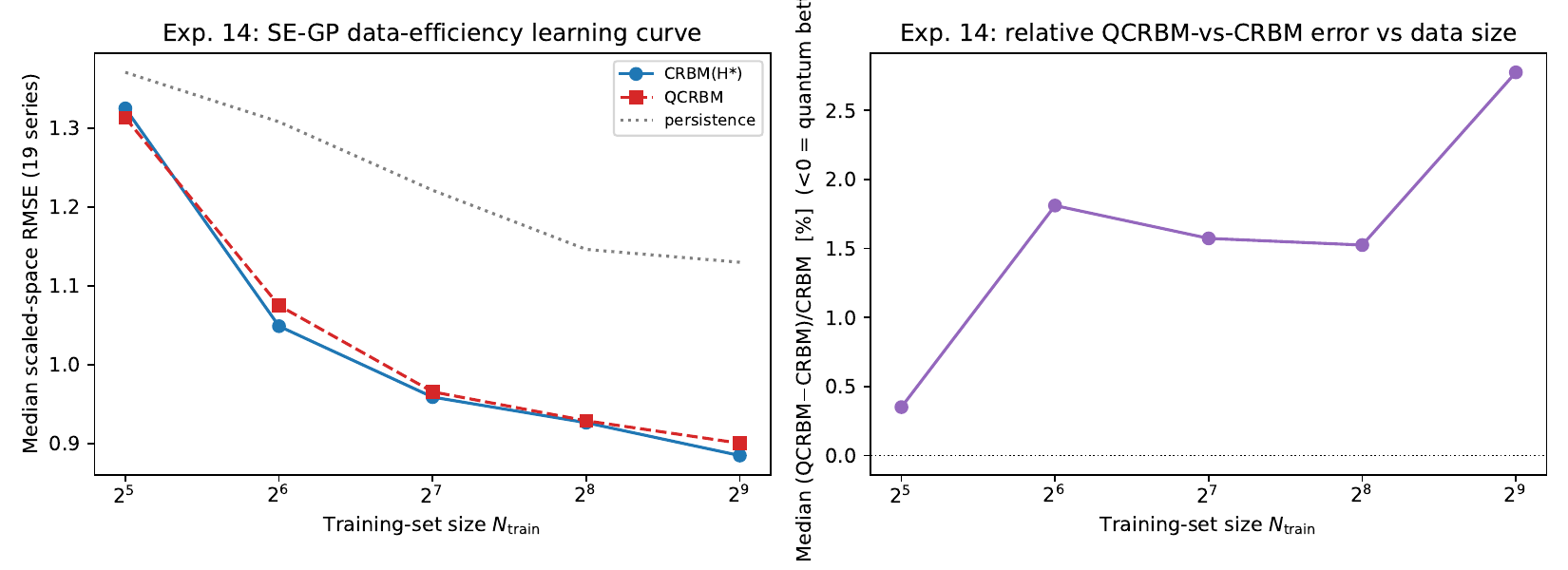}
\caption{Exp.~13: GP data-efficiency learning curve (median over 19 series).
  Left: median scaled-space RMSE vs.\ training size for CRBM$(H^*)$, QCRBM, and
  a persistence baseline.  Right: median per-series relative error
  $(\mathrm{QCRBM}{-}\mathrm{CRBM})/\mathrm{CRBM}$ ($<0$ would indicate a quantum
  advantage); it is near zero at the smallest size and grows positive with data.}
\label{fig:exp13}\end{figure}

The data-efficiency hypothesis is not supported (Table~\ref{tab:exp13}, Fig.~\ref{fig:exp13}).  At the smallest training size
($N_\mathrm{train}{=}32$), where a sample-efficiency advantage would be most
visible, CRBM and QCRBM are statistically tied (Wilcoxon $p{=}0.74$; QCRBM lower
on $9$ of $19$ series; median relative difference $+0.4\%$).  As the training set
grows the classical CRBM pulls slightly ahead: the median relative difference rises
monotonically to $+2.8\%$ at $N_\mathrm{train}{=}512$, with QCRBM \emph{nominally}
worse at $N_\mathrm{train}\ge256$ (uncorrected $p\approx0.02$; Holm-corrected across
the five sizes $p\approx0.08$, not significant at $\alpha{=}0.05$).  The monotone
growth of the deficit ($+0.4\%\to+2.8\%$) is the primary evidence.
This is the \emph{opposite} of a data-efficiency
advantage, which would predict the quantum model ahead at small $N_\mathrm{train}$
and converging as data grow. (At $N_\mathrm{train}{=}32$ both models trail a
persistence baseline neither learns much from $\sim$22 supervised samples, while
both exceed it for $N_\mathrm{train}\ge64$, and a linear baseline remains
competitive throughout, consistent with GP's near-linear character.)  Across 19
series and the full training-size range we therefore find no regime of quantum
advantage, reinforcing the conclusion of Experiment~2.  This 19-series paired
design is also far more sensitive than the $n{=}12$ main protocol (whose
detectable-effect floor is $d_\mathrm{min}{\approx}0.89$): where effects do become
detectable here they are \emph{small}, \emph{negative} (the classical model ahead),
and \emph{growing} with training size---partially answering, in the classical
direction, the ``small advantages cannot be excluded'' caveat of the main analysis.

\section{Discussion}\label{sec:discussion}

\subsection{Absence of Quantum Advantage Across Both Regimes}

The classical baseline remains better than its quantum counterpart in both data regimes; at the available sample size, the quantum models cannot surpass the classical
sharper: both QQRBM ($RMSE=0.119\pm0.028$) and QFeatureQRBM ($RMSE0.105\pm0.009$) are
significantly worse than the classical reference CRBM$(3H^*)$ ($0.056\pm0.013$)
at $p_\mathrm{adj}{<}0.001$. The only quantum model that remains comparable to the classical reference is QCRBM, which is statistically indistinguishable from the classical reference on both datasets ($p_\mathrm{adj}{=}0.187$ on GP, $p_\mathrm{adj}{=}1.0$ on NARMA-10,
matching it to three decimals).  The pattern is therefore consistent rather than
data-dependent: where the quantum pathway can be switched off (QCRBM recovers
a classical CRBM at $\alpha=0$, the quantum term being additive) the model recovers classical performance, and whereas when the quantum pathway is structural and unavoidable (QQRBM, QFeatureQRBM), performance declines.

This is corroborated by the ablation of Experiment~5: on both datasets the
classical-fallback variant ($\alpha{\equiv}{-}\infty$) is the best of the four
QFeatureQRBM variants, and forcing a non-trivial quantum scale only worsens RMSE.
The mechanism is straightforward.  Amplitude encoding maps the current state into
an exponentially large Hilbert space and variation layers produces
nonlinear Pauli-$Z$ corrections to the hidden logits, but on GP the response is
near-linear in the lag, so these corrections add variance rather than signal; on
NARMA-10 the classical CRBM already captures the 10-step memory well at $H^*$, and
the quantum features do not supply additional useful inductive bias.  In both cases
the extra quantum parameters increase model variance without a compensating
reduction in bias.

\subsection{Iso-Parameter (Matched-Budget) Comparison}\label{sec:disc_iso}

Experiment~10 (four budgets $P\in\{84,120,156,192\}$, QQRBM reference at
$P{\approx}51$) directly addresses the central fairness question: does the QCRBM
outperform the CRBM at equal parameter budgets?  The answer is no in any
statistically meaningful sense.  At the three smaller budgets, CRBM is lowest on
both datasets (GP: CRBM $RMSE=777/775/771$ vs.\ QCRBM $RMSE784/795/798$ at
$P{=}84/120/156$; NARMA-10: $0.060$ vs.\ $0.061$ throughout; seed-averaged over
$n{=}8$). At the largest budget the ordering reverses on GP CRBM rises to
$RMSE=823$ (over-fitting at $H_\mathrm{cl}{=}15$) while QCRBM ($813$) and QFeatureQRBM
($RMSE=776$) fall below it, and on NARMA-10 QCRBM ($RMSE=0.0604$) marginally edges CRBM
($RMSE=0.0606$), but \emph{no} CRBM-vs-QCRBM difference is statistically significant at
any budget (Wilcoxon $p\geq0.15$ throughout).  QFeatureQRBM and QQRBM lie far above
both at the three smaller budgets (QQRBM: $RMSE=980/1043/1098$ on GP,
$RMSE=0.122/0.131/0.158$ on NARMA-10).

There is therefore no consistent or statistically significant matched-budget
quantum advantage: at best, QCRBM ties the classical CRBM, and the only budget at
which it numerically wins (the largest) is one where the classical model is
over-fitting and the difference is not significant.  This
contrasts with the theoretical picture from Demidik et al.\ \citep{demidik2025sqrbm},
whose $\sim$3$\times$ capacity-compression guarantee applies to quantum Gibbs-state
sqRBMs rather than the PQC-expectation models studied here; the present results
provide no empirical evidence of a parameter-efficiency advantage for these
variational architectures on either benchmark (Fig.~\ref{fig:exp10b}).

\subsection{The Importance of Symmetric Hyperparameter Search}

A key methodological strength of this study is that, \emph{despite} fully symmetric hyperparameter optimisation, quantum models do not surpass their classical counterparts at the available sample size. 
The grid search identifies $\alpha^*{=}3.0$ for both GP and NARMA-10, the largest
quantum scale in the grid, so the quantum contribution is amplified rather than
suppressed; even so it does not improve test RMSE.  Likewise, the QFeatureQRBM
alignment weight is selected at its largest value $\lambda_a^*{=}1.0$ on both
datasets.  In other words, the absence of advantage cannot be attributed to a
quantum pathway that was tuned too weakly.

Because $\alpha^*{=}3.0$ sits at the top of the search grid, we ran a
supplementary validation-RMSE sweep extending the additive scale to
$\alpha\in\{3,5,10\}$ under the same frozen-$\alpha$ selection protocol used in
Experiment~2 (Table~\ref{tab:alpha_boundary}; $n{=}8$ seed\,$\times$\,series).
Increasing $\alpha$ beyond 3 changes validation RMSE by less than $1\%$ far
inside one standard deviation on both datasets, with no divergence.  Thus
$\alpha^*{=}3.0$ is not a grid-boundary artifact: a larger additive quantum scale
neither helps nor destabilises, so the grid was not extended further.

\begin{table}[ht]\centering
\caption{Supplementary $\alpha$-boundary sweep: QCRBM
  validation RMSE (mean over $n{=}8$; 4 seeds $\times$ 2 series) at fixed
  $\alpha\in\{3,5,10\}$, using the same frozen-$\alpha$ protocol that selected
  $\alpha^*{=}3.0$ in Experiment~2.  No run diverged; increasing $\alpha$ beyond 3
  yields no material improvement ($<1\%$, well within one SD).  GP in raw
  units, NARMA-10 in raw (unstandardised) units.}\label{tab:alpha_boundary}
\small\begin{tabular}{lccc}\toprule
Dataset & $\alpha{=}3$ & $\alpha{=}5$ & $\alpha{=}10$\\\midrule
GP    & $748.2$  & $747.9$  & $746.6$\\
NARMA-10 & $0.0458$ & $0.0457$ & $0.0455$\\
\bottomrule\end{tabular}\end{table}

Experiment~11 (3 learning rate values, 4 weight decay values) refines the classical learning rate.
The optimal $\eta_\mathrm{cl}^*{=}10^{-3}$ on both GP and NARMA-10.
This consistent optimum ensures that the classical baselines are not
under-tuned in the iso-parameter comparison of Experiment~10.

Experiment~12 closes the two remaining quantum-specific asymmetries: the QCRBM
CD-step count $k_Q$ and the QFeatureQRBM PQC depth $n_\mathrm{layers}$.  Selection
by validation RMSE gives $k_Q^*{=}3$ on GP but $k_Q^*{=}1$ on NARMA-10, and
$n_\mathrm{layers}^*{=}2$ on GP but $n_\mathrm{layers}^*{=}1$ on NARMA-10; in
both sweeps the validation optimum does not coincide with the lowest mean test
RMSE, reflecting the noise inherent in the small-sample regime.  Crucially, at no
setting does QFeatureQRBM approach the classical or QCRBM baselines, so its depth
selection does not alter the overall conclusion.

Taken together, Experiments~11 and~12 complete a fully symmetric hyperparameter
search: every tunable scalar that affects training quality in any model family is
now optimised on the same validation criterion.  This allows all twelve experiments
to contribute to a unified and internally consistent comparison.

\subsection{Gradient Landscape: Plateau Onset at High Depth}

Experiment~6 reveals a stricter-than-expected trainability landscape.
Even at $L_Q=1$ the gradient variance decays with $n$ across
$n\in\{2,4,6,8,10\}$, with fitted base $\hat{b}_1{=}0.622\pm0.022$ already below
the local-cost threshold $2^{-1/2}{\approx}0.707$, so the variational layers
ansatz approaches a near-design even at depth~1 for these sizes, in line with the
expressibility analysis of Sim et al.\ \citep{sim2019expressibility}.
At $L_Q=3$, the fitted exponential decay base $\hat{b}_3{=}0.520\pm0.013$
falls \emph{below} the local-cost mitigation threshold $2^{-1/2}{=}0.707$,
indicating that gradient variance decreases faster than the Cerezo et al.\
\citep{cerezo2021cost} $O(2^{-n/2})$ bound predicts for generic local-cost circuits.
The variational layers ansatz with Pauli-$Z$ observables approaches a
near-design more rapidly, making the theoretical bound non-tight for this
specific circuit family.  At $L_Q{=}5$, $\hat{b}_5{=}0.492\pm0.007$, substantially
below the threshold.  Practitioners should treat $L_Q{\leq}2$ as the reliable
trainability regime for $n\leq8$ qubits with this ansatz, rather than the
wider $L_Q\leq4$ regime suggested by the theoretical lower bound alone.

\subsection{The Demidik Bound as Heuristic Reference}

The CRBM$(3H^*)$ baseline from Demidik et al.\ \citep{demidik2025sqrbm}
functions as a useful heuristic upper bound on classical capacity, but
important caveats apply.  The sqRBM equivalence theorem applies to models
using quantum Gibbs states, not PQC expectation values.  The QCRBM studied
here uses a fundamentally different quantum computational resource: pure-state
variational features rather than thermal quantum statistics.  On GP,
CRBM$(3H^*)$ ($RMSE=605\pm521$) shows negligible
difference from CRBM$(H^*)$ ($RMSE=584\pm511$), confirming that capacity is not the
limiting factor on near-linear data.  On NARMA-10, CRBM$(3H^*)$ ($RMSE=0.056\pm0.013$)
is the single best model overall and is statistically indistinguishable from
CRBM$(H^*)$ and QCRBM ($p_\mathrm{adj}{=}1.0$ for both).  No quantum architecture
matches, let alone beats, the classical capacity-upper-bound baseline: QQRBM
($RMSE=0.119$) is significantly \emph{worse} despite its larger parameter count,
providing no evidence of a parameter-efficiency advantage for these variational
quantum representations.

\subsection{Architecture Comparison}

The QCRBM's conservative-extension property (Remark~\ref{rem:cons}) makes it
the most theoretically principled architecture: classical training is unaffected
by quantum parameters, and the quantum contribution is controlled by a single
interpretable scale $\alpha$.  The QFeatureQRBM provides a compact footprint
($n_q\leq3$) suitable for low-qubit hardware but requires careful lag-window
selection and benefits less from quantum features at $h=1$.  The QQRBM is
limited to small dimensions and is most susceptible to barren plateaus at scale.
Given that the classical CRBM remains the strongest model in this setting, it is the practically recommended choice; if a quantum model is nevertheless desired, QCRBM is the safest option because its conservative-extension property enables it to recover classical performance (\(\alpha = 0\)) rather than reduce it, as the structurally quantum QQRBM and QFeatureQRBM do.

\paragraph{Circuit size at the selected $H^*$.}  A natural concern is that the
QCRBM might be ``near-vacuous'' at the parsimonious operating points selected in
Experiment~1 ($H^*{=}1$ on GP, $H^*{=}2$ on NARMA-10).  This is not the case,
because of amplitude encoding requires $2^{H}\ge V{+}U{=}11$, the QCRBM floors its
hidden/qubit count at $H_Q{=}4$ regardless of the classical $H^*$; at the selected
$H^*$ it therefore still instantiates a full $4$-qubit, $16$-amplitude,
$L_Q{=}2$ \texttt{ variational layers} circuit with $25$ trainable quantum
parameters ($84$ parameters in total).  The statistical indistinguishability of
QCRBM from the classical CRBM is thus \emph{not} an artefact of a trivial
circuit: a non-trivial quantum feature map is present and is amplified rather than
suppressed ($\alpha^*{=}3.0$, the largest grid value), yet on near-linear GP
and on the classically well-captured NARMA-10 memory it adds variance without
supplying useful predictive signal (cf.\ Section~\ref{sec:disc_iso}).

\subsection{NARMA-10 in Context: Published Sequence-Model Baselines}\label{sec:narma_context}

Our comparison is internal to the Boltzmann-machine family.  For external context,
Table~\ref{tab:narma_lit} places the CRBM family's NARMA-10 error against
representative \emph{published} results for dedicated sequence and reservoir
models.  These solve the \emph{same} input-driven NARMA-10 task (predicting $y(t)$
from the drive history), so the comparison is like-for-like in task, if not in
protocol: well-tuned echo-state networks reach NRMSE${<}0.01$
\citep{jaeger2004harnessing}, quantum reservoir computers report NRMSE in the
${\approx}0.04$--$0.12$ range
\citep{fujii2017reservoir,nakajima2019boosting,mujal2021opportunities,martinezpena2021dynamical},
and gated recurrent networks (LSTM \citep{hochreiter1997long}) are standard strong
baselines.  To make the metric directly comparable we also report our own
normalised error (Table~\ref{tab:narma_scalefree}): the CRBM family attains
NRMSE${\approx}0.58$
($\mathrm{NRMSE}{=}\mathrm{RMSE}/\sigma(y_\mathrm{test}^\mathrm{raw})$).  How that
figure reads depends strongly on protocol.  The near-perfect reservoir scores
often quoted arise under long washout and large training sets; a recent
controlled head-to-head on the \emph{same} task \citep{kodali2025sustainable}
instead reports NRMSE of $0.19$ (ESN), $0.49$ (QRC), $0.53$ (LSTM) and $1.05$
(a quantum LSTM) a ${\gtrsim}20$-fold spread for ESNs alone relative to
\citet{jaeger2004harnessing}.  Against that controlled baseline the best-tuned
reservoir still leads, but the CRBM family's $0.58$ sits within the band occupied
by strong recurrent and quantum sequence models rather than far behind them.  The
remaining caveats are genuine but narrow: our hard 10-lag
input window versus the unbounded fading memory of a reservoir or recurrent net,
and our shorter transient discard ($50$ vs.\ ${\sim}100$ steps;
cf.\ Section~\ref{sec:data}).  Two points stand out.  First, that a conditional
Boltzmann machine trails purpose-built recurrent and reservoir predictors is a
property of the \emph{model family} and is orthogonal to the question this paper
asks whether \emph{quantum} augmentation helps \emph{within} that family on
which the conclusion (no measurable advantage at the available sample size) is
unchanged.  Second, the same study finds its quantum LSTM to be the \emph{worst}
model despite the largest training cost, an independent echo of our within-family
finding.

\begin{table}[ht]\centering
\caption{Exp.~2 (NARMA-10) scale-free metrics for external comparability:
  normalised RMSE (NRMSE${=}$RMSE$/\sigma(y_\mathrm{test}^\mathrm{raw})$) and
  $R^2{=}1{-}\mathrm{NRMSE}^2$.  Raw-unit RMSE (Table~\ref{tab:exp2}) remains
  primary; cf.\ the GP analogue (Table~\ref{tab:segp_scalefree}).  As there,
  $R^2$ is averaged per run, so $1{-}\mathrm{NRMSE}^2$ of the mean NRMSE need not
  equal the tabulated mean $R^2$ (Jensen's inequality).}\label{tab:narma_scalefree}
\small\begin{tabular}{lcc}\toprule
Model & NRMSE & $R^2$\\\midrule
CRBM$(H^*)$   & $0.59$ & $+0.65$\\
CRBM$(3H^*)$  & $0.58$ & $+0.66$\\
QCRBM     & $0.58$ & $+0.66$\\
QFeatureQRBM  & $1.09$ & $-0.18$\\
QQRBM         & $1.23$ & $-0.51$\\
\bottomrule\end{tabular}\end{table}

\begin{table}[ht]\centering
\caption{NARMA-10 error in context (same input-driven task; protocols still
  differ in window length and transient discard).  Our CRBM family is reported as
  both raw-unit RMSE and NRMSE${=}\mathrm{RMSE}/\sigma(y_\mathrm{test}^\mathrm{raw})$
  (Table~\ref{tab:narma_scalefree}); the literature values are
  NRMSE.}\label{tab:narma_lit}
\small\begin{tabular}{lll}\toprule
Model class & Reported error & Source\\\midrule
CRBM family (this work)         & NRMSE ${\approx}0.58$ & Tables~\ref{tab:exp2},~\ref{tab:narma_scalefree}\\
Quantum reservoir computing     & NRMSE ${\approx}0.04$--$0.12$     & \citep{fujii2017reservoir,nakajima2019boosting}\\
Echo-state network & NRMSE ${<}0.01$                   & \citep{jaeger2004harnessing}\\
ESN/LSTM/QRC/QLSTM & NRMSE $0.19$/$0.53$/$0.49$/$1.05$ & \citep{kodali2025sustainable}\\
\bottomrule\end{tabular}\end{table}

\subsection{Limitations of the study}

The following limitations bound the scope of these conclusions.  Amplitude encoding
requires $O(2^n)$ gates, placing all models outside the near-term hardware
regime. So, data re-uploading \citep{perezSalinas2020} or angle encoding would be required
for hardware deployment at the moment.  Therefore, all results are from classical simulators and without
noise.  We consider only two data classes; în paricular, other nonlinear processes (Lorenz,
Mackey--Glass \citep{mackeyglass1977}) may reveal additional patterns.
Furthermore the QQRBM was evaluated at small dimensions only. Nevertheless, the results show clear indications which 
can be probed when these limitations are lifted. 

\paragraph{Objective mismatch between training and evaluation.}
A more fundamental, architectural limitation is that the quantum pathway is not
trained on the forecast objective.  Its parameters are updated through the
positive-phase \emph{soft surrogate} reconstruction loss
(Eqs.~\eqref{eq:qcrbm_qloss} and~\eqref{eq:qqrbm_qloss}), which fits the
visible hidden joint, whereas the reported metric is one-step forecast RMSE.
Experiment~8 makes the consequence explicit: with further training the
reconstruction MSE keeps falling while the forecast RMSE plateaus or rises
(Section~\ref{sec:exp8}).  The quantum models are therefore optimised for a
different criterion than the one on which they are judged, which plausibly caps
their forecasting performance; training the quantum path \emph{directly} on a
differentiable forecast loss is a concrete and promising direction for future work.

\section{Conclusions}\label{sec:conclusions}

This paper provides a unified formal treatment of four variational quantum
conditional Boltzmann architectures, with complete derivations of all model
components from energy functions and Contrastive Divergence gradients through
pathwise derivative estimators and autoregressive forecasting combined with the
most complete hyperparameter-symmetric evaluation of this model family to date
across thirteen structured experiments on two data regimes.

The primary scientific finding is that the best classical baseline continues to outperform the quantum architectures across both data regimes, at the available sample size of
$n{=}12$ paired observations (sufficient to detect only medium-to-large effects,
$d\gtrsim d_\mathrm{min}{\approx}0.89$; smaller advantages cannot be excluded). On NARMA-10, the classical CRBM$(3H^*)$
($RMSE=0.056\pm0.013$) is the single best model, statistically tied with CRBM$(H^*)$
and the hybrid QCRBM ($p_\mathrm{adj}{=}1.0$), while both fully quantum
architectures are significantly \emph{worse} (QFeatureQRBM $0.105$ and QQRBM
$0.119$, $p_\mathrm{adj}{<}0.001$).  On GP, QQRBM is significantly worse than
CRBM$(H^*)$ ($RMSE=1456\pm803$ vs.\ $RMSE=584\pm511$, $p_\mathrm{adj}{=}0.024$), and no
quantum model improves on the classical baseline.  The one quantum model that is
never worse QCRBM is precisely the one that recovers the classical CRBM at
$\alpha=0$ (the quantum correction is additive and can be zeroed out); it can
therefore match, but does not beat, the classical reference on both datasets.
These findings hold under fully symmetric hyperparameter optimisation including
the classical learning rate search (Experiment~11) and quantum CD-step and depth
searches (Experiment~12), with the quantum scale selected at its largest grid
value $\alpha^*{=}3.0$ ensuring that the result reflects architecture rather than
an under-tuned quantum pathway.  All four quantum architectures
(QCRBM, QFeatureQRBM, QQRBM, and CRBM as the classical baseline) are included in
the primary comparison (Experiment~2), with QQRBM appearing at its natural
full-register architecture size and as a structural reference in Experiment~10.

The iso-parameter matched-budget comparison (Experiment~10; four budgets
$P\in\{84,120,156,192\}$) gives a clear answer to the parameter-fairness question:
the classical CRBM holds the lowest RMSE at three of the four budgets on both
datasets, the only reversals (at the largest budget, where the classical model
over-fits) are not statistically significant, and QFeatureQRBM and QQRBM remain
far above.  No statistically significant quantum advantage is established from
iso-parameter matching at any budget tested.

The methodological finding is equally important: asymmetric hyperparameter search
can suppress detectable quantum performance.  The optimal classical learning rate
is $\eta_\mathrm{cl}^*{=}10^{-3}$ on both GP and NARMA-10 (Experiment~11).
Experiment~12 reveals that both the optimal QCRBM CD-step count and the optimal
QFeatureQRBM depth are dataset-specific ($k_Q^*{=}3$ on GP, $k_Q^*{=}1$ on
NARMA-10; $n_\mathrm{layers}^*{=}2$ on GP, $n_\mathrm{layers}^*{=}1$ on
NARMA-10), and that the validation optimum does not always coincide with the
test-RMSE minimum.
We advocate for the explicit reporting and symmetric optimisation of \emph{as many as feasible}
tunable scalars classical and quantum as a standard practice in QML benchmarking.

The trainability analysis reveals that the variational layers ansatz with
local Pauli-$Z$ observables exhibits gradient decay faster than the
$O(2^{-n/2})$ local-cost bound at $L_Q\geq3$ (fitted $\hat{b}_3{=}0.520\pm0.013$,
$\hat{b}_5{=}0.492\pm0.007$, both below $2^{-1/2}{\approx}0.707$).  The
recommended practical trainability regime is $L_Q{\leq}2$ for $n\leq8$ qubits.
The recommended final configuration for QCRBM is QCRBM with $\alpha$, $\eta_Q$,
$k_Q$, and $\eta_\mathrm{cl}$ all selected by validation RMSE on the target dataset
(e.g.\ $k_Q^*{=}3$ on GP and $k_Q^*{=}1$ on NARMA-10, with
$\eta_\mathrm{cl}^*{=}10^{-3}$ on both in the current evaluation).
Across all $n{=}12$ paired comparisons we find no statistically significant
advantage for any quantum architecture: QCRBM ties the classical CRBM, while
QQRBM is significantly \emph{worse} on both benchmarks (GP $p_\mathrm{adj}{=}0.024$;
NARMA-10 $p_\mathrm{adj}{<}0.001$).

Future work should explore different directions:  First, hardware deployment via
data re-uploading \citep{perezSalinas2020}, which reduces gate complexity from
$O(2^n)$ to $O(n)$ may five indications about practical transpilation and noise issues affect the results.
Second, the implementation of the conditional sqRBM
\citep{demidik2025sqrbm} as a time-series model, would provide a
theoretically grounded quantum-efficient alternative architecture with closed-form gradients
and the 3$\times$ capacity compression guarantee.  Third, the scope of different data sets should be enlarged
to get a better understanding how (non-)linear characteristics influence the a quantum-classical comparison.
Last not least, extending the iso-parameter analysis to the QFeatureQRBM and QQRBM families to
characterise their Pareto frontiers across a wider budget range.

\section*{Software Environment}
All experiments were executed with PennyLane~0.43.1 (with
\texttt{pennylane-lightning}~0.43.0), PyTorch~2.9.1, scikit-learn~1.8.0,
SciPy~1.16.3, statsmodels~0.14.6, and NumPy~2.3.5.

\section*{Declaration of Generative AI and AI-Assisted Technologies}
During the preparation of this work, the authors used 
Claude, Gemini and GitHub Copilot to assist with initial brainstorming, 
source code development, testing optimization, documentation generation, 
and final language polishing. Following the use of this technology, 
the authors thoroughly reviewed, verified, and edited all generated code 
and text. The authors maintain full accountability for the accuracy, 
integrity, and final content of this publication.

\section*{Acknowledgements}

\emph{This work was supported by the German Federal Ministry of Research, Technology and Space within the
funding program "Application orientied quantum computing" under Contract No. 13N17159.
}

\bibliographystyle{plainnat}
\bibliography{refs_v10}

@article{hinton2002cd,
  author  = {Hinton, Geoffrey E.},
  title   = {{Training Products of Experts by Minimizing Contrastive Divergence}},
  journal = {Neural Computation},
  volume  = {14},
  number  = {8},
  pages   = {1771--1800},
  year    = {2002},
  doi     = {10.1162/089976602760128018},
}

@techreport{hinton2010guide,
  author      = {Hinton, Geoffrey E.},
  title       = {{A Practical Guide to Training Restricted Boltzmann Machines}},
  institution = {University of Toronto},
  number      = {UTML TR 2010-003},
  year        = {2010},
}

@inproceedings{taylor2009fcrbm,
  author    = {Taylor, Graham W. and Hinton, Geoffrey E.},
  title     = {{Factored Conditional Restricted Boltzmann Machines for Modeling Motion Style}},
  booktitle = {Proceedings of the 26th International Conference on Machine Learning (ICML)},
  year      = {2009},
}

@inproceedings{sutskever2008rtrbm,
  author    = {Sutskever, Ilya and Hinton, Geoffrey E. and Taylor, Graham W.},
  title     = {{The Recurrent Temporal Restricted Boltzmann Machine}},
  booktitle = {Advances in Neural Information Processing Systems (NeurIPS)},
  volume    = {21},
  year      = {2008},
}

@inproceedings{carreira2005contrastive,
  author    = {Carreira-Perpi{\~n}{\'a}n, Miguel {\'A}ngel and Hinton, Geoffrey E.},
  title     = {{On Contrastive Divergence Learning}},
  booktitle = {Proceedings of the 10th International Workshop on Artificial Intelligence
               and Statistics (AISTATS)},
  year      = {2005},
}

@inproceedings{tieleman2008pcd,
 author = {Tieleman, Tijmen},
title = {Training restricted Boltzmann machines using approximations to the likelihood gradient},
year = {2008},
isbn = {9781605582054},
publisher = {Association for Computing Machinery},
address = {New York, NY, USA},
url = {https://doi.org/10.1145/1390156.1390290},
doi = {10.1145/1390156.1390290},
booktitle = {Proceedings of the 25th International Conference on Machine Learning},
pages = {1064–1071},
numpages = {8},
location = {Helsinki, Finland},
series = {ICML '08}
}

@inproceedings{glorot2010,
  author    = {Glorot, Xavier and Bengio, Yoshua},
  title     = {{Understanding the Difficulty of Training Deep Feedforward Neural Networks}},
  booktitle = {Proceedings of the 13th International Conference on Artificial Intelligence
               and Statistics (AISTATS)},
  pages     = {249--256},
  year      = {2010},
}

@article{aminQBM2018,
    title = {Quantum Boltzmann Machine},
  author = {Amin, Mohammad H. and Andriyash, Evgeny and Rolfe, Jason and Kulchytskyy, Bohdan and Melko, Roger},
  journal = {Phys. Rev. X},
  volume = {8},
  issue = {2},
  pages = {021050},
  numpages = {11},
  year = {2018},
  month = {May},
  publisher = {American Physical Society},
  doi = {10.1103/PhysRevX.8.021050},
  url = {https://link.aps.org/doi/10.1103/PhysRevX.8.021050}
}

@article{benedetti2017qhm,
  author  = {Benedetti, Marcello and Realpe-G{\'o}mez, John
             and Perdomo-Ortiz, Alejandro},
  title   = {{Quantum-Assisted Helmholtz Machines: A Quantum--Classical
               Deep Learning Framework for Industrial Datasets in Near-Term Devices}},
  journal = {Quantum Science and Technology},
  volume  = {3},
  number  = {3},
  pages   = {034007},
  year    = {2018},
  doi     = {10.1088/2058-9565/aabd98},
}

@article{zoufalVQBM2021,
  author  = {Zoufal, Christa and Lucchi, Aurelien and Woerner, Stefan},
  title   = {{Variational Quantum Boltzmann Machines}},
  journal = {Quantum Machine Intelligence},
  volume  = {3},
  pages   = {7},
  year    = {2021},
  doi     = {10.1007/s42484-020-00033-7},
}

@article{paramshift2019,
  title = {Evaluating analytic gradients on quantum hardware},
  author = {Schuld, Maria and Bergholm, Ville and Gogolin, Christian and Izaac, Josh and Killoran, Nathan},
  journal = {Phys. Rev. A},
  volume = {99},
  issue = {3},
  pages = {032331},
  numpages = {7},
  year = {2019},
  month = {Mar},
  publisher = {American Physical Society},
  doi = {10.1103/PhysRevA.99.032331},
  url = {https://link.aps.org/doi/10.1103/PhysRevA.99.032331}
}

@article{mcclean2018barren,
  author  = {McClean, Jarrod R. and Boixo, Sergio and Smelyanskiy, Vadim N.
             and Babbush, Ryan and Neven, Hartmut},
  title   = {{Barren Plateaus in Quantum Neural Network Training Landscapes}},
  journal = {Nature Communications},
  volume  = {9},
  pages   = {4812},
  year    = {2018},
  doi     = {10.1038/s41467-018-07090-4},
}

@article{cerezo2021cost,
  author  = {Cerezo, Marco and Sone, Akira and Volkoff, Tyler and
             Cincio, Lukasz and Coles, Patrick J.},
  title   = {{Cost Function Dependent Barren Plateaus in Shallow Parametrized
               Quantum Circuits}},
  journal = {Nature Communications},
  volume  = {12},
  pages   = {1791},
  year    = {2021},
  doi     = {10.1038/s41467-021-21728-w},
}

@article{cerezo2021vqa,
  author  = {Cerezo, Marco and Arrasmith, Andrew and Babbush, Ryan and
             Benjamin, Simon C. and Endo, Suguru and Fujii, Keisuke and
             McClean, Jarrod R. and Mitarai, Kosuke and Yuan, Xiao and
             Cincio, Lukasz and Coles, Patrick J.},
  title   = {{Variational Quantum Algorithms}},
  journal = {Nature Reviews Physics},
  volume  = {3},
  pages   = {625--644},
  year    = {2021},
  doi     = {10.1038/s42254-021-00348-9},
}

@article{perezSalinas2020,
 doi = {10.22331/q-2020-02-06-226},
  url = {https://doi.org/10.22331/q-2020-02-06-226},
  title = {Data re-uploading for a universal quantum classifier},
  author = {P{\'{e}}rez-Salinas, Adri{\'{a}}n and Cervera-Lierta, Alba and Gil-Fuster, Elies and Latorre, Jos{\'{e}} I.},
  journal = {{Quantum}},
  issn = {2521-327X},
  publisher = {{Verein zur F{\"{o}}rderung des Open Access Publizierens in den Quantenwissenschaften}},
  volume = {4},
  pages = {226},
  month = feb,
  year = {2020}
}

@article{pennylane2018,
  title={PennyLane: Automatic differentiation of hybrid quantum-classical computations},
  author={Bergholm, Ville and Izaac, Josh and Schuld, Maria and Gogolin, Christian and Ahmed, Shahnawaz and Ajith, Vishnu and Alam, M Sohaib and Alonso-Linaje, Guillermo and AkashNarayanan, Bharath and Asadi, Ali and others},
  journal={arXiv preprint arXiv:1811.04968},
  year={2018}
}

@misc{plAmplitude,
  author = {{PennyLane Development Team}},
  title  = {\texttt{qml.AmplitudeEmbedding}},
  year   = {2024},
  url    = {https://docs.pennylane.ai/en/stable/code/api/pennylane.AmplitudeEmbedding.html},
}

@misc{plSEL,
  author = {{PennyLane Development Team}},
  title  = {\texttt{qml.StronglyEntanglingLayers}},
  year   = {2024},
  url    = {https://docs.pennylane.ai/en/stable/code/api/pennylane.StronglyEntanglingLayers.html},
}

@article{mottonen2005,
  author  = {M{\"o}tt{\"o}nen, Mikko and Vartiainen, Juha J. and Bergholm, Ville
             and Salomaa, Martti M.},
  title   = {{Transformation of Quantum States Using Uniformly Controlled Rotations}},
  journal = {Quantum Information and Computation},
  volume  = {5},
  number  = {6},
  pages   = {467--473},
  year    = {2005},
}

@article{plesch2011,
  author  = {Plesch, Martin and Brukner, {\v{C}}aslav},
  title   = {{Quantum-State Preparation with Universal Gate Decompositions}},
  journal = {Physical Review A},
  volume  = {83},
  pages   = {032302},
  year    = {2011},
  doi     = {10.1103/PhysRevA.83.032302},
}

@article{larocca2024review,
author={Larocca, Mart{\'i}n
and Thanasilp, Supanut
and Wang, Samson
and Sharma, Kunal
and Biamonte, Jacob
and Coles, Patrick J.
and Cincio, Lukasz
and McClean, Jarrod R.
and Holmes, Zo{\"e}
and Cerezo, M.},
title={Barren plateaus in variational quantum computing},
journal={Nature Reviews Physics},
year={2025},
month={Apr},
day={01},
volume={7},
number={4},
pages={174-189},
issn={2522-5820},
doi={10.1038/s42254-025-00813-9},
url={https://doi.org/10.1038/s42254-025-00813-9}
}

@article{mitarai2018qnn,
  title = {Quantum circuit learning},
  author = {Mitarai, K. and Negoro, M. and Kitagawa, M. and Fujii, K.},
  journal = {Phys. Rev. A},
  volume = {98},
  issue = {3},
  pages = {032309},
  numpages = {6},
  year = {2018},
  month = {Sep},
  publisher = {American Physical Society},
  doi = {10.1103/PhysRevA.98.032309},
  url = {https://link.aps.org/doi/10.1103/PhysRevA.98.032309}
}

@misc{bowles2024benchmarking,
  author        = {Bowles, Joseph and Ahmed, Shahnawaz and Schuld, Maria},
  title         = {{Better than Classical? The Subtle Art of Benchmarking
                    Quantum Machine Learning Models}},
  year          = {2024},
  eprint        = {2403.07059},
  archivePrefix = {arXiv},
  primaryClass  = {quant-ph},
  doi           = {10.48550/arXiv.2403.07059},
}

@article{abbas2021power,
  title={The power of quantum neural networks},
  author={Abbas, Amira and Sutter, David and Zoufal, Christa and Lucchi, Aur{\'e}lien and Figalli, Alessio and Woerner, Stefan},
  journal={Nature computational science},
  volume={1},
  number={6},
  pages={403--409},
  year={2021},
  doi={10.1038/s43588-021-00084-1},
  publisher={Nature Publishing Group US New York}
}

@article{sim2019expressibility,
  author  = {Sim, Sukin and Johnson, Peter D. and Aspuru-Guzik, Al{\'a}n},
  title   = {{Expressibility and Entangling Capability of Parameterized
               Quantum Circuits for Hybrid Quantum-Classical Algorithms}},
  journal = {Advanced Quantum Technologies},
  volume  = {2},
  pages   = {1900070},
  year    = {2019},
  doi     = {10.1002/qute.201900070},
}

@misc{chittoor2024qultsf,
  author        = {Chittoor, Hari Hara Suthan and Griffin, Paul Robert and Neufeld, Ariel and Thompson, Jayne and Gu, Mile},
  title         = {{QuLTSF: Long-Term Time Series Forecasting with
                    Quantum Machine Learning}},
  year          = {2024},
  eprint        = {2412.13769v2},
  archivePrefix = {arXiv},
  primaryClass  = {quant-ph},
  doi           = {10.48550/arXiv.2412.13769},
}

@article{crawford2018rlqbm,
author = {Crawford, Daniel and Levit, Anna and Ghadermarzy, Navid and Oberoi, Jaspreet S. and Ronagh, Pooya},
title = {Reinforcement learning using quantum boltzmann machines},
year = {2018},
issue_date = {February 2018},
publisher = {Rinton Press, Incorporated},
address = {Paramus, NJ},
volume = {18},
number = {1–2},
issn = {1533-7146},
journal = {Quantum Info. Comput.},
month = feb,
pages = {51–74},
numpages = {24},
}

@article{kieferova2017tomography,
  title = {Tomography and generative training with quantum Boltzmann machines},
  author = {Kieferov\'a, M\'aria and Wiebe, Nathan},
  journal = {Phys. Rev. A},
  volume = {96},
  issue = {6},
  pages = {062327},
  numpages = {13},
  year = {2017},
  month = {Dec},
  publisher = {American Physical Society},
  doi = {10.1103/PhysRevA.96.062327},
  url = {https://link.aps.org/doi/10.1103/PhysRevA.96.062327}
}

@article{holmes2022connecting,
  author  = {Holmes, Zo{\"e} and Sharma, Kunal and Cerezo, Marco and
             Coles, Patrick J.},
  title   = {{Connecting Ansatz Expressibility to Gradient Magnitudes and
               Barren Plateaus}},
  journal = {PRX Quantum},
  volume  = {3},
  pages   = {010313},
  year    = {2022},
  doi     = {10.1103/PRXQuantum.3.010313},
}

@article{demidik2025sqrbm,
  author  = {Demidik, Maria and T{\"u}ys{\"u}z, Cenk and Piatkowski, Nico and
             Grossi, Michele and Jansen, Karl},
  title   = {{Expressive Equivalence of Classical and Quantum Restricted Boltzmann Machines}},
  journal = {Communications Physics},
  volume  = {8},
  pages   = {413},
  year    = {2025},
  doi     = {10.1038/s42005-025-02353-1},
}

@article{coopmans2024sample,
  author  = {Coopmans, Luuk and Benedetti, Marcello},
  title   = {{On the Sample Complexity of Quantum Boltzmann Machine Learning}},
  journal = {Communications Physics},
  volume  = {7},
  pages   = {274},
  year    = {2024},
  doi     = {10.1038/s42005-024-01763-x},
}

@article{le_roux2008representational,
  author  = {{Le Roux}, Nicolas and Bengio, Yoshua},
  title   = {{Representational Power of Restricted Boltzmann Machines and Deep
               Belief Networks}},
  journal = {Neural Computation},
  volume  = {20},
  pages   = {1631--1649},
  year    = {2008},
  doi     = {10.1162/neco.2008.04-07-510},
}

@article{atiya2000narma,
  author  = {Atiya, Amir F. and Parlos, Alexander G.},
  title   = {{New Results on Recurrent Network Training: Unifying the Algorithms
               and Accelerating Convergence}},
  journal = {IEEE Transactions on Neural Networks},
  volume  = {11},
  pages   = {697--709},
  year    = {2000},
  doi     = {10.1109/72.846741},
}

@article{fujii2017reservoir,
  author  = {Fujii, Keisuke and Nakajima, Kohei},
  title   = {{Harnessing Disordered-Ensemble Quantum Dynamics for Machine Learning}},
  journal = {Physical Review Applied},
  volume  = {8},
  pages   = {024030},
  year    = {2017},
  doi     = {10.1103/PhysRevApplied.8.024030},
}

@article{holm1979simple,
  author  = {Holm, Sture},
  title   = {{A Simple Sequentially Rejective Multiple Test Procedure}},
  journal = {Scandinavian Journal of Statistics},
  volume  = {6},
  pages   = {65--70},
  year    = {1979},
}

@article{mackeyglass1977,
  author  = {Mackey, Michael C. and Glass, Leon},
  title   = {{Oscillation and Chaos in Physiological Control Systems}},
  journal = {Science},
  volume  = {197},
  pages   = {287--289},
  year    = {1977},
  doi     = {10.1126/science.267326},
}

@misc{kingma2015adam,
  author        = {Kingma, Diederik P. and Ba, Jimmy},
  title         = {{Adam: A Method for Stochastic Optimization}},
  year          = {2015},
  eprint        = {1412.6980},
  archivePrefix = {arXiv},
  primaryClass  = {cs.LG},
}

@misc{hinton2015distilling,
  title={Distilling the Knowledge in a Neural Network},
  author={Geoffrey E. Hinton and Oriol Vinyals and Jeffrey Dean},
  year          = {2015},
  eprint        = {1503.02531},
  archivePrefix = {arXiv},
  primaryClass  = {stat.ML},
}

@article{havlicek2019supervised,
  author  = {Havl{\'\i}{\v{c}}ek, Vojt{\v{e}}ch and C{\'o}rcoles, Antonio D. and
             Temme, Kristan and Harrow, Aram W. and Kandala, Abhinav and
             Chow, Jerry M. and Gambetta, Jay M.},
  title   = {{Supervised learning with quantum-enhanced feature spaces}},
  journal = {Nature},
  volume  = {567},
  number  = {7747},
  pages   = {209--212},
  year    = {2019},
  doi     = {10.1038/s41586-019-0980-2},
}

@book{schuld2021machine,
  author    = {Schuld, Maria and Petruccione, Francesco},
  title     = {{Machine Learning with Quantum Computers}},
  publisher = {Springer},
  series    = {Quantum Science and Technology},
  year      = {2021},
  doi       = {10.1007/978-3-030-83098-4},
}

@inproceedings{maddison2017concrete,
  author    = {Maddison, Chris J. and Mnih, Andriy and Teh, Yee Whye},
  title     = {{The Concrete Distribution: A Continuous Relaxation of Discrete
               Random Variables}},
  booktitle = {International Conference on Learning Representations (ICLR)},
  year      = {2017},
}

@book{cohen1988statistical,
  author    = {Cohen, Jacob},
  title     = {{Statistical Power Analysis for the Behavioral Sciences}},
  edition   = {2nd},
  publisher = {Lawrence Erlbaum Associates},
  year      = {1988},
}

@article{sweke2020stochastic,
  author  = {Sweke, Ryan and Wilde, Frederik and Meyer, Johannes Jakob and
             Schuld, Maria and F{\"a}hrmann, Paul K. and
             Meynard-Piganeau, Barth{\'e}l{\'e}my and Eisert, Jens},
  title   = {{Stochastic gradient descent for hybrid quantum-classical
               optimization}},
  journal = {Quantum},
  volume  = {4},
  pages   = {314},
  year    = {2020},
  doi     = {10.22331/q-2020-08-31-314},
}

@article{arrasmith2022equivalence,
  author  = {Arrasmith, Andrew and Holmes, Zo{\"e} and Cerezo, M. and Coles, Patrick J.},
  title   = {{Equivalence of quantum barren plateaus to cost concentration and narrow gorges}},
  journal = {Quantum Science and Technology},
  volume  = {7},
  number  = {4},
  pages   = {045015},
  year    = {2022},
  doi     = {10.1088/2058-9565/ac7d06},
}

@article{nakajima2019boosting,
  author  = {Nakajima, Kohei and Fujii, Keisuke and Negoro, Makoto and
             Mitarai, Kosuke and Kitagawa, Masahiro},
  title   = {{Boosting Computational Power through Spatial Multiplexing in Quantum Reservoir Computing}},
  journal = {Physical Review Applied},
  volume  = {11},
  number  = {3},
  pages   = {034021},
  year    = {2019},
  doi     = {10.1103/PhysRevApplied.11.034021},
}

@article{martinezpena2021dynamical,
  author  = {Mart{\'\i}nez-Pe{\~n}a, Rodrigo and Giorgi, Gian Luca and
             Nokkala, Johannes and Soriano, Miguel C. and Zambrini, Roberta},
  title   = {{Dynamical Phase Transitions in Quantum Reservoir Computing}},
  journal = {Physical Review Letters},
  volume  = {127},
  number  = {10},
  pages   = {100502},
  year    = {2021},
  doi     = {10.1103/PhysRevLett.127.100502},
}

@article{mujal2021opportunities,
  author  = {Mujal, Pere and Mart{\'\i}nez-Pe{\~n}a, Rodrigo and Nokkala, Johannes and
             Garc{\'\i}a-Beni, Jorge and Giorgi, Gian Luca and Soriano, Miguel C. and
             Zambrini, Roberta},
  title   = {{Opportunities in Quantum Reservoir Computing and Extreme Learning Machines}},
  journal = {Advanced Quantum Technologies},
  volume  = {4},
  number  = {8},
  pages   = {2100027},
  year    = {2021},
  doi     = {10.1002/qute.202100027},
}

@article{jaeger2004harnessing,
  author  = {Jaeger, Herbert and Haas, Harald},
  title   = {{Harnessing Nonlinearity: Predicting Chaotic Systems and Saving Energy in Wireless Communication}},
  journal = {Science},
  volume  = {304},
  number  = {5667},
  pages   = {78--80},
  year    = {2004},
  doi     = {10.1126/science.1091277},
}

@article{hochreiter1997long,
  author  = {Hochreiter, Sepp and Schmidhuber, J{\"u}rgen},
  title   = {{Long Short-Term Memory}},
  journal = {Neural Computation},
  volume  = {9},
  number  = {8},
  pages   = {1735--1780},
  year    = {1997},
  doi     = {10.1162/neco.1997.9.8.1735},
}

@misc{adachi2015application,
  author        = {Adachi, Steven H. and Henderson, Maxwell P.},
  title         = {{Application of Quantum Annealing to Training of Deep Neural Networks}},
  year          = {2015},
  eprint        = {1510.06356},
  archivePrefix = {arXiv},
  primaryClass  = {quant-ph},
}

@book{RasmussenWilliams2006,
  author    = {Carl Edward Rasmussen and Christopher K. I. Williams},
  title     = {Gaussian Processes for Machine Learning},
  publisher = {MIT Press},
  year      = {2006},
  url       = {https://gaussianprocess.org/gpml/}
}

@inproceedings{GardnerEtAl2018,
  author    = {Jacob R. Gardner and Geoff Pleiss and David Bindel and Kilian Q. Weinberger and Andrew Gordon Wilson},
  title     = {GPyTorch: Blackbox Matrix-Matrix Gaussian Process Inference with GPU Acceleration},
  booktitle = {Advances in Neural Information Processing Systems},
  year      = {2018},
  url       = {https://papers.nips.cc/paper_files/paper/2018/hash/27e8e17134dd7083b050476733207ea1-Abstract.html}
}

@article{Rabiner1989,
  author  = {Lawrence R. Rabiner},
  title   = {A Tutorial on Hidden Markov Models and Selected Applications in Speech Recognition},
  journal = {Proceedings of the IEEE},
  volume  = {77},
  number  = {2},
  pages   = {257--286},
  year    = {1989},
  doi     = {10.1109/5.18626}
}

@article{BaumEtAl1970,
  author  = {Leonard E. Baum and Ted Petrie and George Soules and Norman Weiss},
  title   = {A Maximization Technique Occurring in the Statistical Analysis of Probabilistic Functions of Markov Chains},
  journal = {The Annals of Mathematical Statistics},
  volume  = {41},
  number  = {1},
  pages   = {164--171},
  year    = {1970},
  doi     = {10.1214/aoms/1177697196}
}

@misc{kodali2025sustainable,
      title={Sustainable NARMA-10 Benchmarking for Quantum Reservoir Computing}, 
      author={Avyay Kodali and Priyanshi Singh and Pranay Pandey and Krishna Bhatia and Shalini Devendrababu and Srinjoy Ganguly},
      year={2025},
      eprint={2510.25183},
      archivePrefix={arXiv},
      primaryClass={quant-ph},
      url={https://arxiv.org/abs/2510.25183}, 
}

\end{document}